\documentclass[12pt,review,doublespacing]{elsarticle}
\usepackage[left=1in,right=1in,top=1in,bottom=1in]{geometry}
\journal{...}
\emergencystretch=\maxdimen
\fontfamily{cmss}\selectfont
\usepackage[scaled=0.92]{helvet}
\usepackage{hyperref}
 \hypersetup{
 	plainpages = false,
	bookmarksopen = true,
	bookmarksnumbered = true,
	breaklinks = true,
	linktocpage,
	colorlinks = true,
	linkcolor = purple,
	urlcolor  = black,
	citecolor = purple,
	}
\usepackage{multicol}
\usepackage{multirow}
\usepackage{imakeidx}
\usepackage{nomencl}
\makenomenclature
\usepackage[xindy]{glossaries}
\makeglossaries
\usepackage{graphics}
\usepackage{graphicx}
\usepackage{epsf,psfrag}
\usepackage{epstopdf}
\usepackage[para]{threeparttable}
\usepackage{subfig}
\usepackage{epsfig}
\usepackage{pdflscape}
\usepackage{rotating}
\usepackage{lscape}
\usepackage{float}
\usepackage{stfloats}
\usepackage{textgreek}
\usepackage{longtable}
\usepackage{lscape}
\usepackage{capt-of}
\usepackage{comment}
\usepackage{tablefootnote}
\usepackage{footnote}
\usepackage{caption}
\usepackage[autopunct=true]{csquotes}
\usepackage{rotating}
\usepackage{txfonts}
\usepackage[normalem]{ulem}
\usepackage{setspace}
\usepackage{color,xcolor,soul}
\usepackage{amsfonts,amsmath,amssymb}
\usepackage{latexsym,array}
\usepackage{textcomp}
\usepackage{mathtools}
\usepackage{tikz}
\usetikzlibrary{shapes,arrows}

\newcommand{\RomanNumeralCaps}[1]
\linenumbers
\makesavenoteenv{tabular}
\makesavenoteenv{table}
\emergencystretch=\maxdimen
\usepackage[left,displaymath, mathlines]{lineno}
 
\DeclareMathAlphabet{\mathpzc}{OT1}{pzc}{m}{it}
\def\fig{Figure~}
\def\figs{Figures~}
\def\eqn{Eq.~}
\def\eqns{Eqs.~}

\def\micro{\textmu}
\providecommand\bnabla{\boldsymbol{\nabla}}
\providecommand\p{{\partial}}

\newcommand\cac{Ca_{\text{c}}}
\newcommand\qr{Q_{\text{r}}}
\newcommand\qc{Q_{\text{c}}}
\newcommand\qd{Q_{\text{d}}}
\newcommand\wrr{w_{\text{r}}}
\newcommand\wc{w_{\text{c}}}
\newcommand\wdp{w_{\text{d}}}
\newcommand\rhor{\rho_{\text{r}}}
\newcommand\rhoc{\rho_{\text{c}}}
\newcommand\rhod{\rho_{\text{d}}}
\newcommand\mur{\mu_{\text{r}}}
\newcommand\muc{\mu_{\text{c}}}
\newcommand\mud{\mu_{\text{d}}}

\newcommand\uc{u_{\text{c}}}

\newcommand\rc{R_{\text{c}}}
\newcommand\rcmin{R_{\text{c,min}}}

\def\tsc#1{\csdef{#1}{\textsc{\lowercase{#1}}\xspace}}
\tsc{WGM}
\tsc{QE}
\tsc{EP}
\tsc{PMS}
\tsc{BEC}
\tsc{DE}
\newcommand{\rev}[1]{\textcolor{black}{#1}}

\begin{document}
%
%
%
\setcounter{page}{1}
\begin{frontmatter} 
%
%
%
%
%
\title{\textcolor{blue}{Computational analysis of interface evolution and droplet pinch-off mechanism in two-phase liquid flow through T-junction microfluidic system}}
\author[labela]{Akepogu Venkateshwarlu}\ead{avenkateshwarlu@ch.iitr.ac.in}
\author[labela]{Ram Prakash Bharti\corref{coradd}}\ead{rpbharti@iitr.ac.in}
\address[labela]{Complex Fluid Dynamics and Microfluidics (CFDM) Lab, Department of Chemical Engineering, Indian Institute of Technology Roorkee, Roorkee - 247667, Uttarakhand, INDIA}
\cortext[coradd]{\textit{Corresponding author.}}
%
\begin{abstract}
This work has explored interface evolution and pinch-off mechanism of the droplet formation in two-phase flow through cross-flow microfluidic device. The two-dimensional mathematical model equations have been solved using the finite element method under the squeezing regime ($\cac < 10^{-2}$) for wide range of flow rates ($0.1\leq \qr \leq 10$) and fixed contact angle ($\theta=135^o$). The droplet formation process has been classified into various instantaneous stages as initial, filling, squeezing, pinch-off and stable droplet through microscopic visualization of interface evolution in phase profiles. The dynamics of interface, and point pressure in both phases is further gained and discussed. Maximum pressure in the continuous phase varied linearly with $\qr$. 
The droplet pinch-off mechanism has been thoroughly elucidated by determining the local radius of the curvature ($\rcmin$) and neck width ($2r$) during the squeezing and pinch-off stages. At the pinch-off point, both $\rcmin$ and $2r$ are non-linearly related to $\qr$. 
Further, the topological dynamics of interface has been explored by analyzing the Laplace pressure ($p_{\text{L}}$), acting on the interface curvature, evaluated using (a) pressure sensors in both phases, $p_{\text{L}} =p_{\text{dp}}-p_{\text{cp}}$, (b) local radius of curvature $p_{\text{L}} = \sigma \left(1/R_{\text{f}}+ 1/R_{\text{r}} \right)$, and (c) minimum radius of curvature, $p_{\text{L}} =\sigma \left( 1/{R_{\text {c,min}}}\right)$.
The insights obtained from the present work can reliably be used in designing the model and prototypes of microfluidic devices \rev{for  generating monodispersed droplets in emulsions,} and the droplet breakup mechanism  would help accurate prediction of the pinch-off moment. \rev{The proposed knowledge provides detailed insights of the interface evolution and droplet pinch-off to a precision of 10 $\mu$s and resolution of 10  $\mu$m, equivalent to experimental flow visualization with a high-speed ($10^{5}$ fps) and high-resolution (10  $\mu$m pixel size) camera.}
\end{abstract}
\end{frontmatter}
%
\section{Introduction}
%
\label{sec:intro}
\noindent
Microfluidics is a promising area in the cutting-edge research world and {has} drawn significant attention due to its wide-ranging potential applications in various fields like the food industry, biomedical and drug synthesis, single-cell analysis, and inkjet printing. Microfluidic technology has become a useful platform to investigate interactions between the cells and tissues, drugs formulation, biological organs, biodefence, drug delivery, biomedical diagnosis and chemical analysis \citep{Barnes1994,Whitesides2001,MAAN2011334,Mansard2016,Kaminski2017,KULJU2018,Gerecsei2020,RABIEE2021}.  Enormous progress has been made in developing technologies to miniaturize conventional complex processes and integrate many operations/procedures into tiny micro-electro-mechanical systems (MEMS). These devices are often referred to as lab-on-a-chip (LOC), microfluidic chip, or micro total analysis system ($\mu$TAS). Their main advantages include reduced space, ease of handling and analysis, reduced sample volume, faster analysis, increased rate of transport processes, and cost-effective \citep{rohtua,Bharti20083593,Pamela2013,Shang2017}.

\noindent
An emulsion consists of a mixture of two immiscible fluids where one fluid is dispersed in the form of tiny droplets in another fluid acting as a continuous phase.  Emulsions of lower polydispersity produce a better quality of materials in specific critical technological systems and wherein the orderly structure is highly desired \citep{Umbanhowar2000,MAAN2011334,NISHIMURA2012,Azarmanesh2019,Kang2019}.
For instance, the product quality in the pharmaceutical industry is determined in terms of better drug release maintaining uniform size and stability for an extended time \citep{Liu2020BioTechniques,Sugiura2002}. The production of monodispersed droplets for a sufficiently extended time is essential for many industrial processes and applications. The liquid droplets or emulsions are commonly produced using microfluidic devices having co-flow, cross-flow, or flow focusing arrangements for the two immiscible fluids phases \citep{Whitesides2006,Ulmeanu2008,Hashimoto2010,ABEDI2019}.
The cross-flow microfluidic devices are widely used in droplets generation due to their simplicity, ease of control and manipulation of emulsion hydrodynamics \citep{DELUCA2006,Xu2006,Zhu2017}. The dispersity, frequency and regime of the formation of the droplets depend on several parameters like device geometry, flow rates, viscosity, density, interfacial tension, and contact angle \citep{McClements2004,Crawford2017,YI202052,venkat2021,YI202159}.

\noindent
Excellent review articles have summarized the voluminous research efforts devoted to understanding the dynamics and control mechanisms of droplet formation at the micro-scale level using both experimental and numerical approaches {\citep[e.g.,][etc.]{Pit2015,Anna2016,Chong2016,Martino2016,Zhu2017,Doufene2019,Khojasteh2019,Sohrabi2020,Cai2021,Han2021,Lashkaripour2021,Roy2021,WuAPR2021,Xia2021,venkat2022a,Dhondi2022,Shuvam2022}.}
Since the detailed literature related to the two-phase flow through T-junction cross-flow microfluidic device has been reviewed in our recent study \citep{venkat2021}, only the salient literature is summarized here to avoid replication. For instance,
\citet{Thorsen2001} first introduced an idea of using a T-junction microchannel system to generate monodispersed water droplets in a water-oil two-phase system and reported the dynamic patterns of the droplets at low Reynolds number. Subsequently, various efforts are made to understand the hydrodynamics, control, and manipulation of droplet generation. The two-phase flow regimes are further suitably defined as a function of flow governing parameters such as device type and size, viscosity ($\mu$), density ($\rho$), flow rate ($Q$), interfacial tension ($\sigma$), and contact angle ($\theta$).

\noindent
Broadly, the droplet formation is governed by the relative forces acting on the interface between the two immiscible liquid phases \citep{Anna2016,Roy2021}. At the microscale level, the interfacial tension (IFT), density, viscosity, and flow rate are critical flow controlling variables. The balance of the forces attributed to these flow variables determine liquid-liquid interface stability, droplet formation, and flow regimes. The relative influences of the forces acting on the interface are generally expressed with the relevant dimensionless number\rev{s} such as capillary number ($Ca={F}_{\text{v}}/{F}_{\sigma}$), Reynolds number ($Re={F}_{\text{i}}/{F}_{\text{v}}$), and flow rate  ratio ($\qr$). Here ${F}_{\text{i}}$, ${F}_{\text{v}}$ and ${F}_{\sigma}$ are the magnitudes of the inertial, viscous and interfacial tension forces, respectively.
A recent study \citep{Svetlov2021} \rev{has} modelled droplet formation in a microfluidic system based on the force balance method and axisymmetric formulation of Navier-Stokes equations. Their droplet velocity profiles are influenced by a flow rate of the continuous phase ($\qc$), and the flow regimes are defined by using the viscosity of the dispersed phase  ($\mud$).
The two-phase flow is recently \citep{venkat2021} characterized into a droplet and non-droplet regimes for wide ranges of conditions.
Further, several two-phase flow regimes like squeezing, dripping, jetting, and parallel flows are observed and reported in literature \citep{Garstecki2006,Demenech2008,Christopher2008,Bashir2011,venkat2021} for T-junction microchannel operated at varied ranges of $Ca$ and $\qr$, under otherwise identical conditions.
The major difference between these flow regimes is the degree of confinement of the droplet during the formation \citep{Volkert2009,Glawdel2012a}.
While the confinement effects play a vital role in the droplet formation under squeezing regime ($\cac < 10^{-2}$), the interplay of interfacial tension force (pressure-driven) and the viscous force (shear-driven) govern the droplet formation under the dripping regime ($\cac > 10^{-2}$).
The present study is focused on the squeezing flow ($\cac < 10^{-2}$) regime wherein the highly desired monodispersed droplets are generated with precisely controlled size and frequency. The length ($L$) of the generated plug-type droplets, in this regime, is linearly dependent on channel width ($\wc$) and flow rates of the continuous and dispersed phases \citep{Garstecki2006, Gupta2010, VanSteijn2010,Liu2011b, Liu2011a,  venkat2021} as $L/\wc =(\alpha + \beta\qr)$.
\rev{The junction angle between the main and vertical channels significantly affects the droplet size and channel confinement \citep{Garstecki2006, VanSteijn2007, Glawdel2012a}. For instance, the droplet achieves maximum size when the junction angle is $90^{\circ}$ \citep{Jamalabadi2017, Li2019,Schroen2020}. The confinement effects on the droplet formation are also studied \citep{Garstecki2005,Leshansky2009} in-depth by varying the channel widths ($\wrr$). The droplet size is independent of the flow rate ratio ($\qr$) when the width ratio is small ($\wrr<1$), indicating that the squeezing pressure developed upstream is insignificant. However, the squeezing pressure becomes extremely important when the width ratio is large ($\wrr>1$) \citep{Garstecki2006,VanSteijn2010,Glawdel2012a}.}
	
\noindent Understanding the breakup mechanism is essential for accurate prediction of the moment of pinch-off during the droplet formation.
The droplet formation consists of filling, squeezing (or necking), and breakup stages. In the filling stage, the dispersed phase enters into the main channel and, as time progresses, reaches close to the top wall of the channel. Subsequently, the gap between the wall and the dispersed phase becomes minimum. Eventually, it restricts the flow of the continuous phase. Further, the dispersed phase starts experiencing the shearing by the continuous phase either through the pressure developed in the upstream region or viscous shear. It is called the squeezing or necking stage \citep{Volkert2009,Pang2020,venkat2021}. In the squeezing stage, the pressure developed upstream by the continuous phase accelerates the dispersed phase to push downstream. Subsequently, the squeezing stage is followed by the pinch-off and breakup stage, wherein the interface neck collapses and splits into two domains. The pinch-off of the droplet is delayed until the radius of the interface curvature becomes negligibly small \citep{CHEN2021}. Thus, the evolution of the interface shape and curvature is the key during droplet formation.
Recently, \citet{Volkert2009,VanSteijn2010} have conducted experimental and theoretical modeling under the squeezing regime ($\cac < 10^{-2}$) to predict the moment of pinch-off of gas bubble based on the interface curvature evolution and geometric description of the neck during the collapse. \citet{Glawdel2012a} have developed a model \rev{for liquid-liquid system} to predict the necking time and reported another additional stage, namely, the lag stage, during the droplet formation when the droplet formation cycle repeats.

\noindent
Further, pressure is an easily measurable quantity, amongst all forces responsible, during the cyclic process of continuous droplet formation. It is, therefore, essential to understanding the synchronization between different stages of the droplet formation process and the corresponding local pressure fluctuations \citep{Cristini2004,Xu_Ke_Tostado}. The pressure fluctuations can be detected by  (a) installing the sensors, (b) utilizing the force balance between the radius of the interface curvature, (c) the differential pressure across the interface, and (d) the interfacial tension as described by Young - Laplace equation \citep{Garstecki2006,Abate2012,Bashir2014,LECACHEUX2022}. The in-depth knowledge still lacks to understand the physics of interface evolution during droplet formation for a wide range of the parameters \citep{VanSteijn2007,Volkert2009,Glawdel2012a}. Although several attempts have been made to elucidate the mechanism of droplet pinch-off in a cross-flow microfluidic system, it is challenging to accurately capture the rapidly changing instantaneous topology of the interface due to the highly non-linear coupled physics.

\noindent
Hence, the present work has aimed to understand the interface evolution and droplet breakup dynamics to meet monodispersity and high throughput demand. The efforts have been made to systematically explain the dynamics and droplet pinch-off mechanism for the broader ranges of operating conditions ($\mur$, $\qr$, $\sigma$) of the cross-flow microfluidic device. The mathematical model based on transient Navier-Stokes (N-S) equations and the conservative level set method (CLSM) is solved numerically using the finite element method (FEM) based computational fluid dynamics (CFD) solver COMSOL multiphysics for a wide range of flow rate ratios ($\qr$) in the squeezing regime ($\cac < 10^{-2}$). The droplet pinch-off mechanism is explained with the help of the instantaneous local radius of interface curvature, instantaneous neck width or thickness (i.e., the shortest distance from the junction point to the nearest point on the interface), the instantaneous evolution of the local pressure profiles, and instantaneous Laplace pressure acting on the interface. A new insight into the droplet formation has been brought by comparing the local radius of interface curvature and neck width. The evolutions of the local pressure profiles in the continuous and dispersed phases are further captured and analyzed to relate with the droplet formation process.
%
\section{Physical and mathematical modelling}
\begin{figure}[h]
	\centering
	\subfloat[]{\includegraphics[width=1\linewidth]{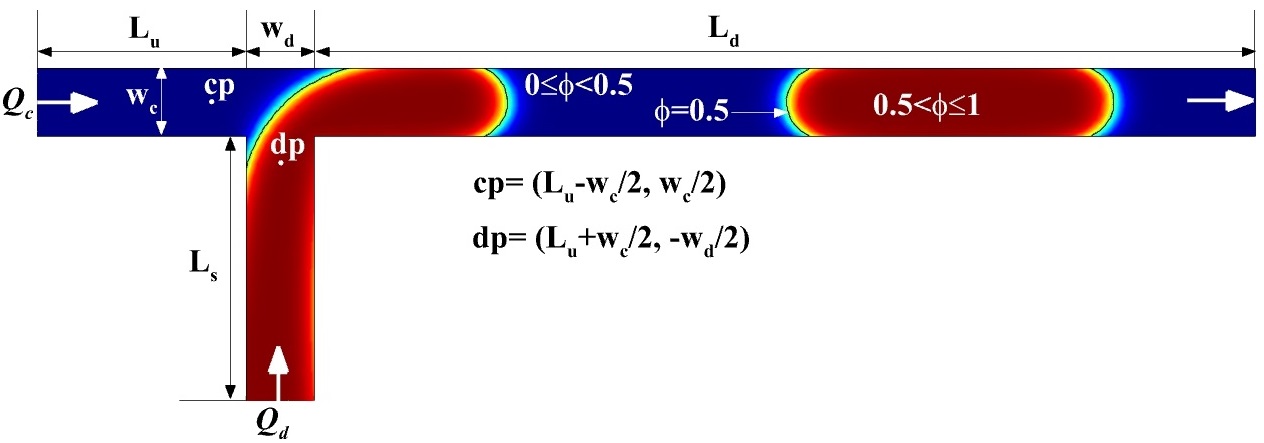}\label{fig:1a}}\\
	\subfloat[]{\includegraphics[width=0.5\linewidth]{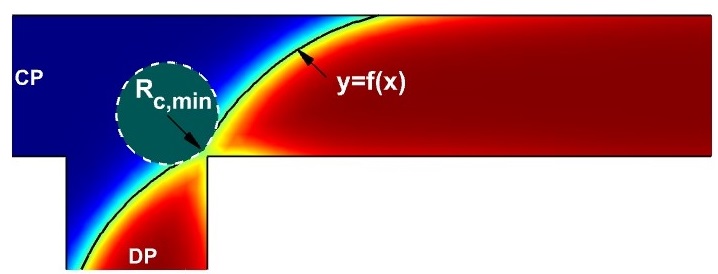}\label{fig:1b}}
	\subfloat[]{\includegraphics[width=0.49\linewidth]{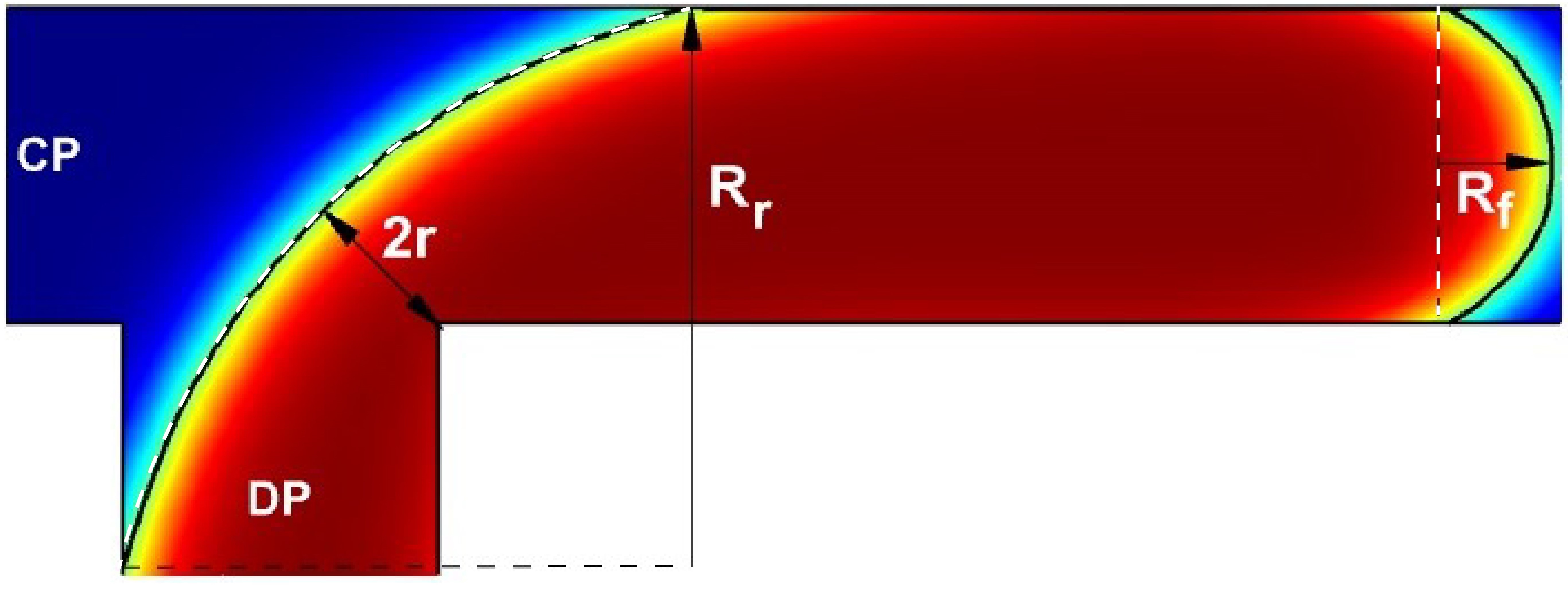}\label{fig:1c}}
	\caption{Schematics of the (a) two-phase flow through T-junction microfluidic device, (b) the local minimum radius of curvature ($\rcmin$), and (c) neck width or the interface thickness ($2r$).}
	\label{fig:1}
\end{figure}

%
\noindent
Consider the two-dimensional (2D) laminar cross-flow of two fluids through T-junction rectangular microfluidic device, as shown in \fig\ref{fig:1a}. T-junction is constructed by a vertical channel intersecting perpendicular ($\perp$) to the horizontal primary channel.
The vertical channel is placed at $L_{\text{u}}$ and $L_{\text{d}}$ distances from the inlet and outlet of the primary channel, respectively. The dimensions (i.e., length and width) of the vertical and primary channels are ($L_{\text{s}}$ \micro m and $\wdp$ \micro m) and ($L_{\text{m}} = L_{\text{u}} + \wdp +L_{\text{d}}$ \micro m and $\wc$ \micro m), respectively.

\noindent
The flow is assumed to be fully developed at the inlets and outlet of the channels. Both liquids are assumed to be non-reacting, immiscible, isothermal, incompressible, and Newtonian.  The physical properties (density $\rho$ kg/m$^3$ and dynamic viscosity $\mu$ Pa.s) are considered to be linearly related to the fluid phase composition ($\phi$). The wall surface of channels is assumed to be hydrophobic. The effects of Marangoni stresses and dynamic interface on the surface tension are ignored herein. Therefore, the interfacial tension does not change during the flow.

\noindent
The continuous phase (CP, $\rhoc$ and $\muc$) and dispersed phase (DP, $\rhod$ and $\mud$) are injected at volumetric flow rates of $\qc$ \micro L/s and $\qd$ \micro L/s through the inlets of the primary and vertical channels, respectively.
The two immiscible fluid phases interact at the junction. Subsequently, both phases (CP \& DP) flow downstream of the device as droplets emulsion or parallel/jet depending on the operating conditions. The outlet of the device is open to the ambient, i.e., $p=0$ atm, and the no-slip boundary condition is applied to the solid rigid impenetrable walls of the device.
%
\subsection{Governing equations}\label{sec:goeq}
%
\noindent
The present physical model can mathematically be expressed by the following set of equations governing the conservation of mass and momentum.
\begin{gather}
\bnabla \cdot \mathbf{u} = 0\label{eqn:l} \\
\rho(\phi)\left[\dfrac{\p \mathbf{u}}{\p t}+\mathbf{u} \cdot \bnabla\mathbf{u}\right]= -\bnabla p+\bnabla \cdot \boldsymbol{\tau}+\mathbf{F}_{\sigma} \label{eqn:2}
\end{gather}
where $\mathbf{u}$, $\rho(\phi)$, $t$, $p$, $\boldsymbol{\tau}$, and $\mathbf{F}_{\sigma}$ are the velocity vector, density, time, pressure, deviatoric stress tensor, and  interfacial tension force, respectively.
The stress ($\boldsymbol{\tau}$) is related with the rate of deformation ($\mathbf{D}$) as follows.
\begin{gather}
\boldsymbol{\tau}=2\mu (\phi)\mathbf{D} \qquad\text{where}\quad
\mathbf{D}=\frac{1}{2}\left[(\bnabla \mathbf{u})+(\bnabla \mathbf{u})^{T}\right] \label{eq:tauD}
\end{gather}
The physical properties ($\rho$ and $\mu$) of the two fluid phases (CP and DP) are defined as follows.
\begin{gather}
\rho(\phi)=\rhoc+(\rhod-\rhoc)\phi,\qquad \text{and}\qquad
\mu(\phi)=\muc+(\mud-\muc)\phi
\end{gather}
where the subscripts ‘c’ and ‘d’ denote CP and DP, respectively.
An extra body force term ($\mathbf{F}_{\sigma}$), included in the Navier–Stokes (N-S) equation, is correlated with interfacial tension (IFT) between the two fluids adopted from the continuum surface force (CSF) model \citep{Brackbill1992} as follow.
\begin{gather}
	\mathbf{F}_{\sigma}=\sigma \kappa \delta_{\phi} \mathbf{n}
	\quad\text{where}\quad
	\kappa={R}^{-1}=- (\bnabla \cdot \mathbf{n})
\label{eq:ift}
\end{gather}
where $\sigma$, $\kappa$, $\delta_{\phi}$, and $\mathbf{n}= ({\mathbf{\bnabla \phi}}/{|\mathbf{\bnabla \phi}|})$ are the interfacial tension,  interface curvature, Dirac Delta function, and  unit normal to interface, respectively.

\noindent
Further, a smooth step function tracks the fluid-fluid interface in the two-phase flow by the scalar level set function ($\phi$) as follows for the continuous phase (CP, $0\le\phi < 0.5$), the liquid-liquid interface (LLI, $\phi = 0.5$), and the dispersed phase (DP, $0.5 < \phi\le1$).
The transport of the fluid phases and liquid-liquid interface topology is governed by the conservative level set method (CLSM) \citep{Sethian2003} based on the Lagrangian approach as follows.
\begin{gather}
\dfrac{\p \phi}{\p t}+\mathbf{u} \cdot {\bnabla} \phi=\gamma {\bnabla} \cdot \left[\epsilon_{\text{ls}}{\bnabla} \phi- \phi (1-\phi){\mathbf{n}}\right]
\label{eqn:lsm}
\end{gather}
where the left side accounts for the advection of liquid phases, whereas the right side maintains conservativeness and numerical stability.
Here, $\epsilon_{ls}$, and $\gamma$ are the interface thickness controlling parameter, and numerical stabilization parameter of $\phi$, respectively.
%
\subsection{Boundary conditions}
%
\noindent
The governing equations (Sec. \ref{sec:goeq}) are subjected to the following physically consistent boundary conditions:
\begin{enumerate}
\item[(a)] The constant volumetric flow rates $\qc$ and $\qd$ are maintained at the inlets of cross-sectional widths of $\wc$ and $\wdp$ respectively for both continuous and dispersed phases.
\item[(b)] The outlet of the primary channel is open to ambient ($p=0$). Neumann condition is also imposed to maintain fully developed velocity and phase profiles.
\item[(c)] The no-slip condition ($\mathbf{u}=0$) is implemented on the rigid, impermeable channel walls.
\end{enumerate}
\noindent
The numerical solution of the mathematical model produces in the instantaneous phase, velocity, and pressure fields as a function of the flow governing parameters.  These fields are analyzed to present the interface development and droplet pinch-off mechanism.
%
\subsection{Relevant definitions and dimensionless parameters}
%
\noindent
The essential definitions and dimensionless parameters used in subsequent sections are detailed as follows.

\noindent
In the level set method, the interface curvature and the vector normal to the interface are readily determined using the level-set function ($\phi$) \citep{Lervag2013}. The interface evolution and droplet pinch-off mechanism are gained by analyzing the interface's curvature and normal vector. The \textit{curvature} is defined as a tangent passing through a point on the interface between two immiscible fluids.  It is thus a vector pointing to the center of the tangent circle. The radius of the circle, or the reciprocal of curvature ($\kappa$), is the radius of the curvature ($R=\kappa^{-1}$ \micro m), as shown in \fig\ref{fig:1b}.
The point/region of the highest curvature experiences maximum pressure and results in the most stretched interfaces. Vice versa, the lowest curvature region has minimum pressure. It creates an imbalance in the Laplace pressure, and the difference in pressure drives the liquid to attain a perfectly spherical droplet.
The droplet phase does not occupy the whole cross-section of the microchannel but leaves gutters (i.e., a space between the droplet interface and channel surface/corners) with curvature \citep{Volkert2009}.

\noindent
The pinch-off mechanism of a droplet is elucidated by using the pressure fluctuations in two phases, the neck width or thickness ($2r$ \micro m) during the squeezing stage of the droplet formation \citep{Glawdel2012a}, and the local radius of the interface curvature ($\rc$ \micro m) in the subsequent sections.

\noindent
The placement of the pressure sensor is vital to measure pressure fluctuations during the droplet formation process \citep{Xu_Ke_Tostado,Abate2012}. In this work, two points `cp' and `dp' are chosen (\fig\ref{fig:1a}) at locations $(L_{\text u}-\wc/2, \wc/2)$ and   $(L_{\text u}+\wc/2, -\wdp/2)$ to measure the pressure in CP and DP, respectively.
The pressure in upstream or continuous phase at point `cp' is denoted as $p_{\text{cp}}$, pressure in dispersed phase at point `dp' is denoted as $p_{\text{dp}}$, and pressure drop across the interface in between points `dp' and 'cp' is defined as $\Delta p_{\text{dc}} = (p_{\text{dp}} - p_{\text{cp}})$.

\begin{figure}[t]
	\centering
	\includegraphics[width=0.8\linewidth]{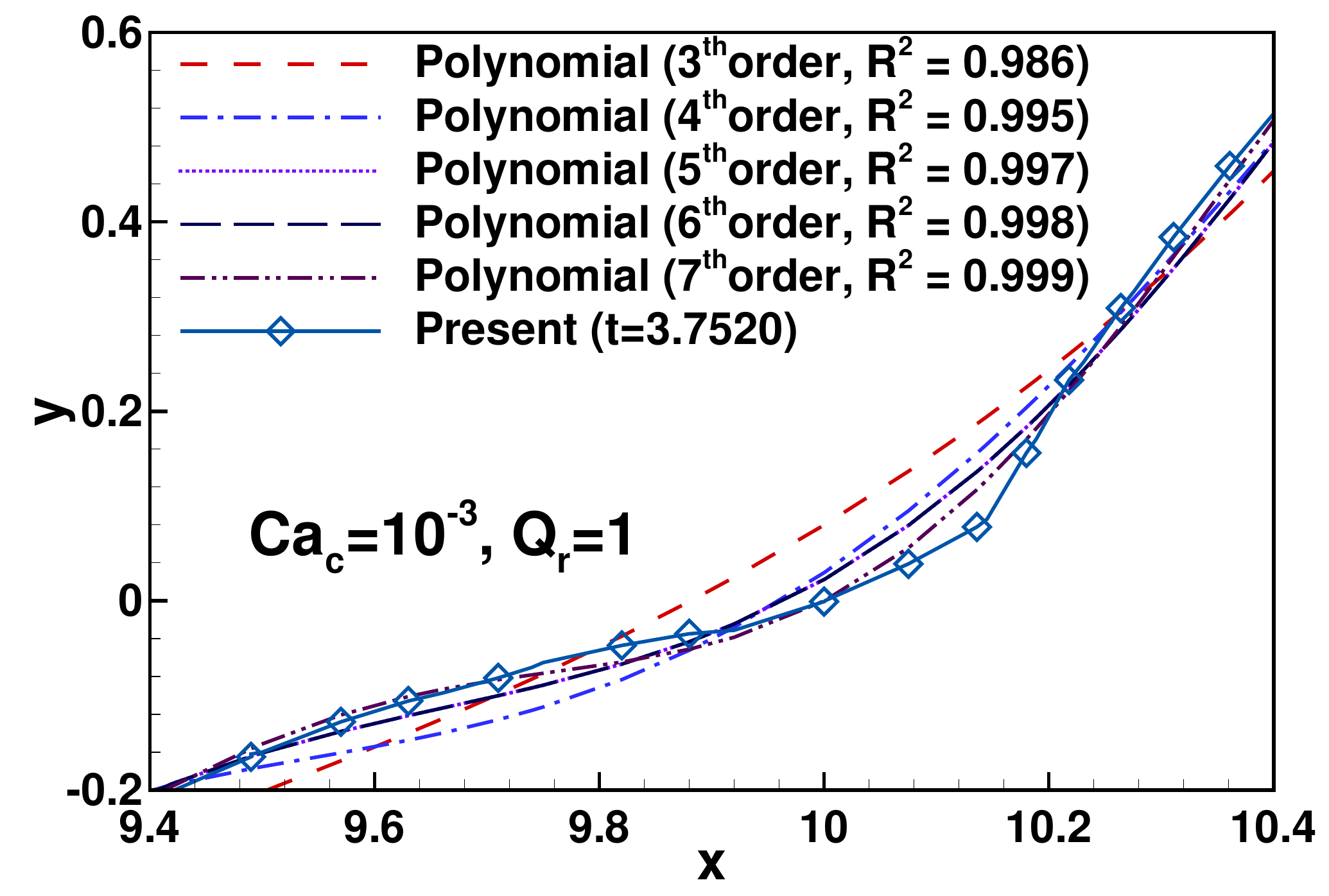}
	\caption{A polynomial fitting for the interface curvature profile.}
	\label{fig:2}
\end{figure}

\noindent
Further, the neck thickness ($2r$) is measured as the shortest distance from the receding interface neck $(x_{i},y_{i})$ to the lower-right corner of the junction $(x_{j},y_{j})$, as shown in \fig\ref{fig:1c}. It is determined by fitting a quarter of a circle to the interface  \citep{Deremble2006,VanSteijn2010} as follows.
\begin{gather}
	2r=\sqrt{(x_{i}-x_{j})^{2}+(y_{i}-y_{j})^{2}}
\end{gather}
\begin{figure}[!t]
\centering
\scalebox{0.8}{
\begin{tikzpicture}
    \node[draw=red, ultra thick, 
        text=black,
         text width = 0.9\linewidth,
        minimum height=1cm] at (0,4) (block0){Perform the time-dependent CFD simulations using level set and finite element methods};

    \node[draw=red, ultra thick, 
        text=black,
	    fill=yellow!50,
         text width = 1\linewidth,
        minimum height=1cm] at (0,2) (block1){Extract the numerical data for instantaneous interface ($\phi=0.5$) shape ($x,y$) profiles at all time ($t$) in the squeezing stage until droplet pinch-off};

    \node[draw=red, ultra thick, 
        text=black,
	    fill=yellow!50,
         text width = 0.9\linewidth,
        minimum height=1cm] at (0,0) (block11){Import the instantaneous interface curve profiles ($x,y,t$) into MATLAB};

    \node[draw=red, ultra thick, 
	    fill=yellow!50,
         text width = 0.9\linewidth,
        minimum height=1cm] at (0,-2) (block2){Use `polyfit' function to best fit ($R^2\approx 1$) the interface profiles by the $n^{\text{th}}$ order polynomial curve $y=f(x)=\sum_{i=0}^{n}a_ix^i$ for all $t$};

    \node[draw=red, ultra thick, 
	    fill=yellow!50,
         text width = 0.9\linewidth,
        minimum height=1cm] at (0,-4) (block3){Calculate 1st and 2nd derivatives of the fitted polynomial, i.e., $y_{\text{x}}=dy/dx$ and $y_{\text{xx}}=d^2y/dx^2$};

    \node[draw=red, ultra thick, 
	    fill=yellow!50,
         text width = 0.9\linewidth,
        minimum height=1cm] at (0,-6) (block4){Calculate the local radius of curvature at every point along the interface, $R_{\text{c}}(x,y,t)=({A}/{B})$ where $A=(1+y_{\text{x}}^2)^{3/2}$ and $B=|y_{\text{xx}}|$};

    \node[draw=red, ultra thick, 
	    fill=yellow!50,
         text width = 0.9\linewidth,
        minimum height=1cm] at (0,-8) (block5){Determine minimum radius of curvature, $\rcmin(t)=min(R_{\text{c}}(x,y,t))$};

    \node[draw=red, ultra thick, 
	    fill=yellow!50,
         text width = 0.9\linewidth,
        minimum height=1cm] at (0,-10) (block6){Determine minimum $R_{\text{c}}$ at droplet pinch-off as $R_{\text{cmin}}=min(\rcmin(t))$};

\draw[-latex, blue, ultra thick] (block0) edge (block1)
	(block1) edge (block11)
	(block11) edge (block2)
    (block2) edge (block3)
    (block3) edge (block4)
    (block4) edge (block5)
    (block5) edge (block6);
\end{tikzpicture}
}
	\caption{Procedure for calculation of $\rcmin$.}
	\label{fig:3}
\end{figure}
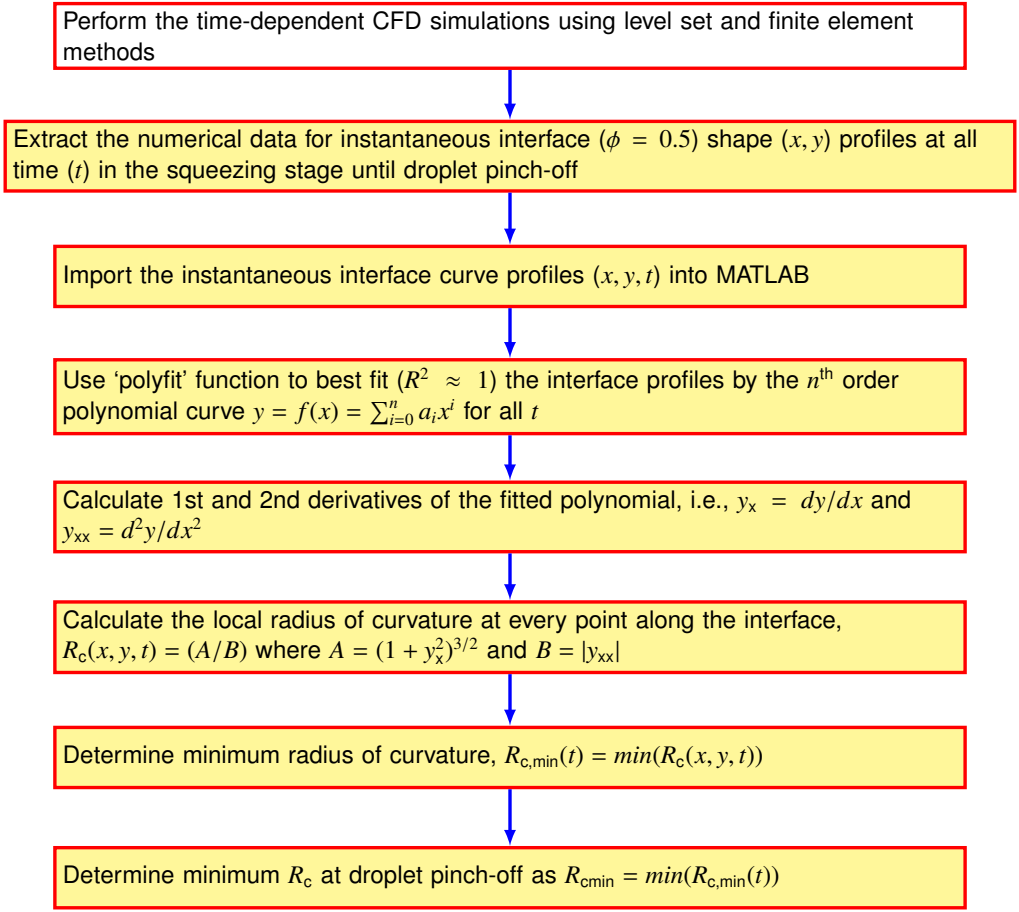
\noindent 
The local instantaneous radius of the interface curvature $\rc(x,y,t)$, the minimum instantaneous radius of curvature $\rcmin(t)$, and the minimum radius of curvature at the droplet pinch-off ($\rcmin$) are determined by fitting the interface curvature profiles ($x,y,t$) into a polynomial function $y=f(x)$ for all time ($t$) instants under the squeezing stage by using  MATLAB program developed as per the algorithm depicted in \fig\ref{fig:3}. The interface evolution profiles are best fitted with seventh order polynomial, $y=\sum_{i=0}^{7}(a_ix^i)$, as represented in \fig\ref{fig:2} for one flow condition ($\cac=10^{-3}$ and $\qr=1$).
%
%

\noindent
The important dimensionless parameters used hereafter are defined as follows.
\begin{gather}
\cac = \frac{\uc^{*} \muc^{*}}{\sigma^{*}},\quad
Re_{\text{c}} = \frac{\rhoc^{*} \uc^{*} \wc^{*}}{\muc^{*}},\quad
\qr = \frac{\qd^{*}}{\qc^{*}}, \quad
\mur = \frac{\mud^{*}}{\muc^{*}}, \quad
\rhor = \frac{\rhod^{*}}{\rhoc^{*}}, \nonumber\\
\wrr = \frac{\wdp^{*}}{\wc^{*}}, \quad
2r=\frac{2r^{*}}{\wc^{*}}, \quad
\rcmin=\frac{\rcmin^{*}}{\wc^{*}},\quad
t=t^{*}\left(\frac{\uc^{*}}{\wc^{*}}\right), \quad
p={p^{*}}\left(\frac{\wc^{*}}{\uc^{*} \muc^{*}}\right)
\label{eq:8}
\end{gather}
where $\cac$ is capillary number, $Re_{\text{c}}$ is Reynolds number, and  $\rcmin$ the minimum local radius of the curvature.
The subscripts `$\text{c}$', `$\text{d}$', and `$\text{r}$' refer to the CP, DP and ratio, respectively. The variables with ${`*'}$ superscript in \eqn(\ref{eq:8}) are dimensional, however, used without ${`*'}$ superscript before \eqn(\ref{eq:8}).
The subsequent section details the solution approach and numerical parameters used in this work.
%
\section{Solution approach and numerical parameters}\label{sec:sanp}
\noindent
In this work, a mathematical model, based on the coupled Navier-Stokes (N-S) and conservative level set method (CLSM), governing the two-phase flow through T-junction microfluidic device has been solved numerically. The detailed numerical solutions are obtained using the finite element method (FEM) based on computational fluid dynamics (CFD) solver COMSOL multiphysics.

\noindent
The two-dimensional (2D) mathematical model has been represented by “fluid flow $\rightarrow$ multiphase flow $\rightarrow$ two-phase flow, level set $\rightarrow$ laminar flow” modules of COMSOL. The computational domain has been discretized by the linear, non-uniform, triangular, unstructured mesh.
\rev{The} finite element method is used to transform the transient PDEs into  ODEs. The polynomials of order p and q (i.e., P$_{\text{p}}$ + P$_{\text{q}}$), i.e., the shape functions with p$^{\text{th}}$ and q$^{\text{th}}$ order elements, are used for velocity and pressure fields, respectively, for spatial discretization. In this work, both p and q are taken as 1. Further, the variable order (i.e., 1st for Euler method to 5th) accurate implicit backward difference formula (BDF) selected for the temporal discretization of ODEs results in stable differential-algebraic equations (DAEs) with variable time steps ($\Delta t$) for complex dynamics problems \citep{Bashir2011,Sartipzadeh2020}. A trade-off between the accuracy of the solution and robust, stable convergence is generally set based on the higher-order and lower-order accurate approximations of BDF, respectively. Further, the discretized equations have been solved using a fully coupled  PARDISO and Newton's non-linear solvers.

\noindent
In all simulations, a sufficiently lower time step ($\Delta t=10$ \micro s) is used to obtain the fully converged (relative tolerance = $5\times 10^{-3}$) iterative solution of two phase flow (\rev{$\mathbf{u}$}, $p$ and $\phi$) field for the following numerical parameters and flow conditions:
%
(a) \textit{Geometrical parameters}: $\wc=\wdp=100$ \micro m; $L_{\text{u}}=L_{\text{s}}=9\wc$; $L_{\text{d}}=30\wc$; and $L_{\text{m}}=40\wc$.
(b) \textit{Mesh characteristics}: linear, non-uniform, triangular, unstructured mesh; maximum size of mesh element, $\Delta_{\text{max}}=10$ \micro m; total number of mesh element, $N_{\text{e}}=13,766$; and degrees of freedom, DoF = 53,029.
(c) \textit{Level set parameters}: $\gamma=1$ m/s; and $\epsilon_{ls}= \Delta_{\text{max}}/2=5$ \micro m.
(d) \textit{Flow governing parameters}: $\rhor=1$; $\theta =135^{\circ}$; $\mud=10^{-3}$ Pa.s; $\qd=0.14$ \micro L/s; $Re_{\text{c}}=0.1$; $\cac\rev{<}10^{-2}$; $0.1\le \qr\le 10$; {$7.143\times 10^{-3}\le\mur\le 7.143\times 10^{-1}$; and $1.96\times 10^{-3}\le\sigma \le 1.96\times 10^{5}$ mN/m}.
Note that the geometrical and mesh parameters and time step have been tested elsewhere \citep{venkat2021} for their independence on the numerical results.
\section{Results and discussion}
%
\noindent
In this work, the interface evolution and droplet pinch-off mechanism are presented and discussed through pressure ($p$) profiles, neck width ($2r$), and radius of interface curvature ($R_{\text{c}}$) for the broad range of conditions under the squeezing regime for the two-phase flow through the T-junction microfluidic device. However, the reliability and accuracy of the present numerical approach have been established before presenting new results.
\subsection{Validation of the results}
\noindent
The present numerical solution procedure for two-phase flow has been validated in detail in our previous study \citep{venkat2021} to ensure its reliability and accuracy.
\begin{figure}[h]
	\centering
	\includegraphics[width=0.8\linewidth]{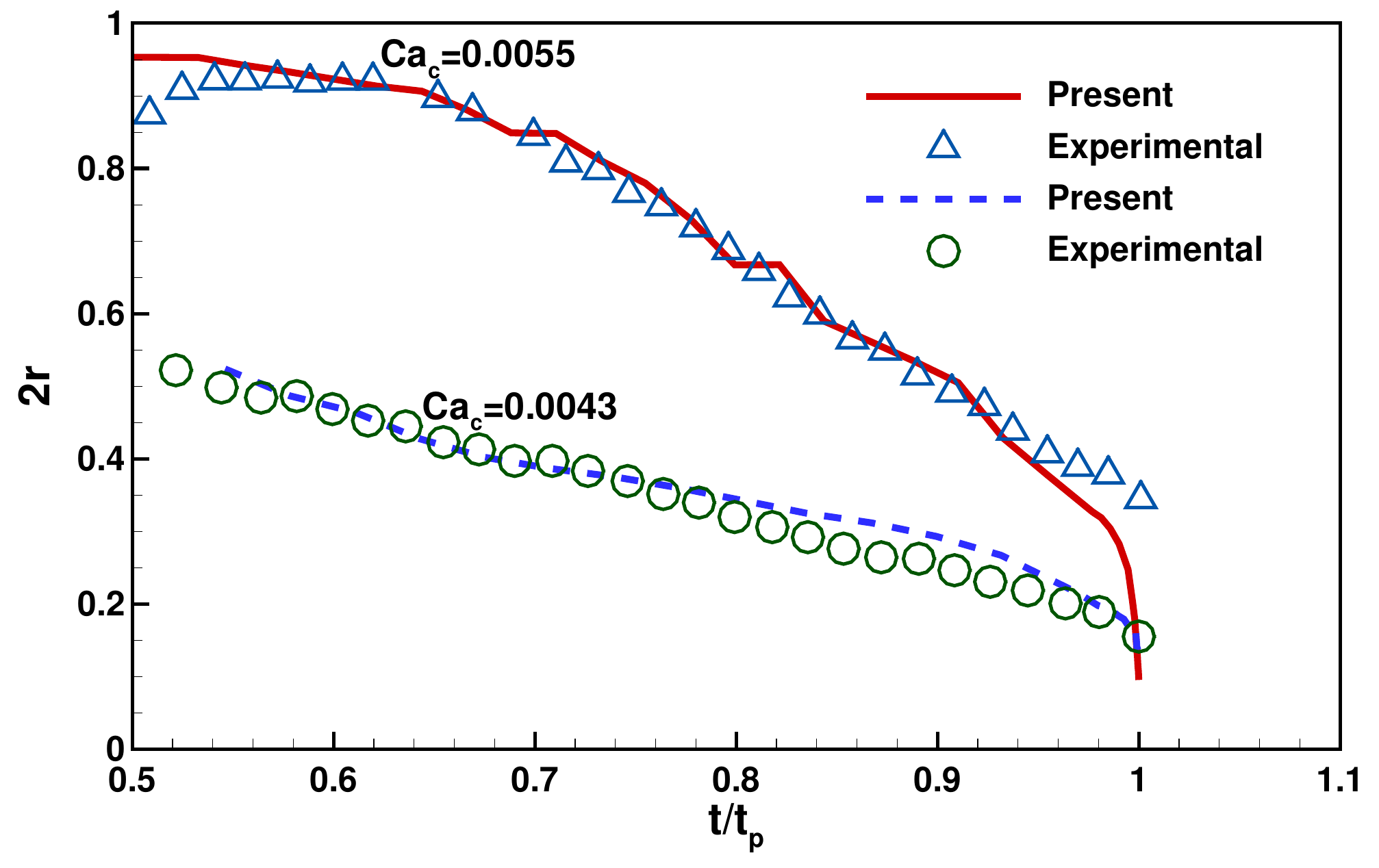}
	\caption{Comparison between the present and experimental \citep{Glawdel2012a} results for neck width ($2r$) as a function of time ($t/t_{\text{p}}$).}
	\label{fig:4}
\end{figure}
The additional comparison of the present and experimental \citep{Glawdel2012a} values of the neck width ($2r$) variation with time ($t/t_{\text{p}}$, where $t_{\text{p}}$ is droplet pinch-off cycle time) has been made in \fig\ref{fig:4} for two capillary numbers (a) $\cac=0.0055$ for $\qr=0.483$ and $\wrr=0.35$, and (b) $\cac=0.0043$ for $\qr=0.4143$ and $\wrr=0.90$. The values of $2r$ decrease in the squeezing or necking stage and approach a minimum value wherein the interface collapse rapidly, resulting in a pinch-off of the droplet.
Both experimental and numerical values have shown excellent correspondence, thereby establishes the accuracy of present modelling and simulation approaches. The results presented hereafter are believed to have an excellent ($\pm 1\%$) accuracy.
%
\subsection{Stages of droplet formation}\label{sec:stages}
\begin{figure}[!htbp]
	\centering
	(I)\hspace{0.5in} (II)\hspace{1in} (III)\hspace{1in} (IV)\hspace{1in} (V)\qquad$~$\\\vspace{-1em}
	\subfloat[$\qr=10$]{\includegraphics[width=\linewidth]{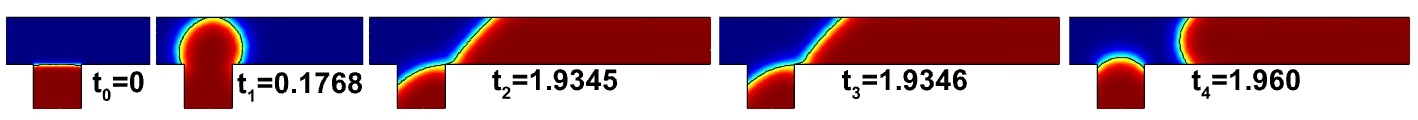}\label{fig:5a}}\\\vspace{-1em}
	\subfloat[$\qr=5$]{\includegraphics[width=\linewidth]{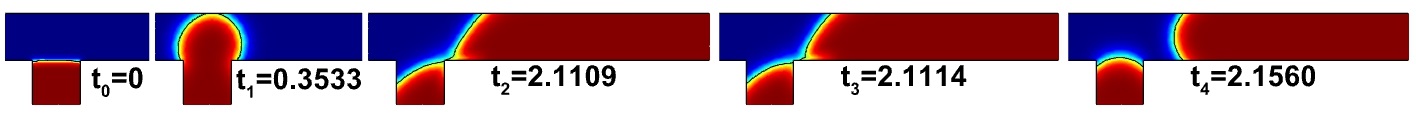}\label{fig:5b}}\\\vspace{-1em}
	\subfloat[$\qr=2$]{\includegraphics[width=\linewidth]{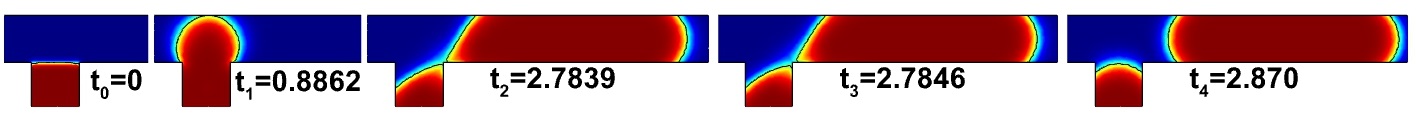}\label{fig:5c}}\\\vspace{-1em}
	\subfloat[$\qr=1$]{\includegraphics[width=\linewidth]{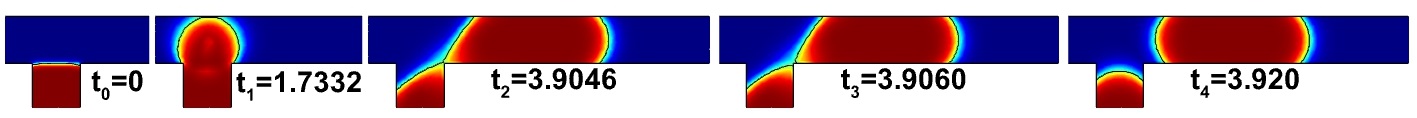}\label{fig:5d}}\\\vspace{-1em}
	\subfloat[$\qr=1/2$]{\includegraphics[width=\linewidth]{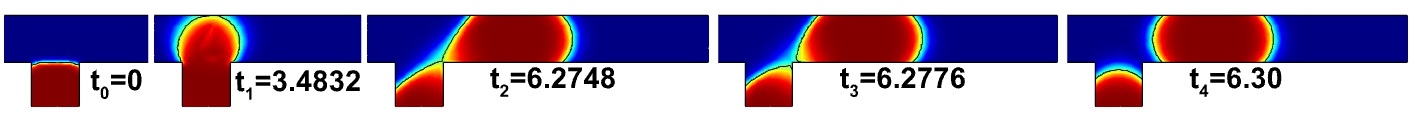}\label{fig:5e}}\\\vspace{-1em}
	\subfloat[$\qr=1/4$]{\includegraphics[width=\linewidth]{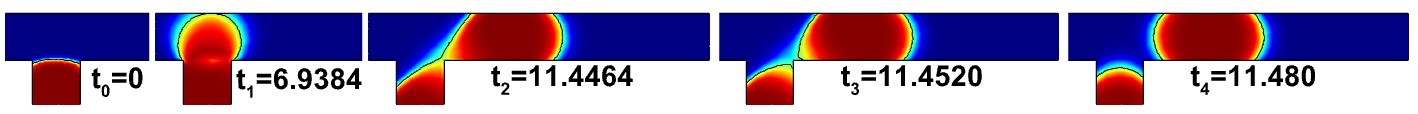}\label{fig:5f}}\\\vspace{-1em}
	\subfloat[$\qr=1/8$]{\includegraphics[width=\linewidth]{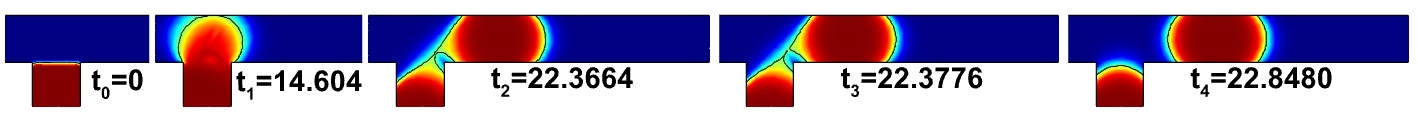}\label{fig:5g}}\\\vspace{-1em}
	\subfloat[$\qr=1/10$]{\includegraphics[width=\linewidth]{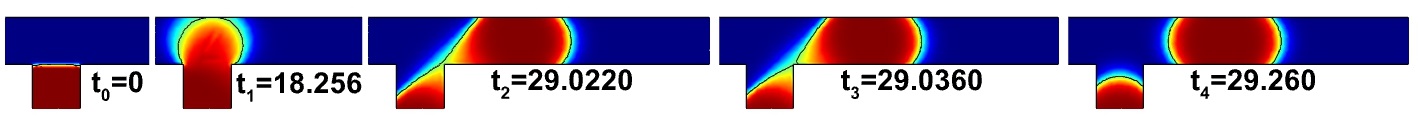}\label{fig:5h}}
	\caption{Instantaneous stages of the droplet generation as a function of $\qr$ at $\cac=10^{-4}$.}
	\label{fig:5}
\end{figure}

%
\noindent
The phase flow contours depicting the stages of droplet formation based on the instantaneous interface ($\phi=0.5$) evolution and movement are shown in \fig\ref{fig:5} for $0.1 \le \qr\le 10$ at $\cac=10^{-4}$.
Based on the evolution and movement of the interface curvature, the droplet generation cycle is classified into the following stages: (a) S-0: initial (or leg), (b) S-1: filling (or growing or expansion), (c) S-2: squeezing (or necking), (d) S-3: pinch-off (or breakup), and (e) S-4: stable droplet.  The process continues, i.e., the droplet formation cycle (stages 1 to 4) repeats at regular time intervals wherein the viscous force due to \rev{channel} confinement plays a crucial role.

\noindent
In the initial (S-0) stage, both continuous and dispersed phases start to flow through the primary and vertical channels, respectively. This stage ends when the vertical channel (i.e., T leg) is completely filled by the dispersed phase (DP) at time $t_0$, as shown in \fig\ref{fig:5}(I). Depending on the value of the relative flow rates ($\qr$), both phases may come in contact with each other. The time taken by the initial (S-0) stage is noted as $t_{\text{i}}=t_{0}$.

\noindent
Further, in the filling (S-1) stage, the dispersed phase (DP) continues to flow into the main channel at the junction and starts to grow, in a convex shape, until the evolving interface reaches the closest possible to the top wall of the main channel.
It indicates the completion of the filling  (S-1) stage, as shown in \fig\ref{fig:5}(II), and the initiation of the squeezing  (S-2) stage at time $t_{1}$.
The time taken by the filling  (S-1) stage $t_{\text{f}}=(t_1-t_0)$ is best correlated to flow rate ratio ($0.1\le\qr\le 10$) and capillary number ($10^{-4}\le \cac \le 10^{-3}$) as follows.
\begin{gather}
t_{\text{f}}= a\qr^{b}
\label{eqn:tf}
\end{gather}
where, the correlation coefficients are statistically obtained, \rev{by performing a non-linear regression analysis using MATLAB and DataFit (trial version) tools,} as $a=1.7587$, and $b=-1.006$ with $R^2=0.9995$, $\delta_{\text{min}}=0.39\%$, $\delta_{\text{max}}=2.45\%$, and $\delta_{\text{avg}}=1.37\%$.

\noindent
Subsequently, the liquid-liquid interface continually grows in the squeezing (S-2) stage, and the dispersed phase (DP) gradually starts obstructing the continuous phase (CP) flow.
Simultaneously, the pressure in the upstream continues to rise to maximum (or critical) value. The interface gradually starts to experience the shearing from the streaming of the continuous phase (CP).
The geometrical wall confinement acts against the perturbations caused due to the pressure differences during the flow, and \rev{thus}, the instabilities are suppressed. In turn, the front side (i.e., downstream) of the interface remains intact to its position. However, the rear  (i.e., upstream) side of the interface is evolving and taking different shapes mainly due to the interplay of the forces acting on the interface.
The dispersed phase (DP) is thus forced and sheared downstream of the primary channel. Since the interface deformation due to shear is negligible, the pressure gradient drives the dispersed phase (DP) downstream. It, thereby, shrinks the neck size from maximum (at time $t_1$) to minimal (or critical at time $t_2$) value of the dispersed phase (DP) near the junction. It indicates completion of the squeezing (S-2) stage, as shown in \fig\ref{fig:5}(III),  and initiation of the droplet pinch-off (S-3) stage.
The time taken by the squeezing stage $t_{\text{s}}=(t_2-t_1)$ is best correlated for  flow rate ratio ($0.1\le\qr\le 10$) and capillary number ($10^{-4}\le \cac \le 10^{-3}$) as follows.
\begin{gather}
t_{\text{s}} = X_{\text{s}}t_{\text{f}} \label{eqn:ts}\\
\text{ where } X_{\text{s}} = a+bx_1+c\qr+dx_1^2+e\qr^{2}+f(\qr x_1)+gx_1^3+h\qr^{3}+i(\qr^{2}x_1)+j\qr x_1^{2} \nonumber
\end{gather}
where, $x_1=(1/\cac)$ and the correlation coefficients are statistically obtained as $a=-1353.6956$, $b=782.6003$, $c=1.6133\times 10^{5}$, $d=-0.8593$, $e=0.0749$, $f=-177.4681$, $g=7.8111\times 10^{-5}$, $h=-1.5834\times 10^{-3}$, $i=-4.2112 \times 10^{-6}$ and $j=0.0161$ with $R^2=0.9987$, $\delta_{\text{min}}=0.0084\%$, and $\delta_{\text{max}}=4.9170\%$.

\noindent
As time progresses, the interface curve slowly shrinks and attains a concave shape before it pinches off. The necking (S-2) transits to pinch-off (S-3) stage when the Laplace pressure difference reverses its direction. The additional flow due to confinement triggers the rate of the thinning of the interface collapse \citep{Volkert2009}. The pinch-off  (S-3) stage initiates, at time $t_2$, as the interfacial tension and viscous forces resist the pressure force due to continued flow of both phases. The droplet detachment or pinch-off subsequently triggers spontaneously, at time $t_3$, as a result of a balancing of interfacial tension ($\mathbf{F}_{\sigma}$), viscous  ($\mathbf{F}_{\text{v}}$), and pressure  ($\mathbf{F}_{\text{p}}$) forces at the neck.
The time taken by the spontaneous pinch-off (S-3) stage is $t_{\text{b}}=(t_3 - t_2)\lll 1$, and  best correlated for flow rate ratio ($0.1\le\qr\le 10$) and capillary number ($10^{-4}\le \cac \le 10^{-3}$) as follows.
\begin{gather}
t_{\text{b}}= X_{\text{b}}t_{\text{s}} \label{eqn:tb}\\
\text{ where }
X_{\text{b}} = a+bx_1+cx_2+dx_1^2+ex_2^2+f(x_1x_2)+gx_1^{3}+hx_2^3+i(x_1x_2^2)+j(x_1^{2}x_2)\nonumber
\end{gather}
where, $x_1=(1/\cac)$, $x_2=(\log\qr)$, and the correlation coefficients are statistically obtained as  $a=-40.882$, $b=0.1645$, $c=3456.4656$, $d=-1.356\times 10^{-4}$, $e=1.743 \times 10^{-4}$, $f=-3.8021$, $g=1.1958$, $h=3.878\times 10^{-5}$, $i=-9.8612$ and $j=3.456\times 10^{-4}$ with $R^2=0.9999$, $\delta_{\text{min}}=0.0006\%$, and $\delta_{\text{max}}=0.1452\%$.

\noindent
After the droplet pinch-off (S-3) stage, stable droplet (S-4) stage initiates at $t_3$ as the detached droplet flows downstream of the primary channel. It subsequently forms a stable droplet whose shape and size do not alter with time by balancing the forces acting over the droplet. The development time of stable droplet (S-4) in two-phase flow is recorded as $t_4$. The time taken by the stable droplet (S-4) stage is thus  $t_{\text{sd}}=(t_4 - t_3)$,  and best correlated for flow rate ratio ($0.1\le\qr\le 10$) and capillary number ($10^{-4}\le \cac \le 10^{-3}$) as follows.
\begin{gather}
t_{\text{sd}} = X_{\text{sd}}t_{\text{b}} \label{eqn:tsd}
\\
\text{where} \qquad X_{\text{sd}} = a+bx_1+cx_2+dx_2^2+ex_2^3+fx_2^4+gx_2^5 \nonumber
\end{gather}
$x_1=(\log\cac)$, $x_2=(\log\qr)$, and the correlation coefficients are statistically obtained as  $a=19.3741$, $b=1.5538\times 10^{-4}$, $c=13.5748$, $d=-1.2299$, $e=-5.4876$, $f=0.4098$, and $g=0.7571$ with $R^2=0.9612$, $\delta_{\text{min}}=0.0565\%$, and $\delta_{\text{max}}=6.5742\%$.

\noindent
The total time taken by the one cycle of the droplet pinch-off is $t_{\text{p}}=(t_{\text{f}} + t_{\text{s}}  + t_{\text{b}})$, and that for the formation of stable droplet is $t_{\text{d}}=(t_{\text{p}} +t_{\text{sd}})$. Note that during the whole process of droplet formation, both CP and DP continue to flow. The time taken by the individual stages of the droplet formation depend on the flow rates ratio ($\qr$) and capillary number ($\cac$), in addition to other parameters ($\rhor$, $\mur$, $\theta$, $\wrr$). Further understandings of droplet formation dynamics are gained and discussed in terms of the instantaneous interface evolution profiles in the next section.
\begin{figure}
	\centering
	\subfloat[$\qr=10$]{\includegraphics[width=0.44\linewidth]{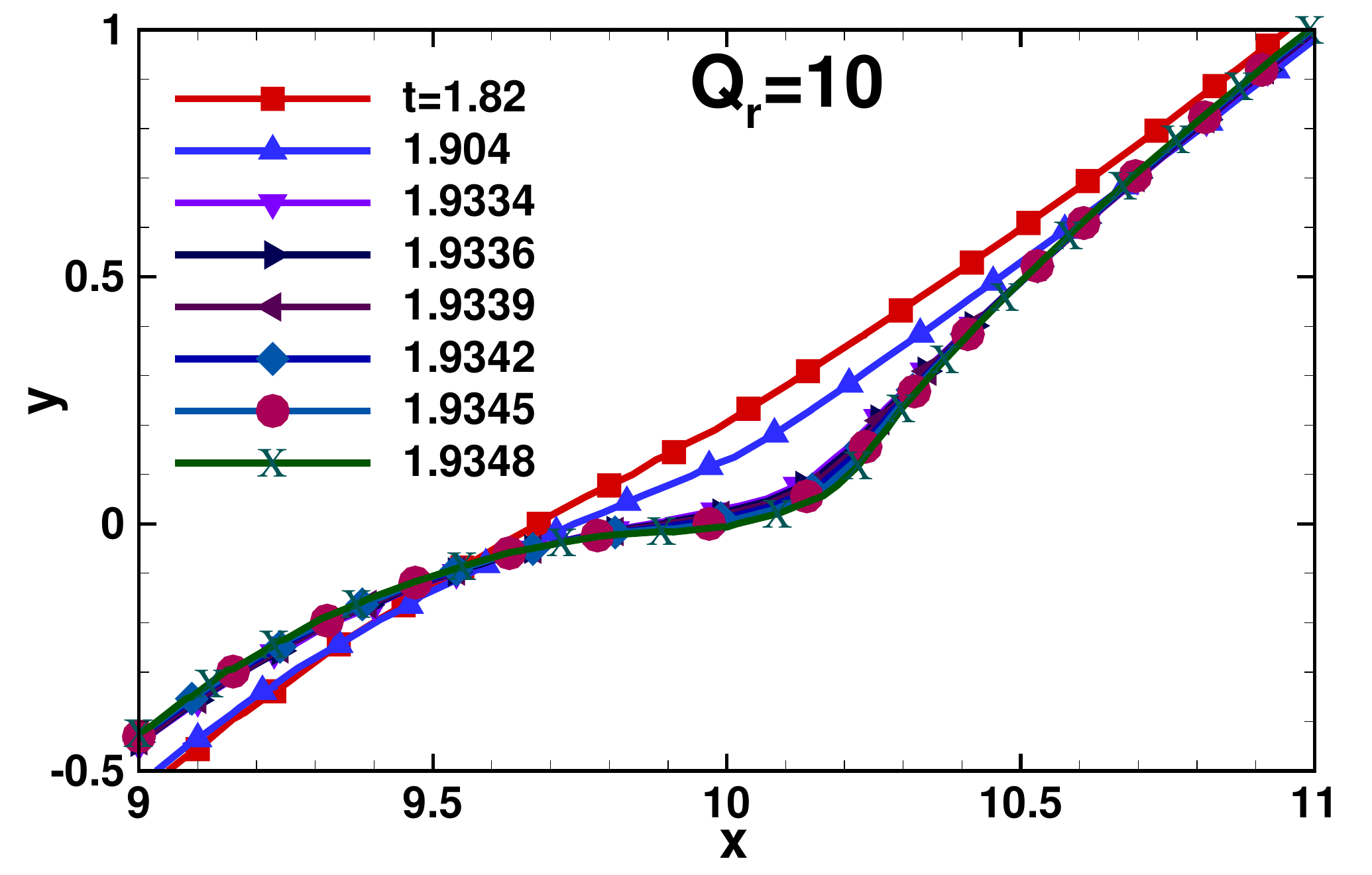}}
	\subfloat[$\qr=5$]{\includegraphics[width=0.44\linewidth]{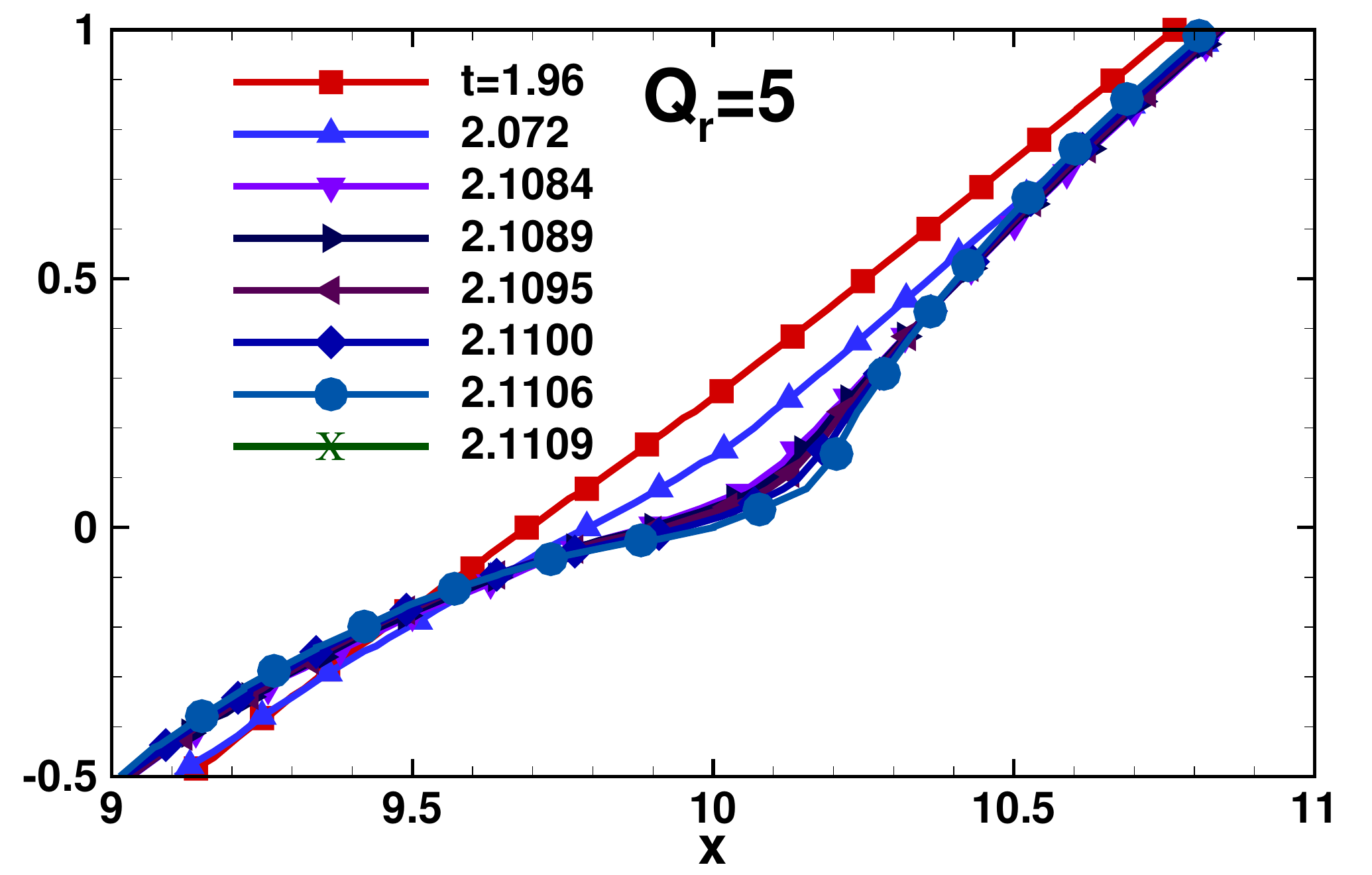}}\\
	\subfloat[$\qr=2$]{\includegraphics[width=0.44\linewidth]{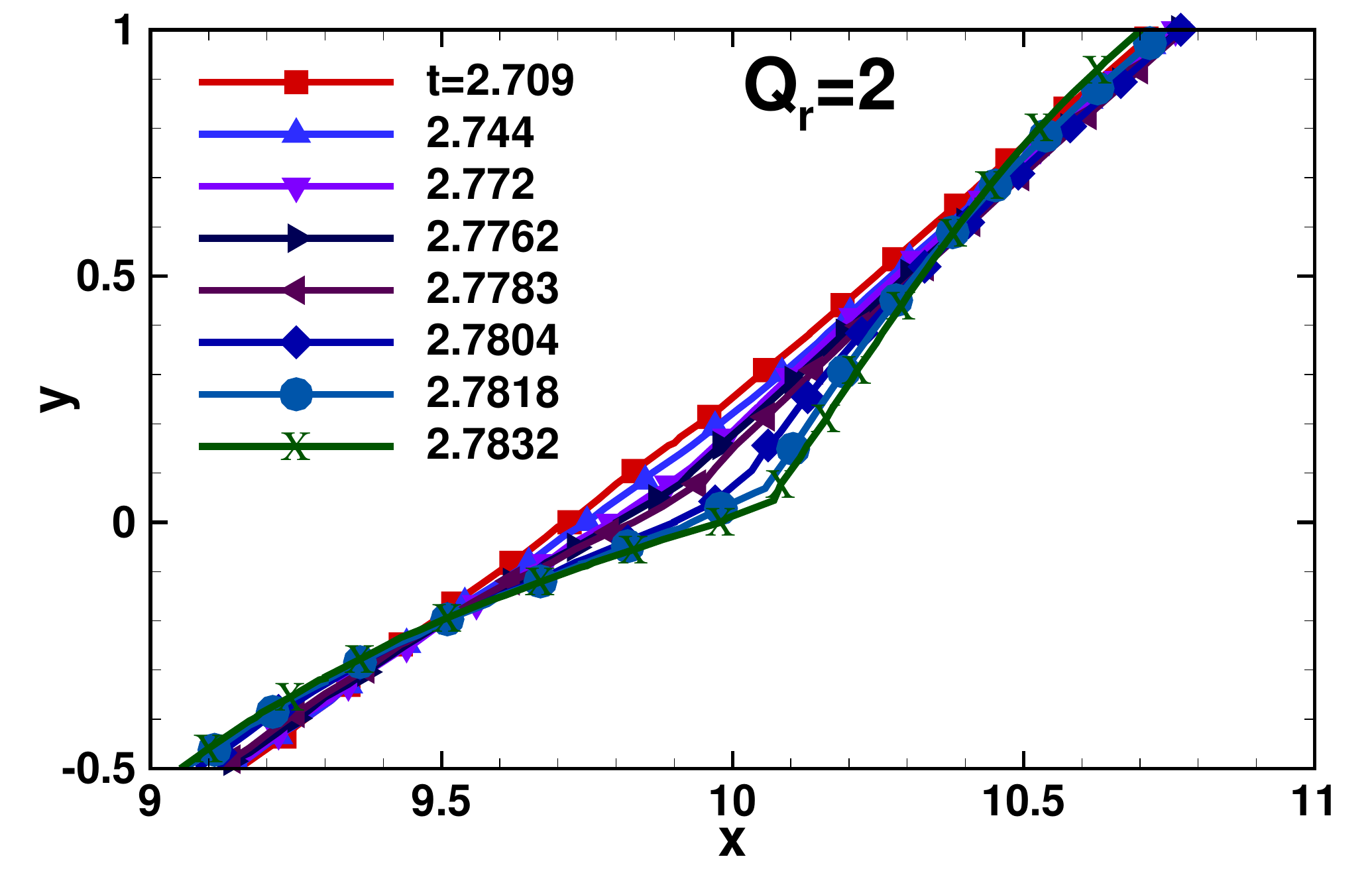}}
	\subfloat[$\qr=1$]{\includegraphics[width=0.44\linewidth]{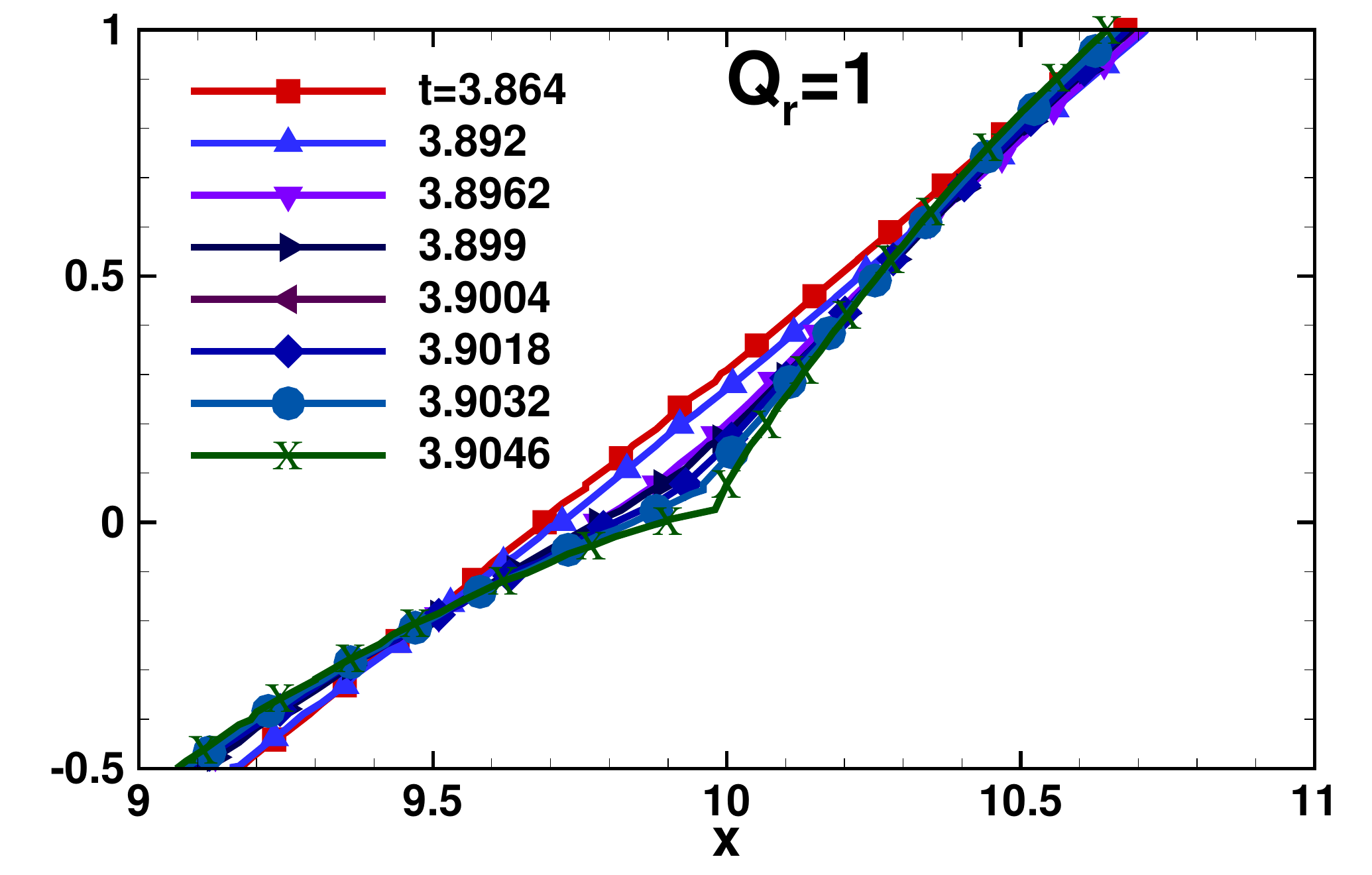}}\\
	\subfloat[$\qr=1/2$]{\includegraphics[width=0.44\linewidth]{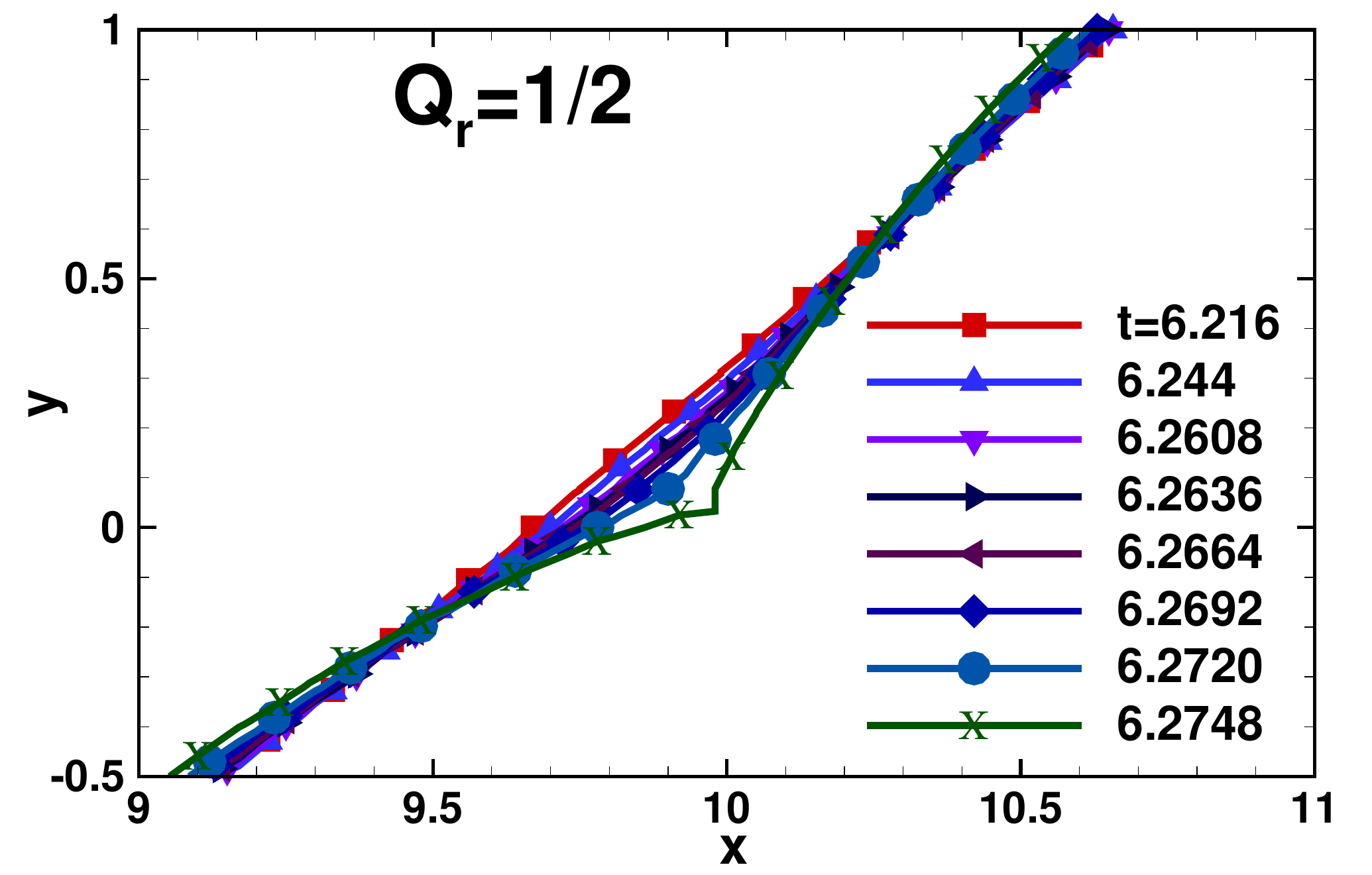}}
	\subfloat[$\qr=1/4$]{\includegraphics[width=0.44\linewidth]{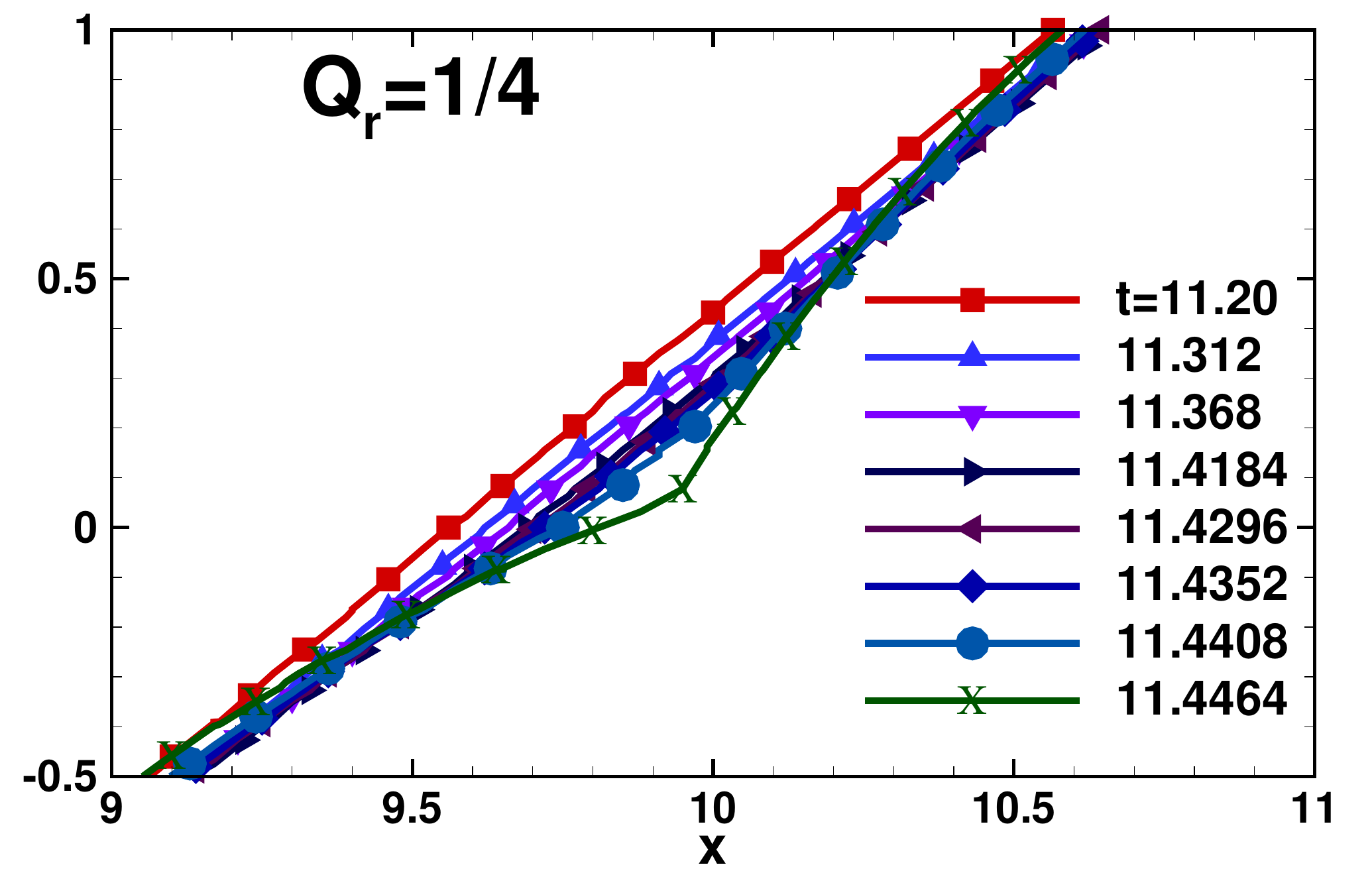}}\\
	\subfloat[$\qr=1/8$]{\includegraphics[width=0.44\linewidth]{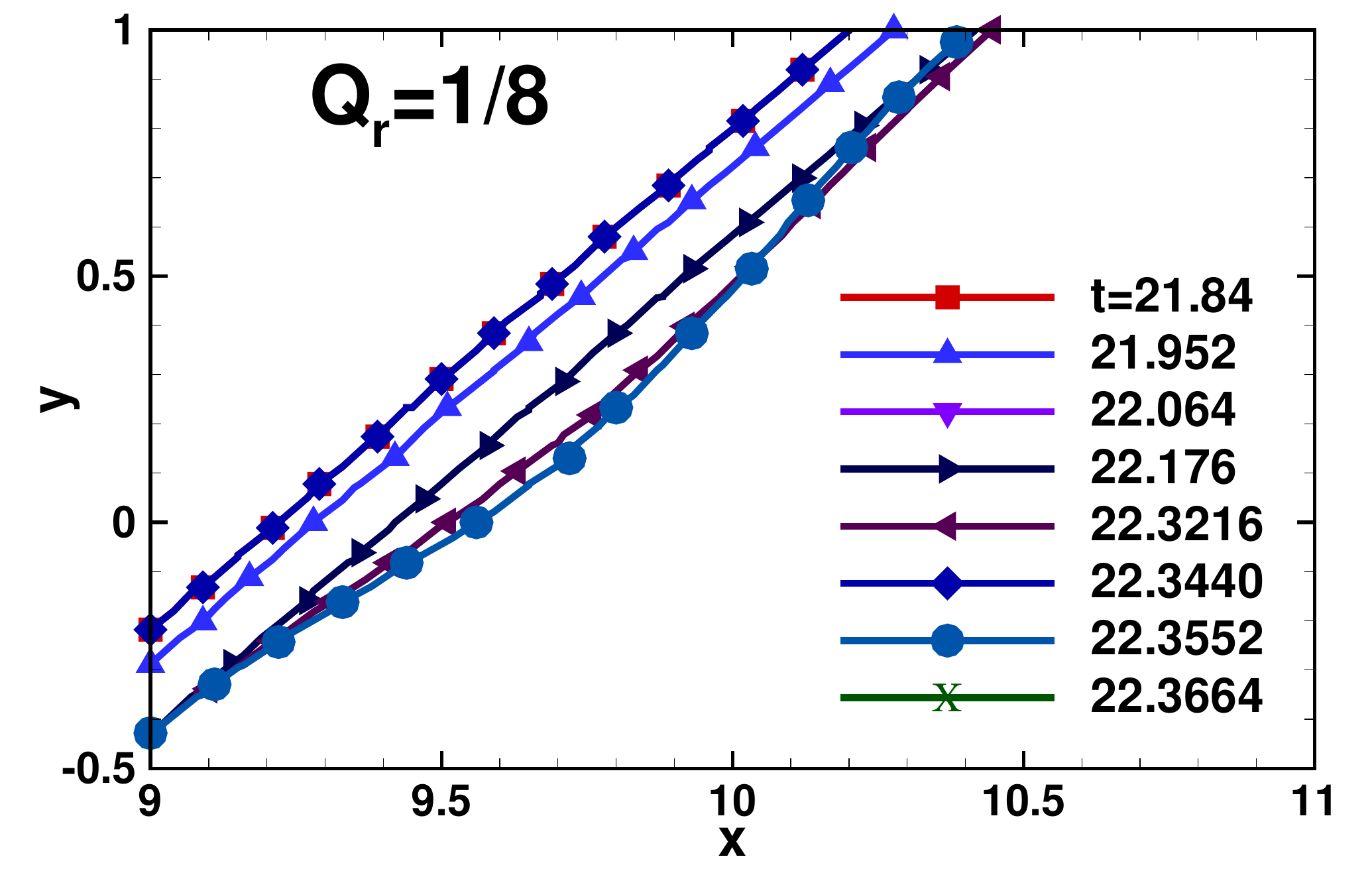}}
	\subfloat[$\qr=1/10$]{\includegraphics[width=0.44\linewidth]{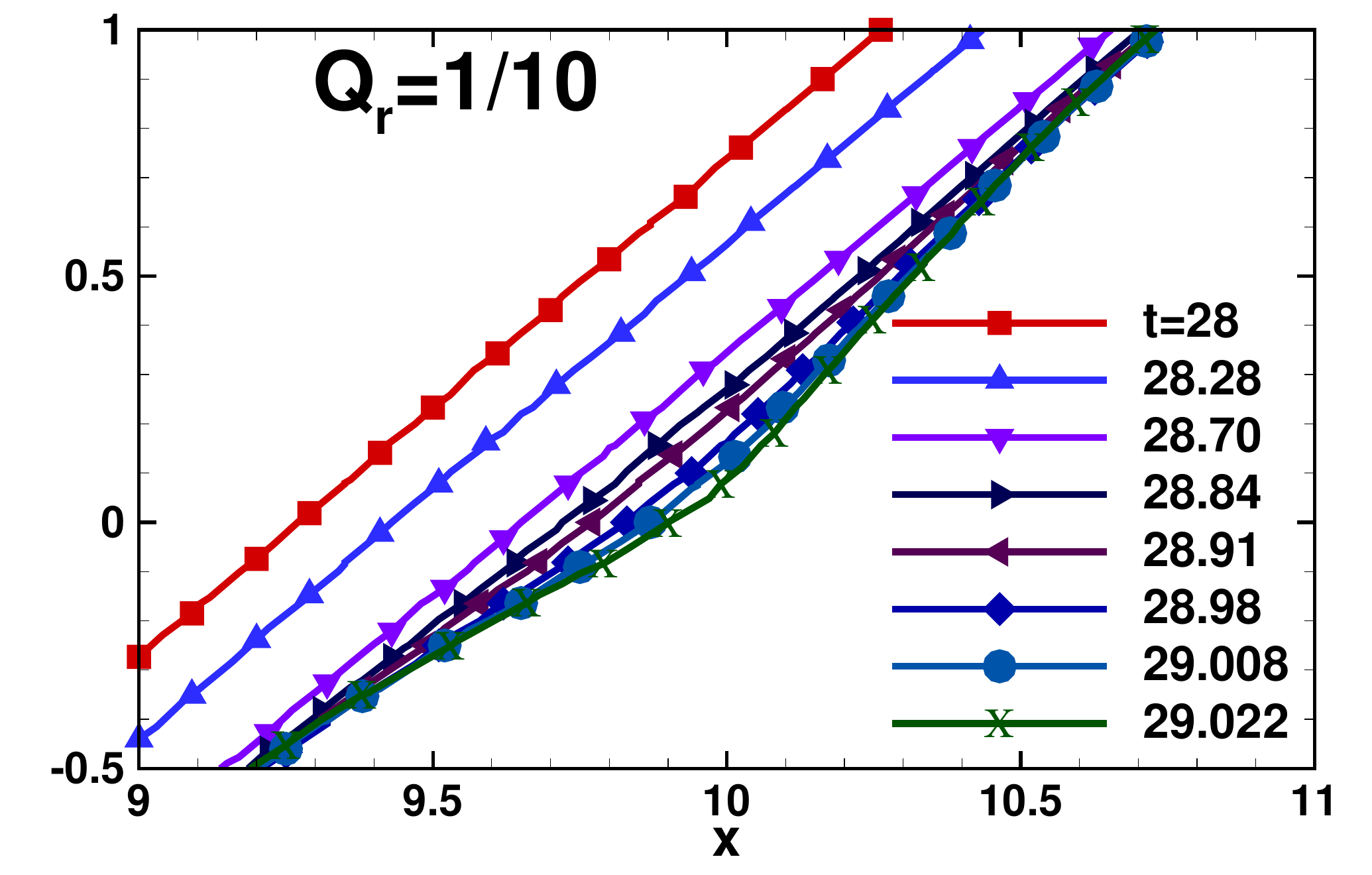}}
	\caption{Instantaneous interface (i.e., $x-y$ position) evolution profiles in S-2 stage for $\cac=10^{-4}$.}
	\label{fig:8}
\end{figure}
%
\subsection{Instantaneous interface evolution}\label{sec:evoi}
%
\noindent
In this section, the instantaneous evolution of interface curvature profiles is recorded as an essential feature to analyze and understand the mechanism of droplet formation and breakup.
The numerical data for the evolutions of the rear side interface curvature are extracted and plotted as a function of time and flow rate ratio.
The sequence of evolution of the liquid-liquid interface during the necking stage, near the pinch-off stage, and its geometrical coordinates ($x-y$) position on the primary channel are recorded in \fig\ref{fig:8} at different time ($t$) instants for wide range of the flow rate ratio ($0.1\le \qr\le 10$) under squeezing ($\cac\rev{<} 10^{-2}$) regime.
At higher flowrate ratios ($\qr \ge 5$), the interface curve shows a smooth bending at the pinch-off point, as shown in \figs\ref{fig:8}(a, b). For instance, the interface evolution has smooth bending like a precise concave shape at pinch-off time $t=1.9348$ for $\qr=10$. However, the interface curve attains a sharp bending before it pinches off for $\qr$ range from $2$ to $1/4$, as shown in \figs\ref{fig:8}(c-f).
The interface curvature has shown sharp V-shape at the pinch-off point ($t=3.9046$) for $\qr=1$, as shown in \fig\ref{fig:8}(d).
In contrast, the interface curves have not shown smooth bending at lower values of $\qr\le 1/8$. For example, the interface evolves as a straight line at pinch-off point ($t=22.3664$) for $\qr=1/8$ in \fig\ref{fig:8}(g).
Since the interfacial tension force contributes more and resists the shear imposed by the continuous phase, the interface does not show bending-like behaviour on its rear side, as observed at the higher flow rate ratios ($1/4 \leq \qr \leq 10$). The pressure acting on the concave side is always higher \citep{Abate2012}. Hence, the interface shows sharp fluctuations near the pinch-off stage as the interfacial force is balanced by the shear force exerted by the surrounding continuous phase due to the pressure.
It can be concluded from all these observations that the droplet shape is finally determined by the interface evolution only. Moreover, it is interesting to note that the pinch-off is happening precisely at the right corner of the junction point. The subsequent section further elaborates the above-discussed features to the instantaneous evolution of pressure.
%
\subsection{Instantaneous evolution of pressure}\label{sec:evop}
\begin{figure}[htbp]
	\centering
	\subfloat[$2\le \qr\le 10$]{\includegraphics[width=0.7\linewidth]{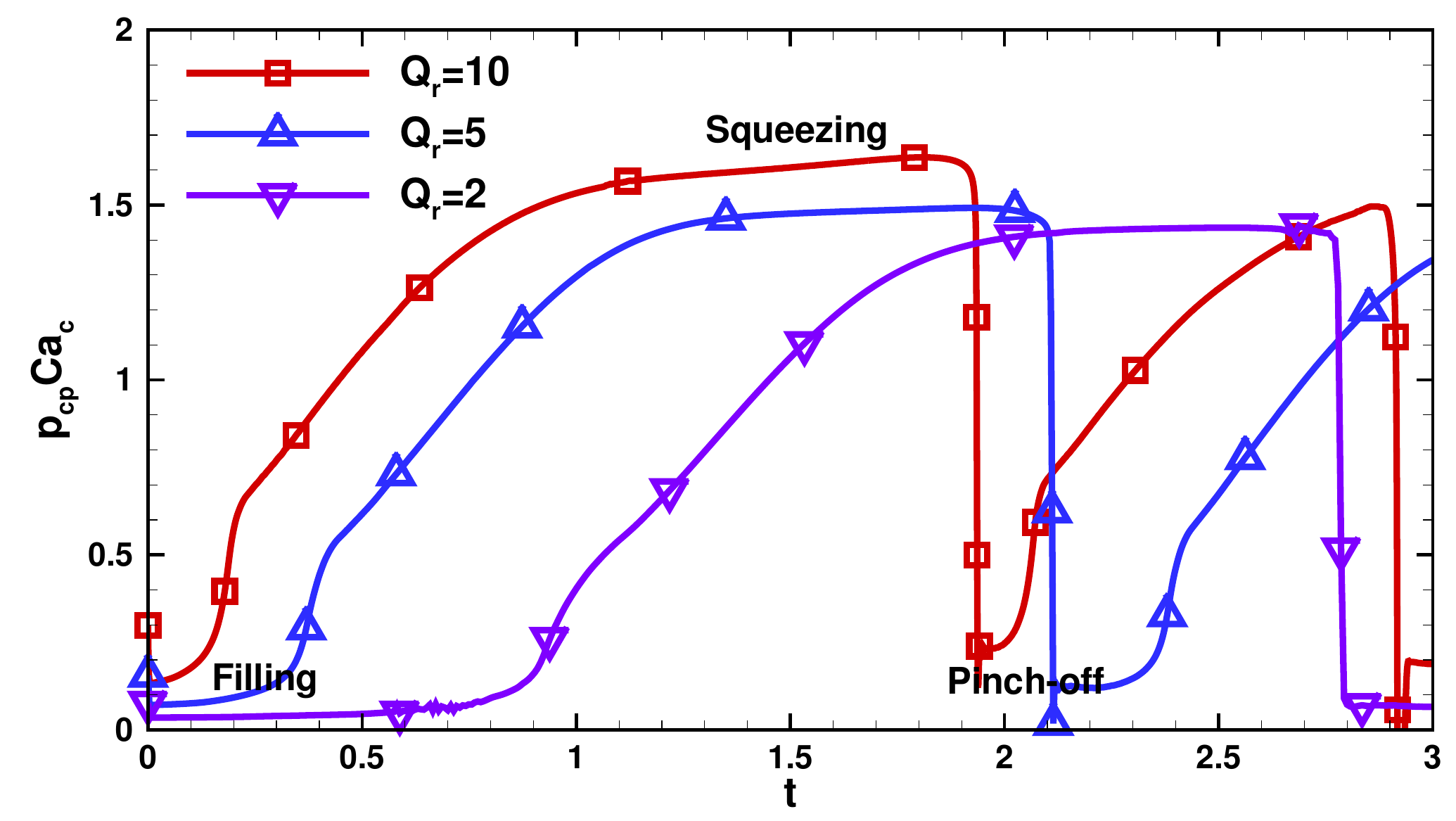}\label{fig:9a}}\\
	\subfloat[$1/4\le \qr\le 1$]{\includegraphics[width=0.7\linewidth]{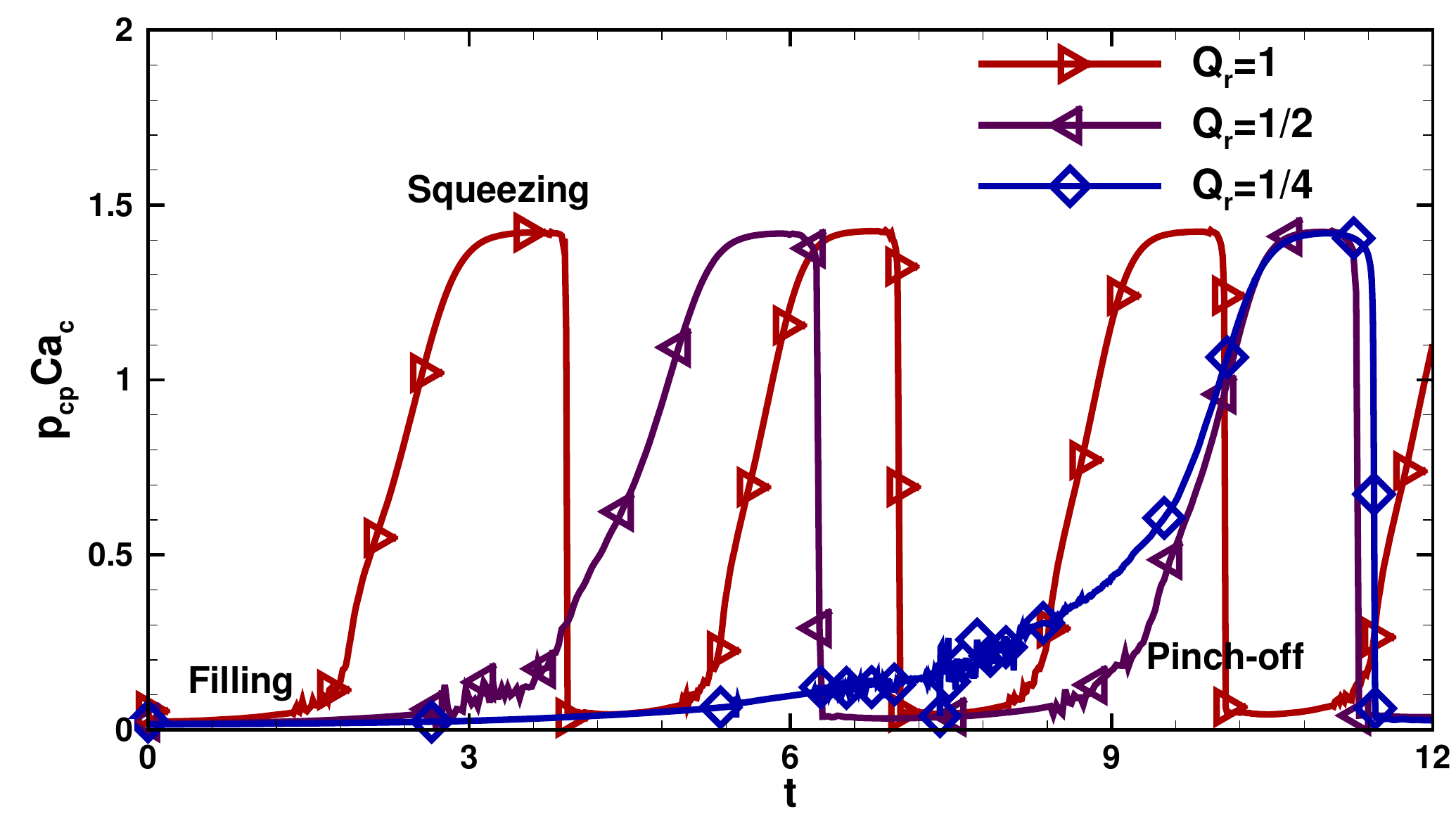}\label{fig:9b}}\\
	\subfloat[$1/10\le \qr\le 1/8$]{\includegraphics[width=0.7\linewidth]{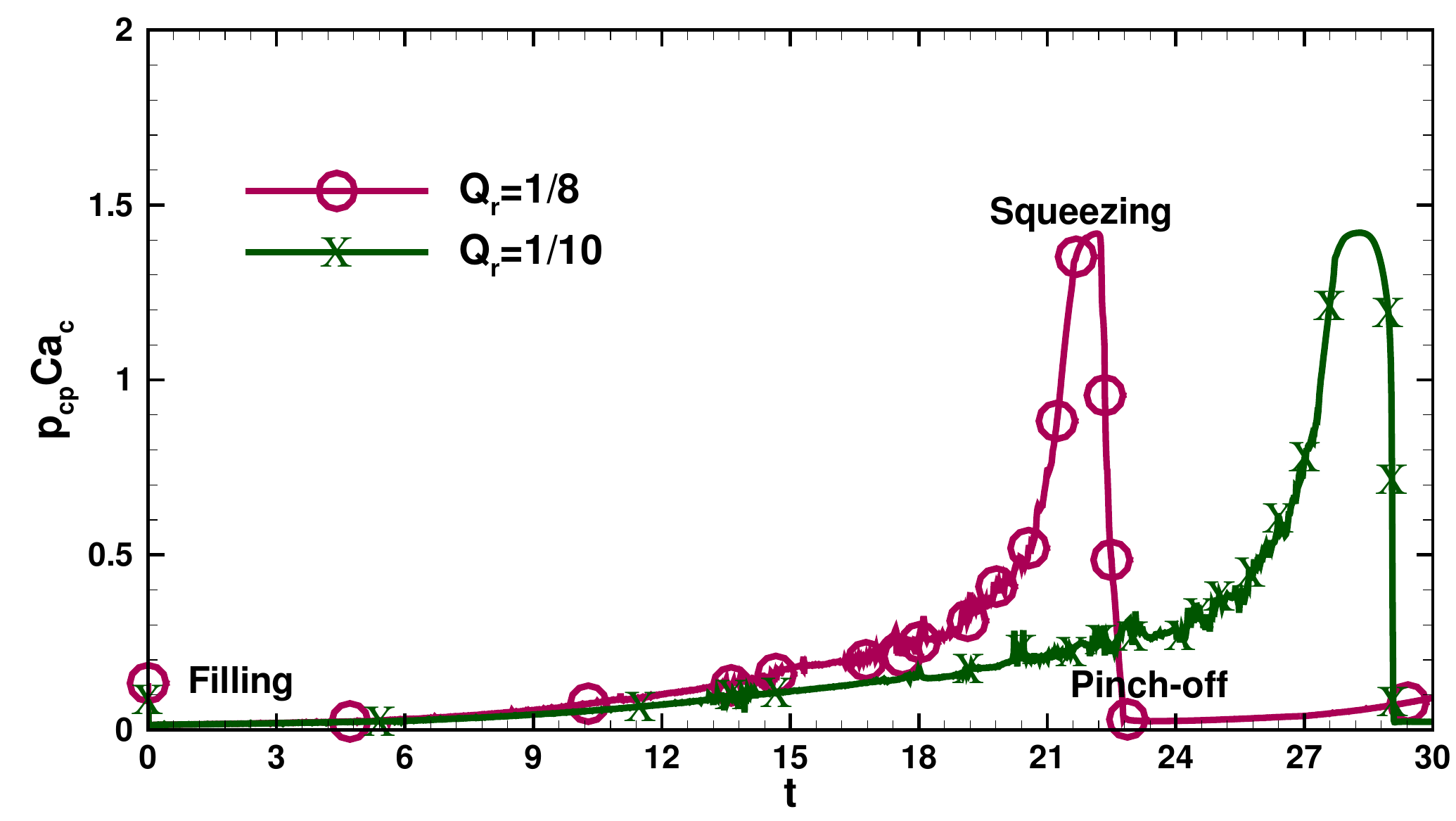}\label{fig:9c}}
	\caption{Instantaneous pressure evolution in the upstream ($p_{\text{cp}}$) for $\cac=10^{-4}$.}
	\label{fig:9}
\end{figure}

%
\noindent
In this section, the instantaneous pressure sensitivity is provided to elucidate the droplet pinch-off mechanism. In this work, two points `cp' and `dp' are chosen (\fig\ref{fig:1a}) at locations $(L_{\text u}-\wc/2, \wc/2)$ and   $(L_{\text u}+\wc/2, -\wdp/2)$ to measure the time-history of the pressure in CP and DP, respectively. \fig\ref{fig:9} depicts the instantaneous evolution of upstream (or continuous phase) pressure ($p_{\text{cp}}$) and its influence on the droplet formation for wide-ranging conditions ($1/10\le\qr\le10$, and $\cac < 10^{-2}$).
\noindent
Initially, the dispersed phase (DP) penetrates the primary channel at the T-junction through the vertical channel. In the filling stage, the upstream pressure ($p_{\text{cp}}$) is minimum ($p_{\text{cp,min}}$) and almost constant. It shows minor variation for all the values of $\qr$, as the dispersed phase has an insignificant restriction on the flow of continuous phase in the upstream. As the dispersed phase starts filling the cross-section of the primary channel, the upstream pressure gradually starts to rise due to continuously increasing obstruction to CP flow.  The filling stage ends when the tip of the dispersed phase reaches the top wall of the primary channel. As a result, CP flow is obstructed by filling DP. It consequently leads to a gradual followed by a sharp increase in the upstream pressure. In the next (i.e., squeezing) stage, the upstream pressure grows slowly and reaches the maximum value ($p_{\text{cp,max}}$) at which the width of the neck becomes minimum (i.e., $2r\approx 0$).
Due to immense pressure, the interface curvature becomes infinitely large, at which the width of the neck becomes the minimum (i.e., $2r\approx 0$).
It, subsequently, results in spontaneous pinch-off or detachment of the droplet from the dispersed phase.
During the spontaneous droplet pinch-off, there is a sudden fall in upstream pressure to its minimum value and an indication of completion of the droplet formation cycle.  The minimum upstream pressure ($p_{\text{cp,min}}$) attained at the pinch-off stage closely matches the pressure at the starting of the filling stage and, thereby, the repetition of the next droplet formation cycle  \citep{Garstecki2006}. It is further noticeable that the pressure buildup in the continuous phase (or upstream) of the primary channel is highly sensitive with time and strongly influencing the squeezing stage.
\noindent
This pressure variation cycle in the upstream CP flow repeats for each stage of the droplet formation. While the trends of pressure variation are qualitatively similar over the range of $\qr$, noticeable quantitative differences are observed. For instance, the maximum upstream pressure ($p_{\text{cp,max}}$) is achieved in comparatively lesser time for higher values of $\qr \ge 2$  (\fig\ref{fig:9a}) and higher time for lower values of $\qr\le 1$ (\figs\ref{fig:9b}-\ref{fig:9c}). This behaviour of attaining $p_{\text{cp,max}}$ is attributed to the relative flow rate of two phases, i.e., the pressure build-up is faster for $\qr>1$ (i.e., $\qd > \qc$) due to faster completion of the filling stage, and slower for  $\qr\le 1$ (i.e., $\qd \le \qc$).

\begin{figure}[htbp]
	\centering
	\includegraphics[width=0.9\linewidth]{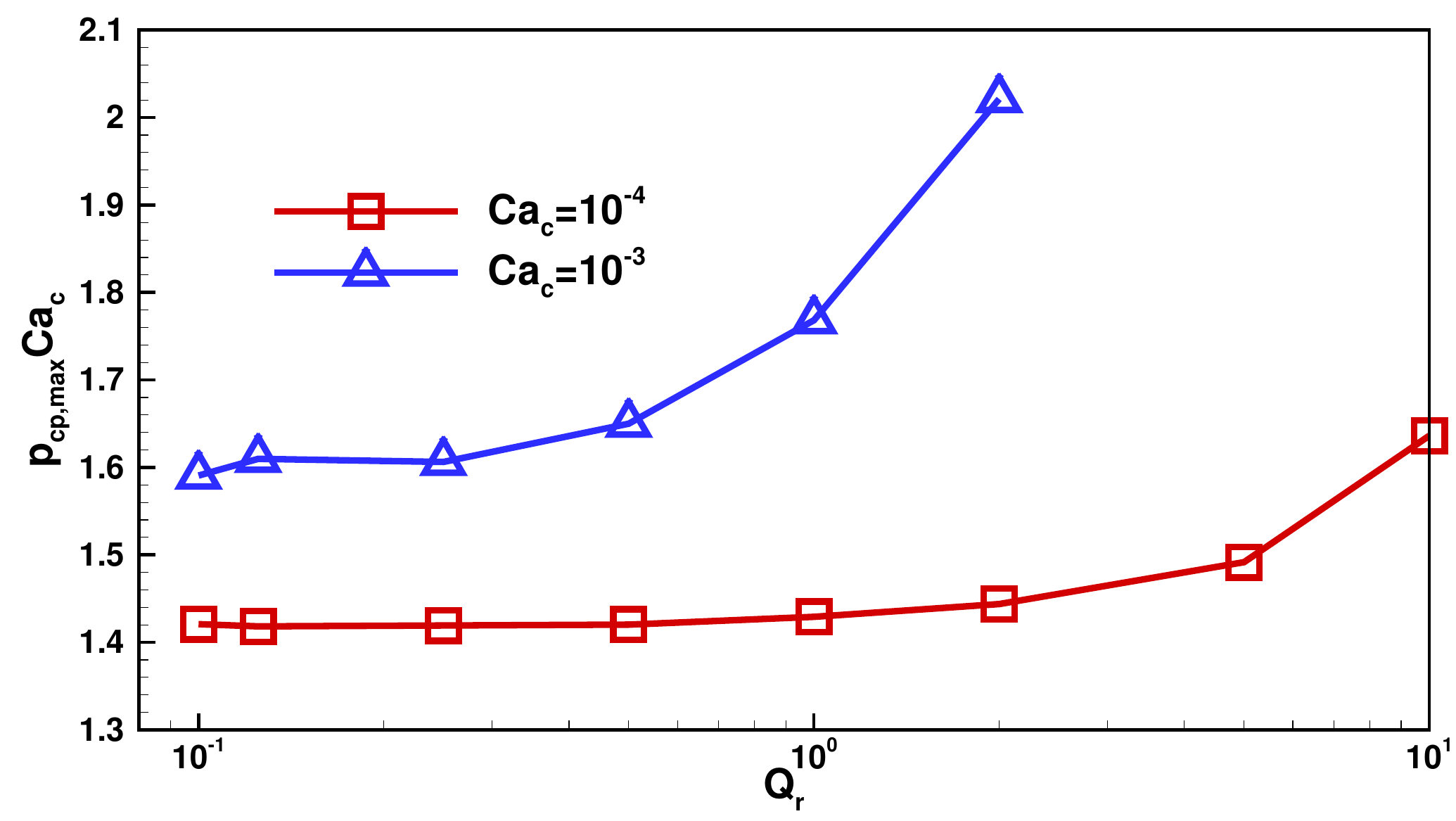}
	\caption{Maximum upstream pressure ($p_{\text{cp,max}}$) as a function of $\qr$ and $\cac$.}
	\label{fig:10}
\end{figure}
\begin{figure}[htbp]
	\centering
	\subfloat[$\qr=10$]{\includegraphics[width=0.45\linewidth]{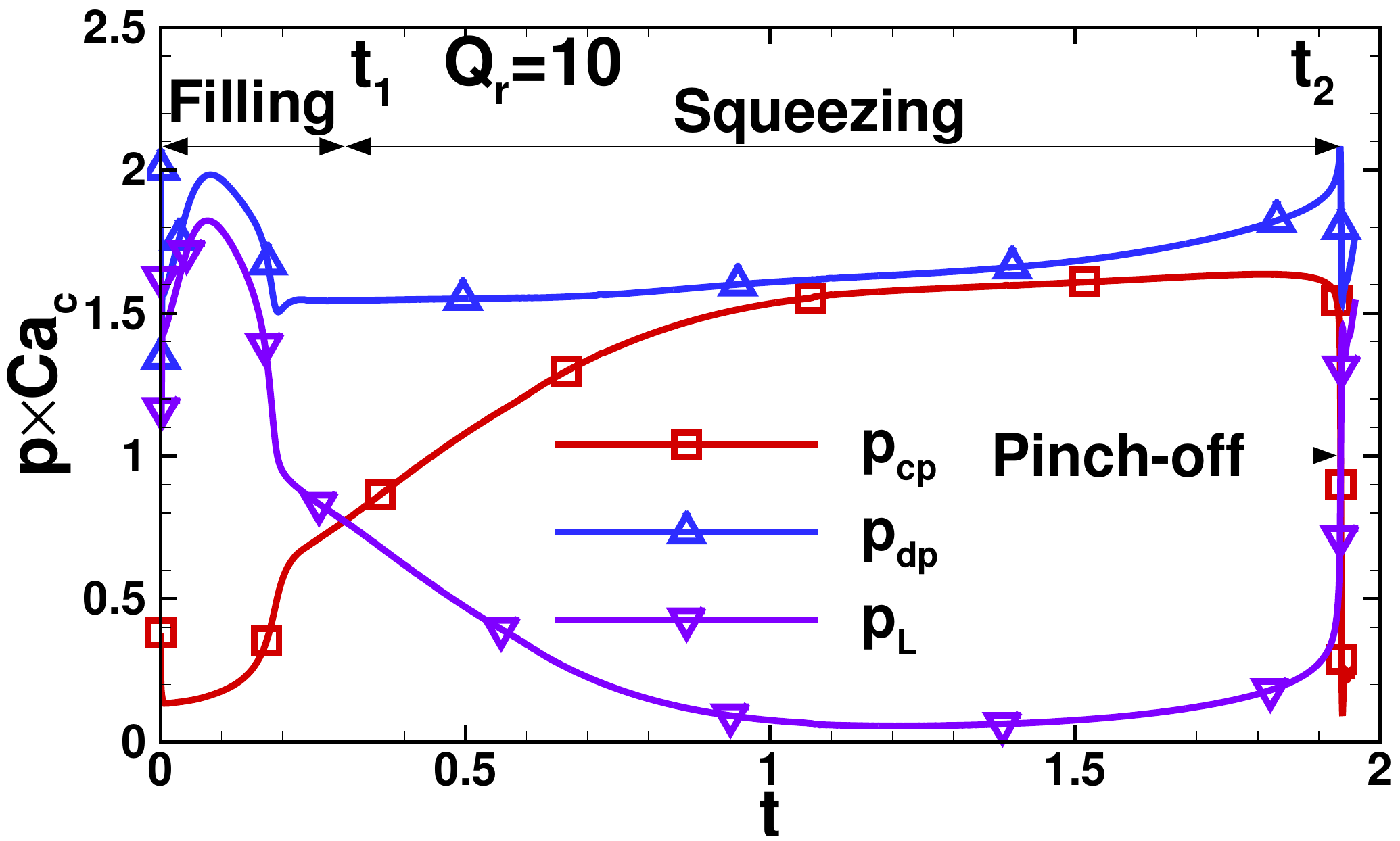}}
	\subfloat[$\qr=5$]{\includegraphics[width=0.45\linewidth]{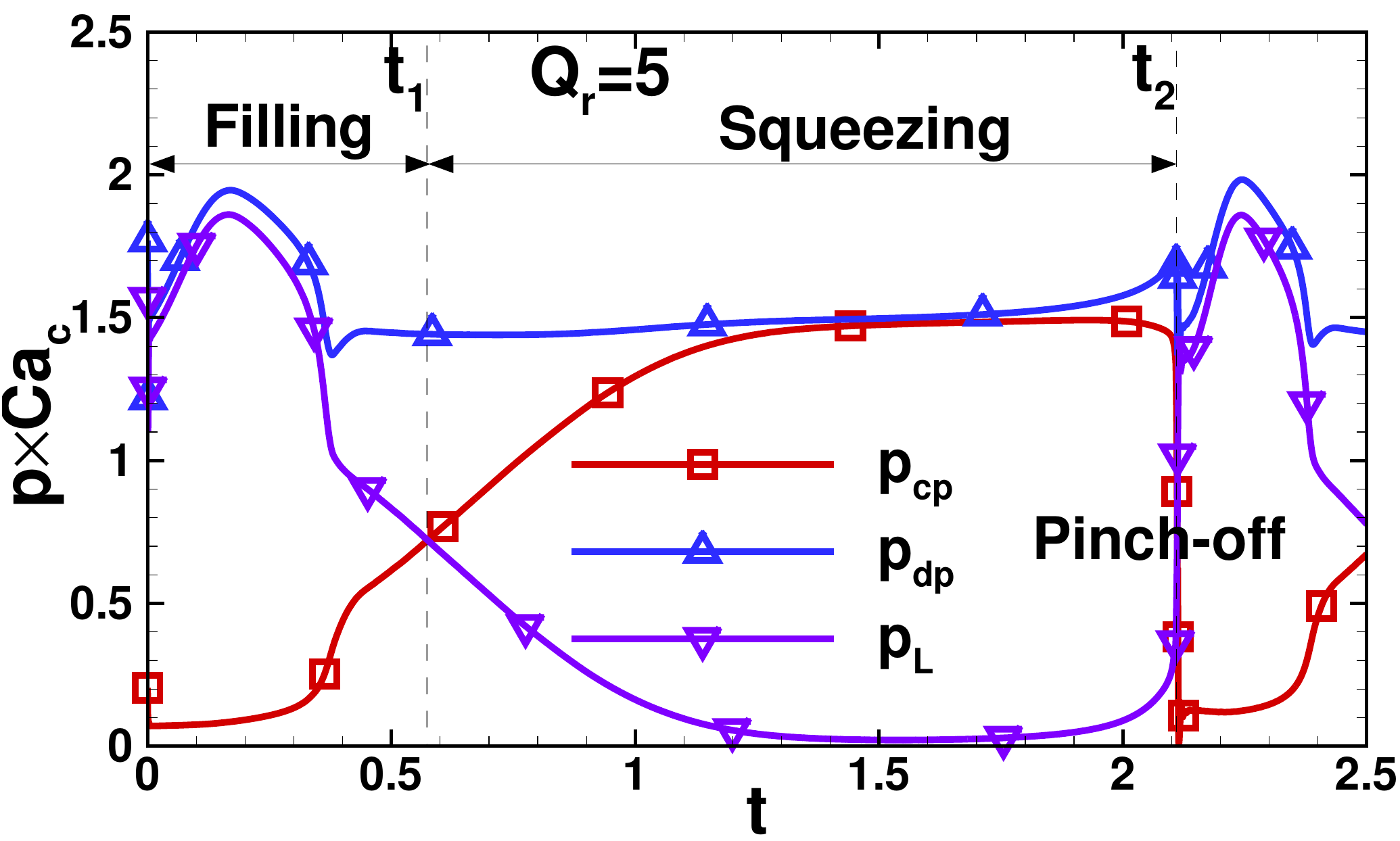}}\\
	\subfloat[$\qr=2$]{\includegraphics[width=0.45\linewidth]{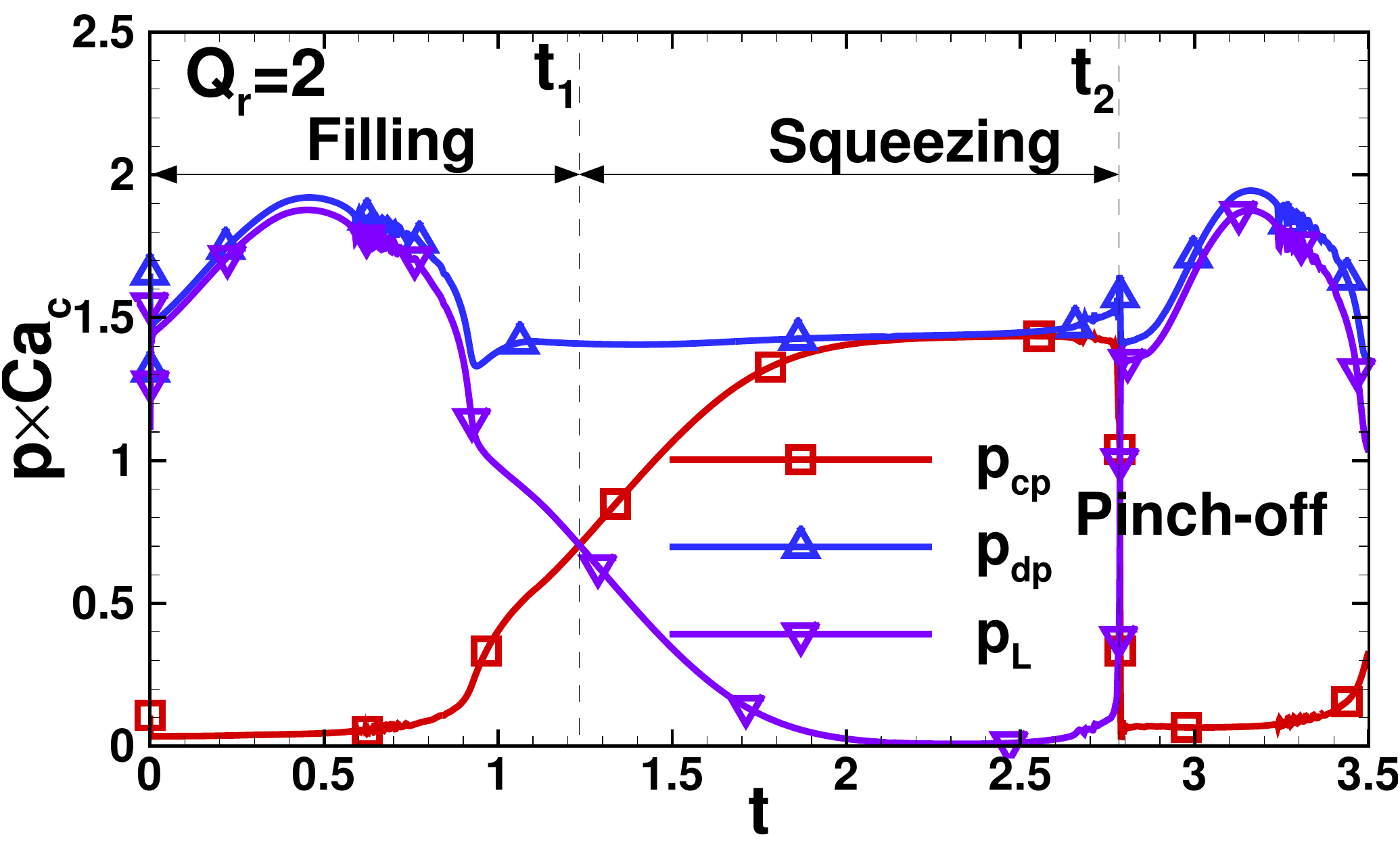}}
	\subfloat[$\qr=1$]{\includegraphics[width=0.45\linewidth]{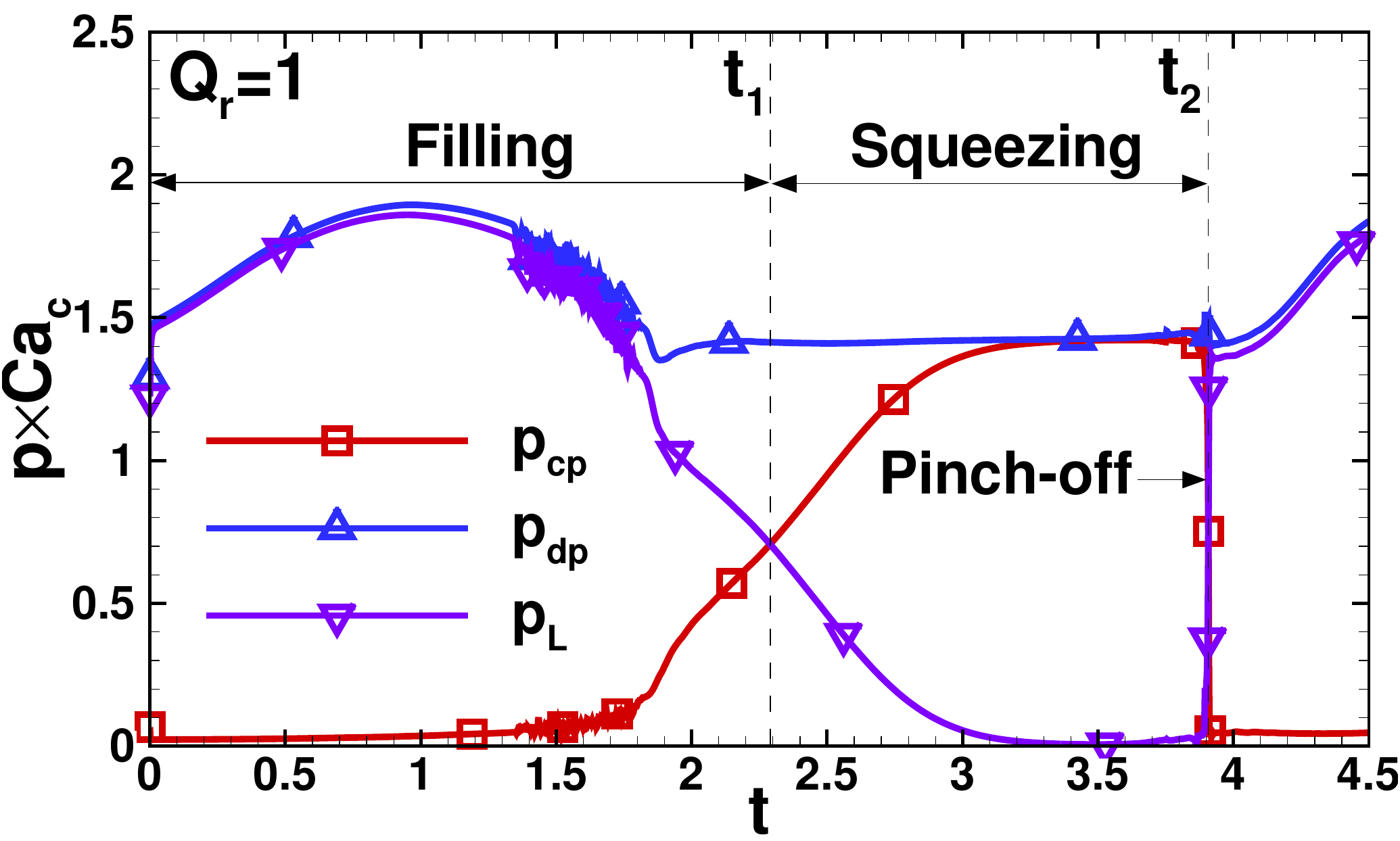}}\\
	\subfloat[$\qr=1/2$]{\includegraphics[width=0.45\linewidth]{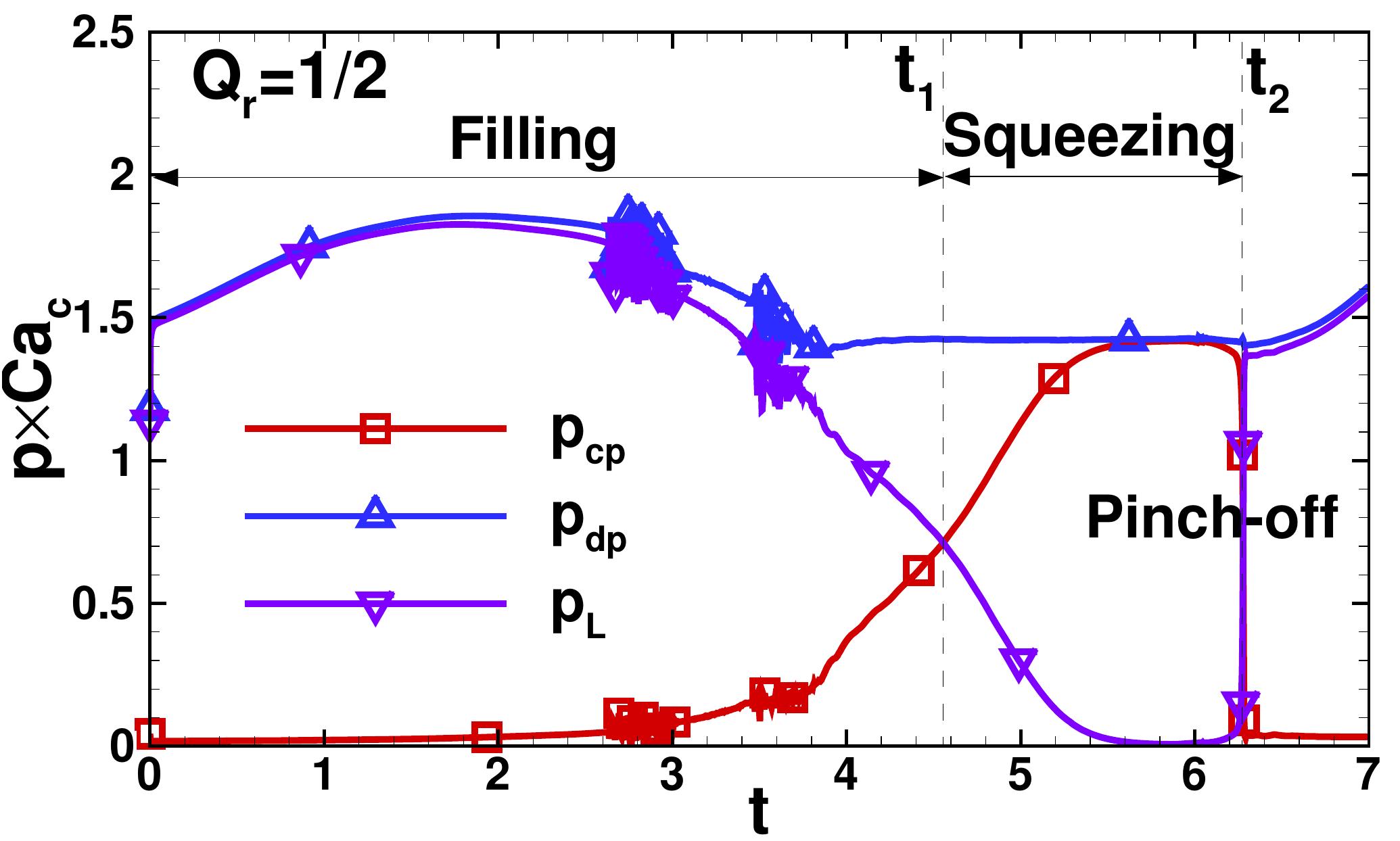}}
	\subfloat[$\qr=1/4$]{\includegraphics[width=0.45\linewidth]{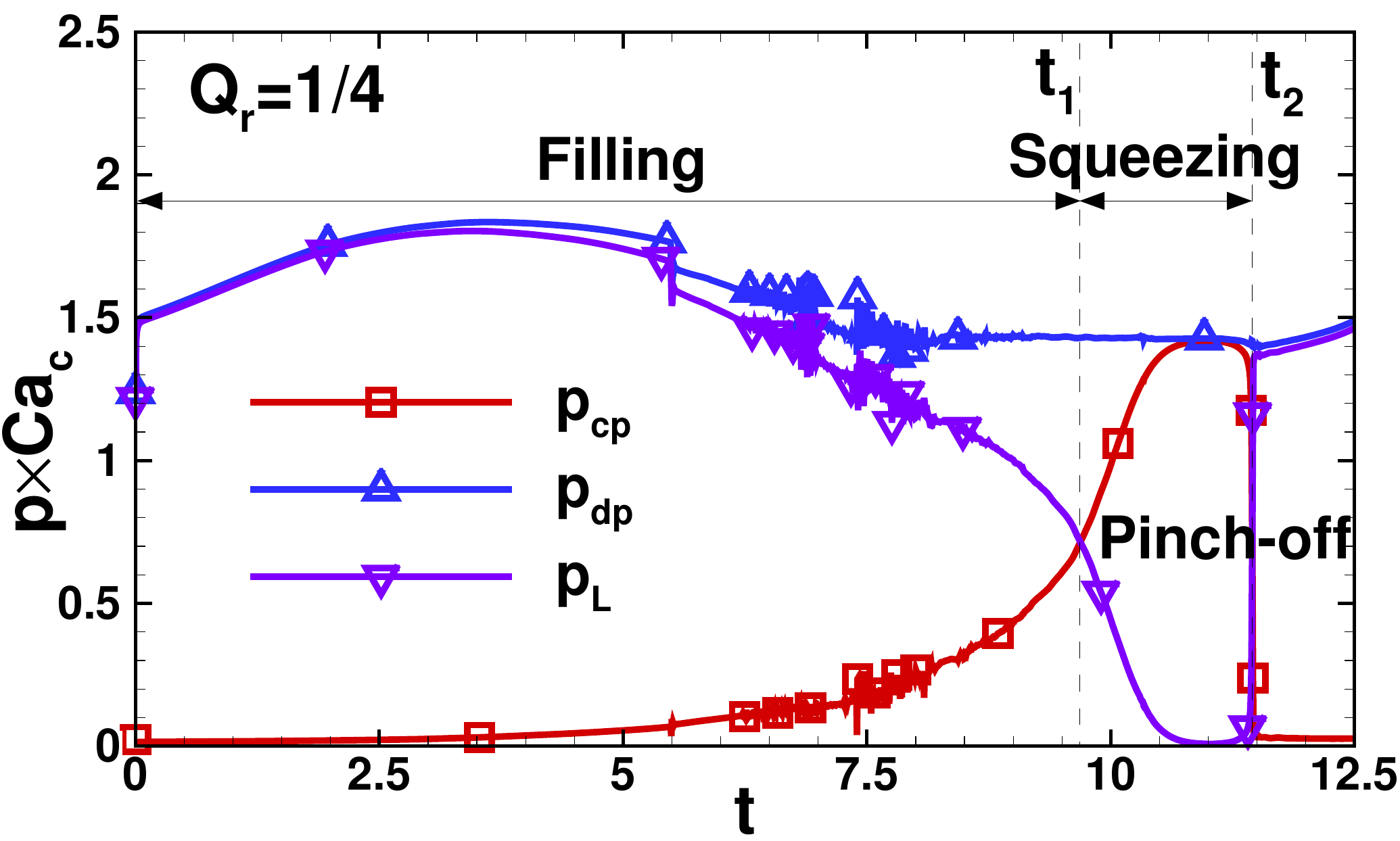}}\\
	\subfloat[$\qr=1/8$]{\includegraphics[width=0.45\linewidth]{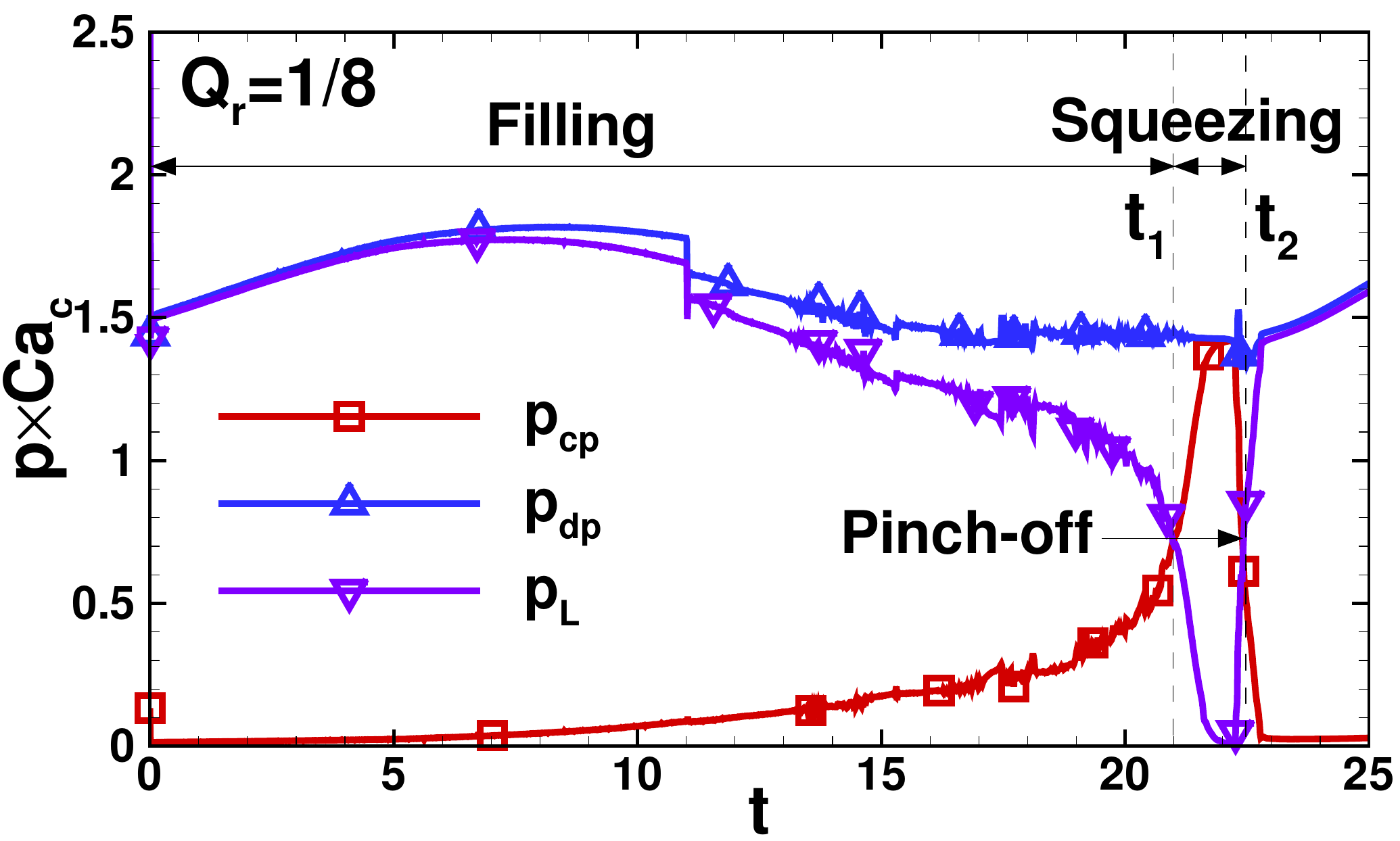}}
	\subfloat[$\qr=1/10$]{\includegraphics[width=0.45\linewidth]{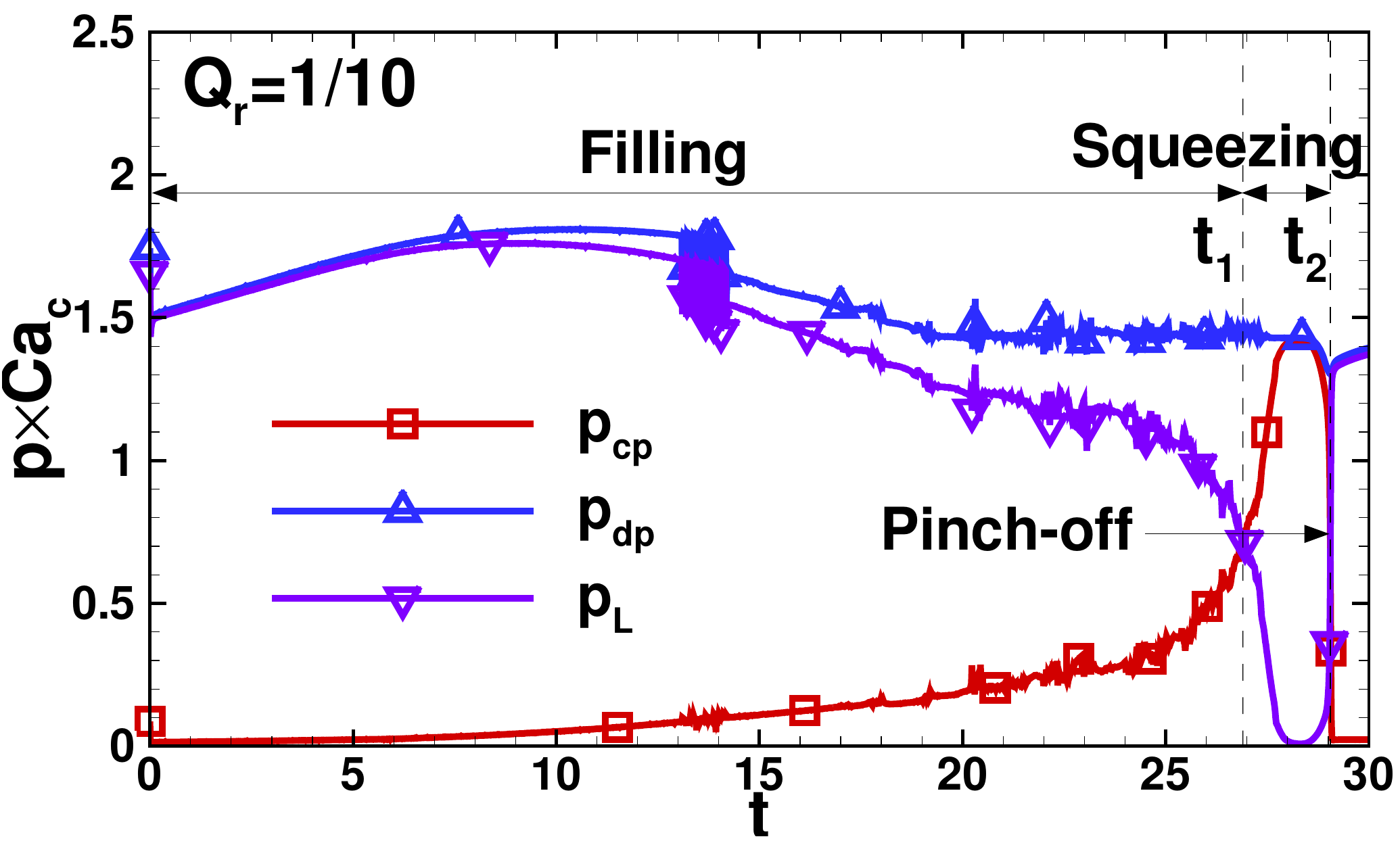}}\\
	\caption{The instantaneous evolution of pressure profiles at the different locations near the junction at $\cac=10^{-4}$.}
	\label{fig:11}
\end{figure}

\noindent
\fig\ref{fig:10} shows the maximum upstream pressure ($p_{\text{cp,max}}$) in continuous phase as a function of the flow rate ratio ($\qr$) and capillary number ($\cac$).  For a given $\cac$, the maximum pressure ($p_{\text{cp,max}}$) display proportional variation with $\qr$.  The numerical values of $p_{\text{cp,max}}$ are best correlated linearly with $\qr$ as follows.
\begin{gather}
	p_{\text{cp,max}}=A+B\qr
	\label{eqn:Pmax}
\end{gather}
where, the statistical analysis results $A=(211.93 + 13400\cac)$ and $B = (15487 -\rev{1 \times} 10^{7}\cac)$ with $R^{2}=0.98$.
\noindent
\fig\ref{fig:11} shows the profiles of the upstream or the continuous phase pressure ($p_{\text{cp}}$), dispersed phase pressure ($p_{\text{dp}}$), and the Laplace pressure ($p_{\text{L}}$, i.e., the pressure difference across the interface between DP and CP, \eqn\ref{eqn:Pdc}).
\begin{gather}
	p_{\text{L}} = \Delta p_{(\text{dc})}=p_{\text {dp}}-p_{\text {cp}}
	\label{eqn:Pdc}
\end{gather}
The pressure in the dispersed phase is always higher than the pressure in the continuous phase, i.e., $p_{\text {dp}}>p_{\text {cp}}$, under otherwise identical conditions \citep{Bashir2014}.
Further, both $p_{\text{cp}}$ and $p_{\text{dp}}$ profiles have shown converse trends (i.e., when one is increasing, other is decreasing) with time ($t$) during the droplet formation cycle.

\noindent
During the filling stage, the pressure in the dispersed phase ($p_{\text{dp}}$) gradually increases and then decreases with time ($t$). Afterward, it remains almost constant from the squeezing to the pinch-off stage, as it is unaffected by the shear of the continuous phase. The growth of $p_{\text{dp}}$ in the filling stage is strongly related to $\qr$.
A sharp variation in pressure seen for higher $\qr$, i.e., less time taken by the filling stage, flattens with decreasing $\qr$. The region of increasing  $p_{\text{dp}}$ is primarily attributed to the balancing of the interfacial tension (IFT) force with a pressure force, whereas   $p_{\text{dp}}$ decreases due to the balancing of frictional shear and interfacial tension forces.  The maximum  $p_{\text{dp}}$ is achieved at the point where viscous force start influencing the filling of the dispersed phase.

\noindent
\fig\ref{fig:11} also depicts the variation of the Laplace pressure ($p_{\text{L}} $, \eqn\ref{eqn:Pdc})  with time ($t$). The Laplacian pressure ($p_{\text{L}}$) shows a sharp increase and decrease, followed by a gradual reduction with time in the filling stage.  The blocking or filling stage ends at time $t_1$ where $p_{\text{L}} = p_{\text{cp}}$, i.e., continuous phase pressure ($p_{\text{cp}}$)
and Laplacian pressure ($p_{\text{L}}$) curves intersect to each other.  The Laplacian pressure ($p_{\text{L}}$) is continuously dropping to its minimum value, followed by an incremental rise in the squeezing stage.  The behaviour of $p_{\text{L}}$ is attributed to the DP resisting the significant shear exerted from the developing CP streaming pressure ($p_{\text{cp}}$) and the frictional resistance by the channel wall confinement on the liquid-liquid interface.
Subsequently, $p_{\text{cp}}$ attains maximum whereas $p_{\text{dp}}$ simultaneously attains minimum value and thereby resulting in minimum neck width \rev{($2r_{\text{min}}$)} at the end of sqeezing stage at time $t_2$.
After that, during the droplet breakup stage, DP pressure ($p_{\text{dp}}$) build-up due to minimum neck width and CP pressure ($p_{\text{cp}}$) demeaning due to larger available flow area, both simultaneously and spontaneously,  results in spontaneous increase of the pressure $p_{\text{L}}$. Finally, the condition $p_{\text{L}} = p_{\text{cp}}$ repeats at time $t_3$ for the second time where the droplet detachment from the dispersed phase takes place spontaneously.
The sharp change (i.e., increase or decrease) seen at higher $\qr$ transit to gradual increase/decrease with decreasing $\qr$. The Laplace pressure ($p_{\text{L}}$) cycle also repeats with droplet formation cycle.

\noindent
As discussed earlier (in Section \ref{sec:stages}), the time duration of various stages of the droplet formation is strongly dependent on the flow rate ratio ($\qr$) and capillary number ($\cac$). For a given $\cac$, the squeezing time ($t_{\text{s}}=t_2-t_1$) reduces with decreasing value of $\qr$, mainly due to the confinement effects and strengthening of the interfacial tension force.
\begin{figure}[htbp]
	\centering
	\subfloat[$\qr=10$]{\includegraphics[width=0.45\linewidth]{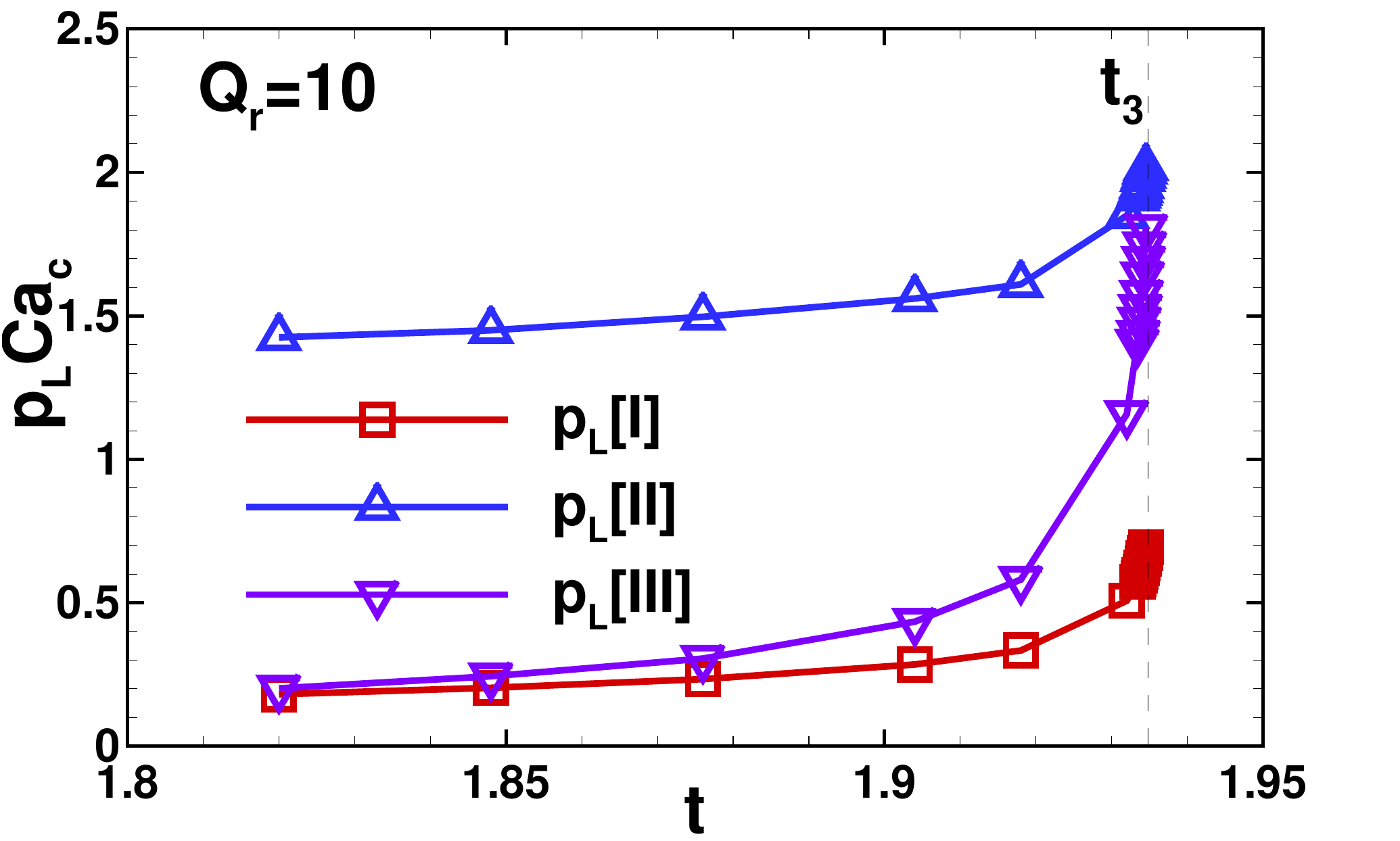}}
	\subfloat[$\qr=5$]{\includegraphics[width=0.45\linewidth]{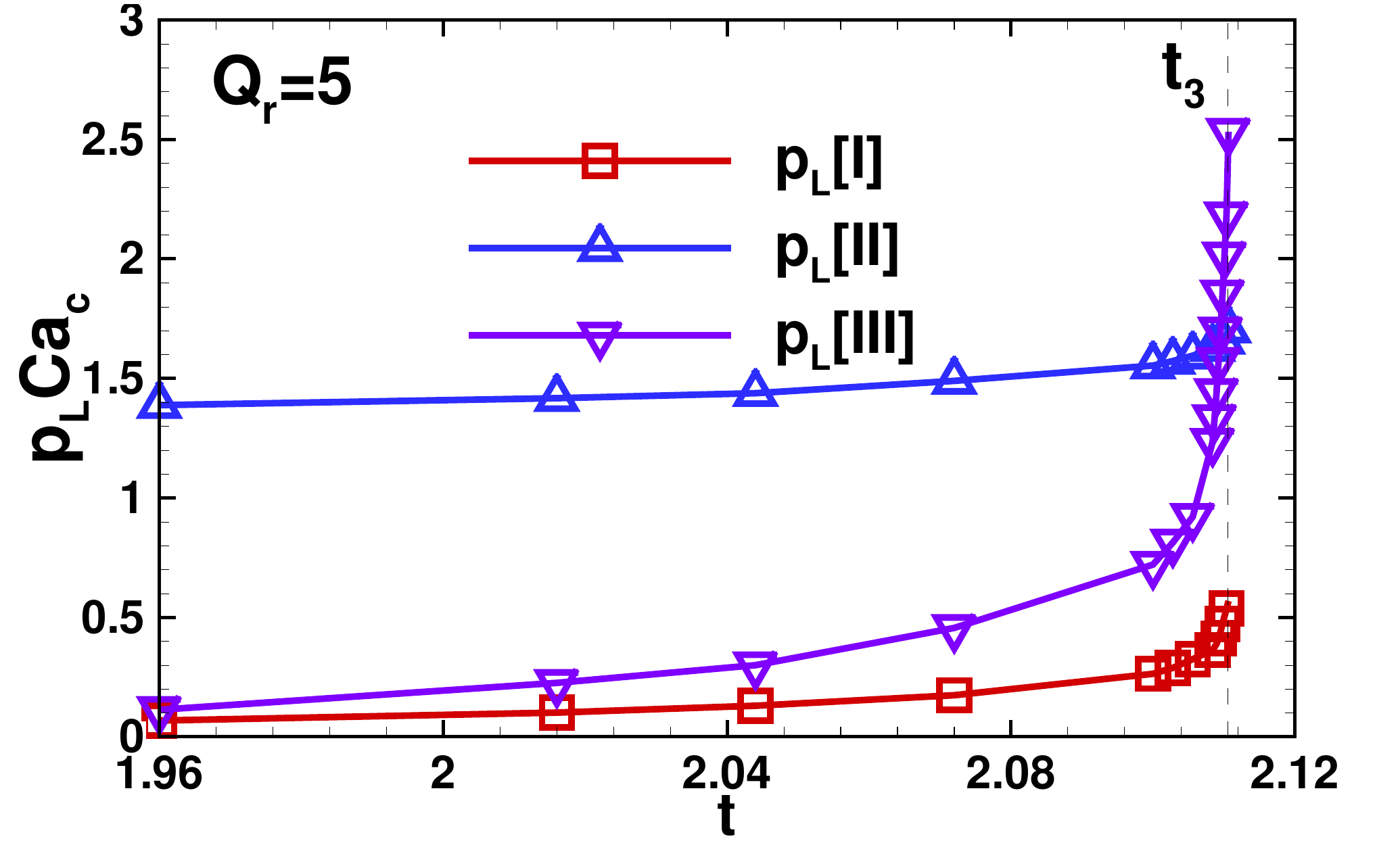}}\\
	\subfloat[$\qr=2$]{\includegraphics[width=0.45\linewidth]{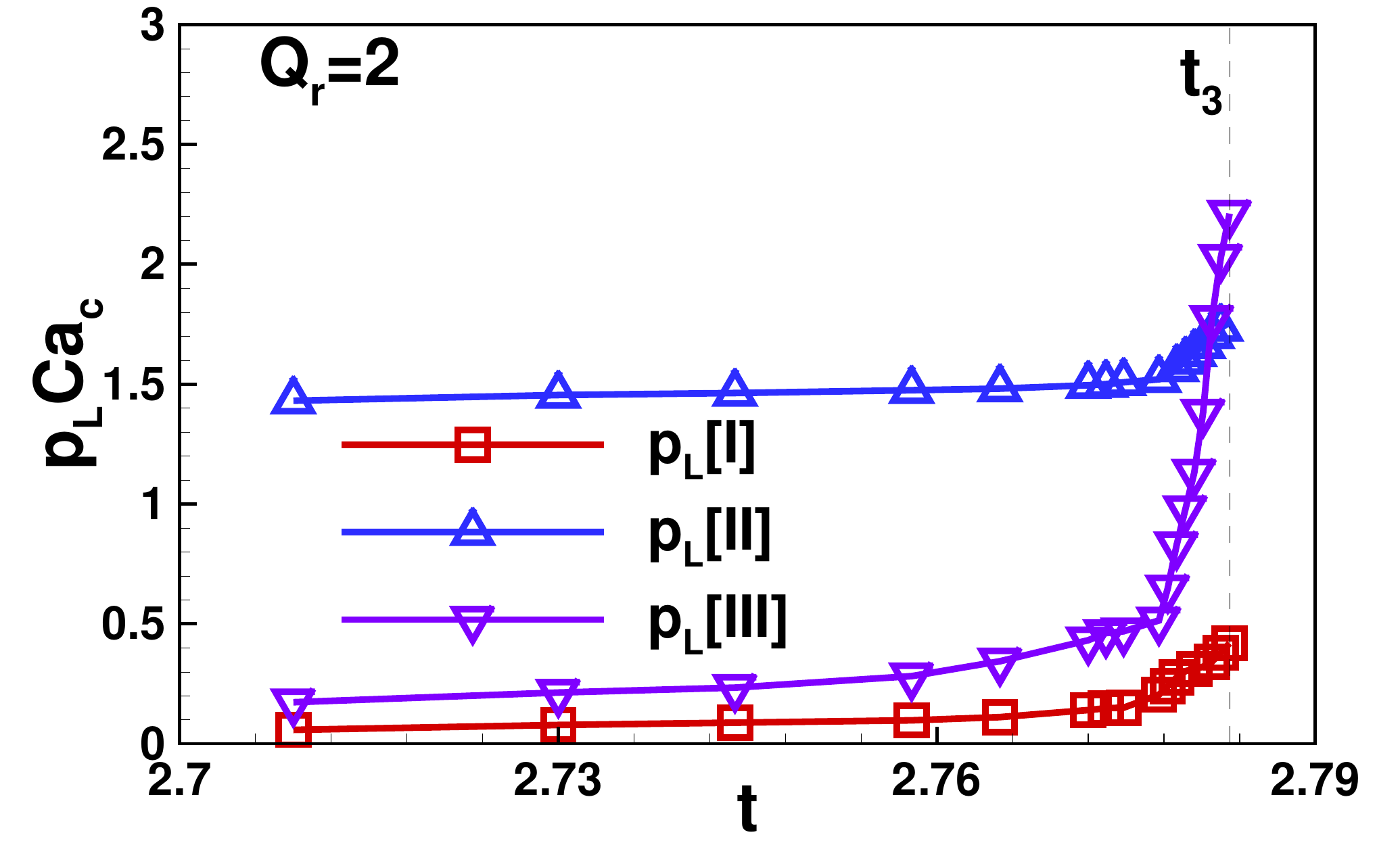}}
	\subfloat[$\qr=1$]{\includegraphics[width=0.45\linewidth]{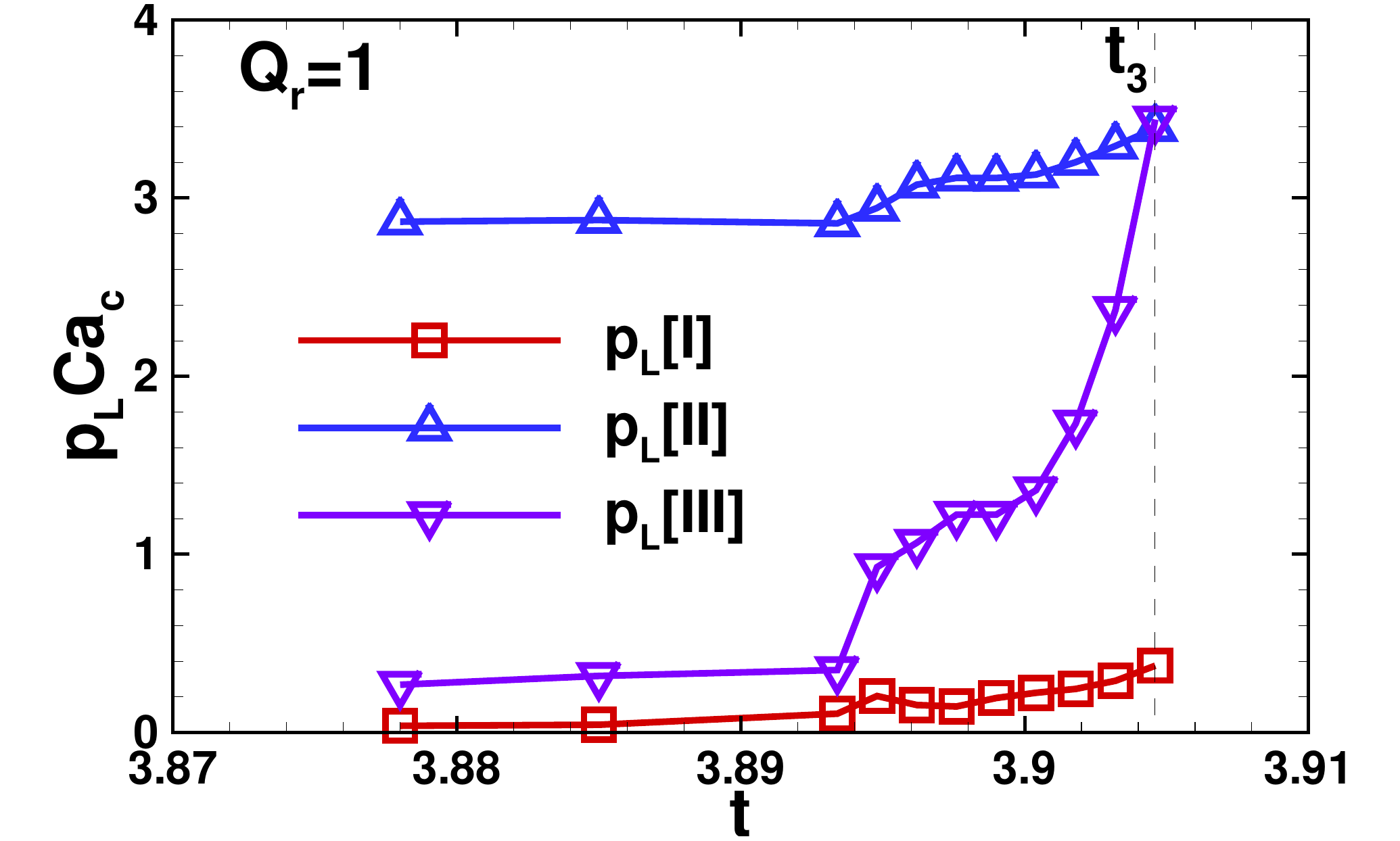}}\\
	\subfloat[$\qr=1/2$]{\includegraphics[width=0.45\linewidth]{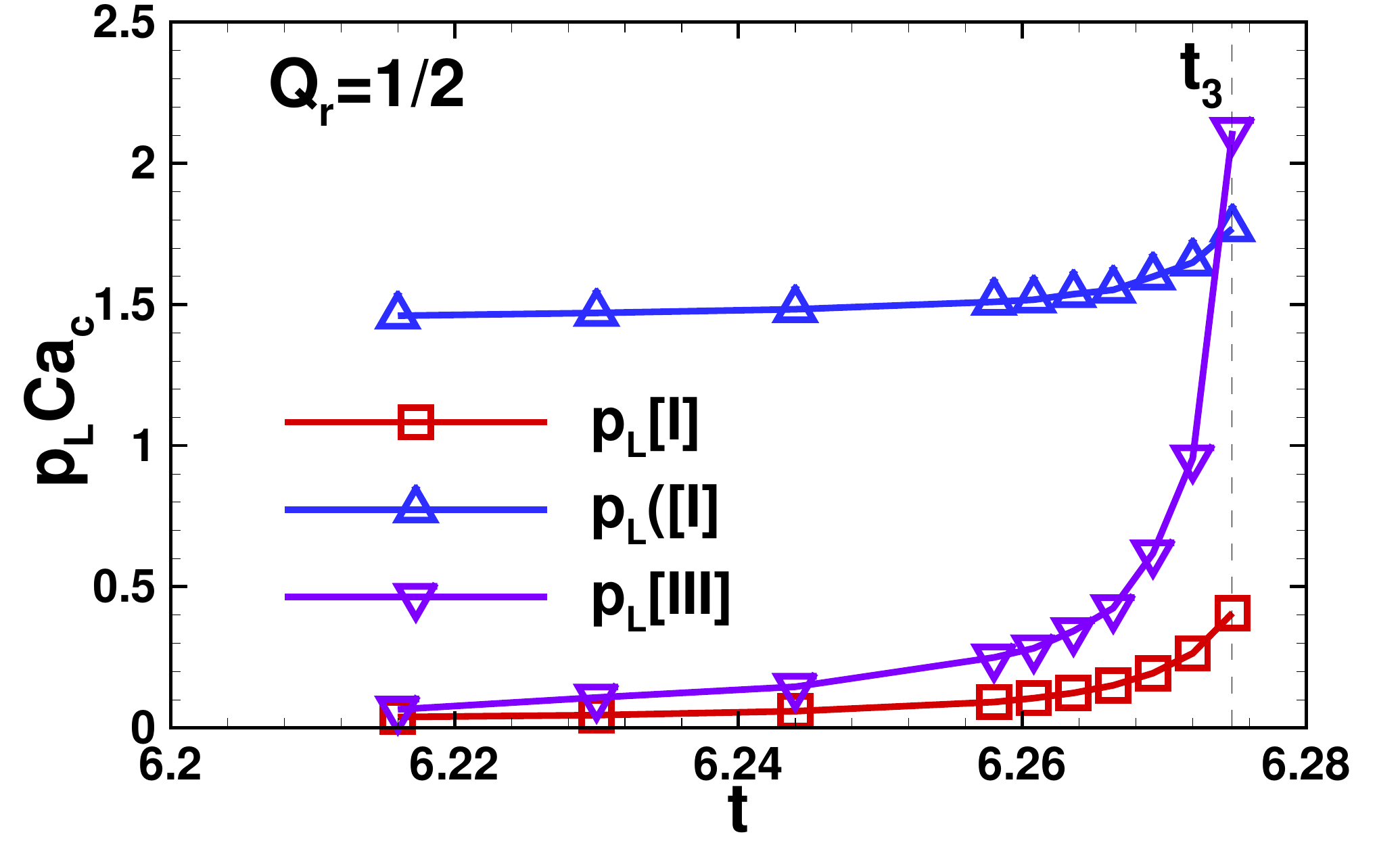}}
	\subfloat[$\qr=1/4$]{\includegraphics[width=0.45\linewidth]{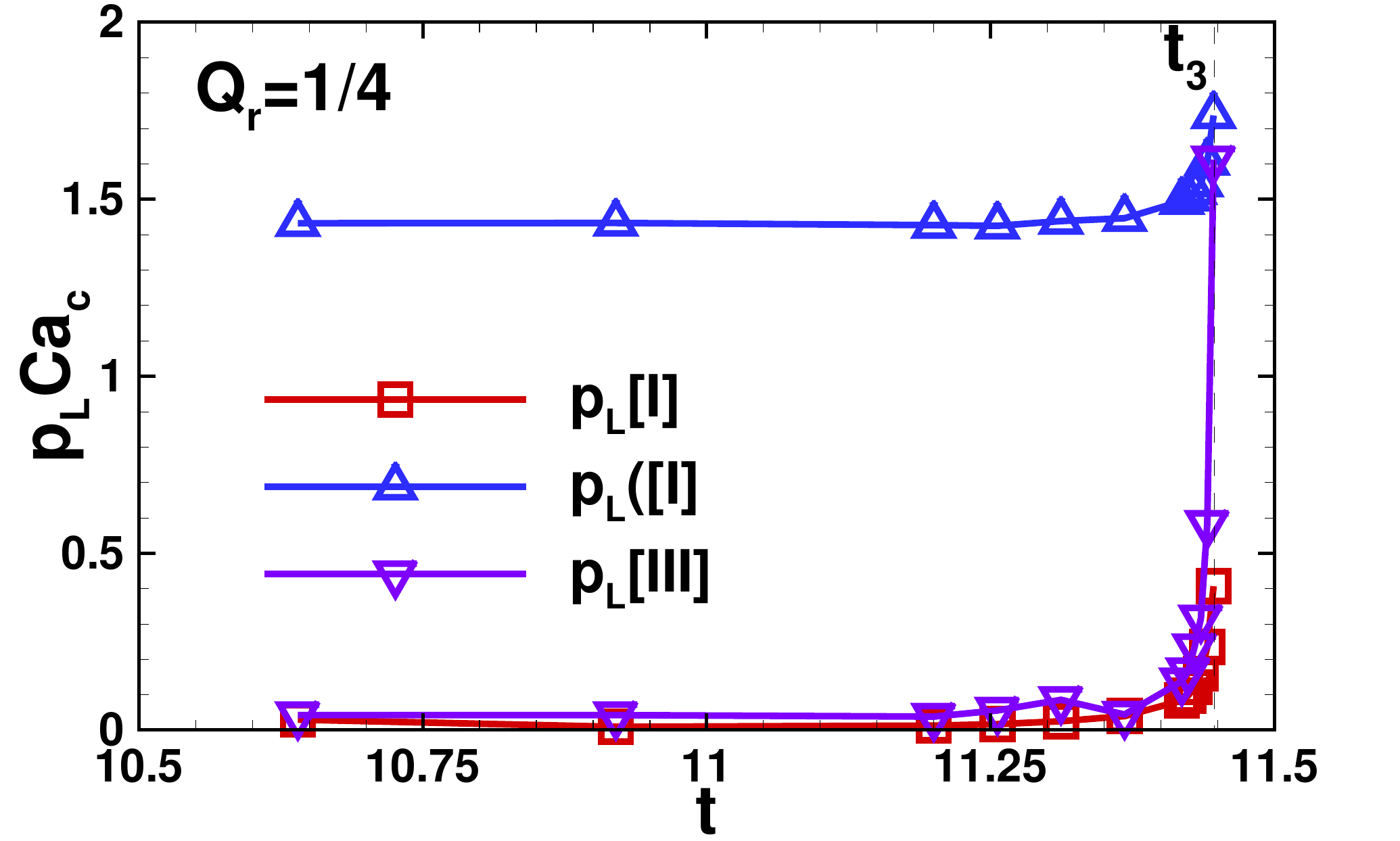}}\\
	\subfloat[$\qr=1/8$]{\includegraphics[width=0.45\linewidth]{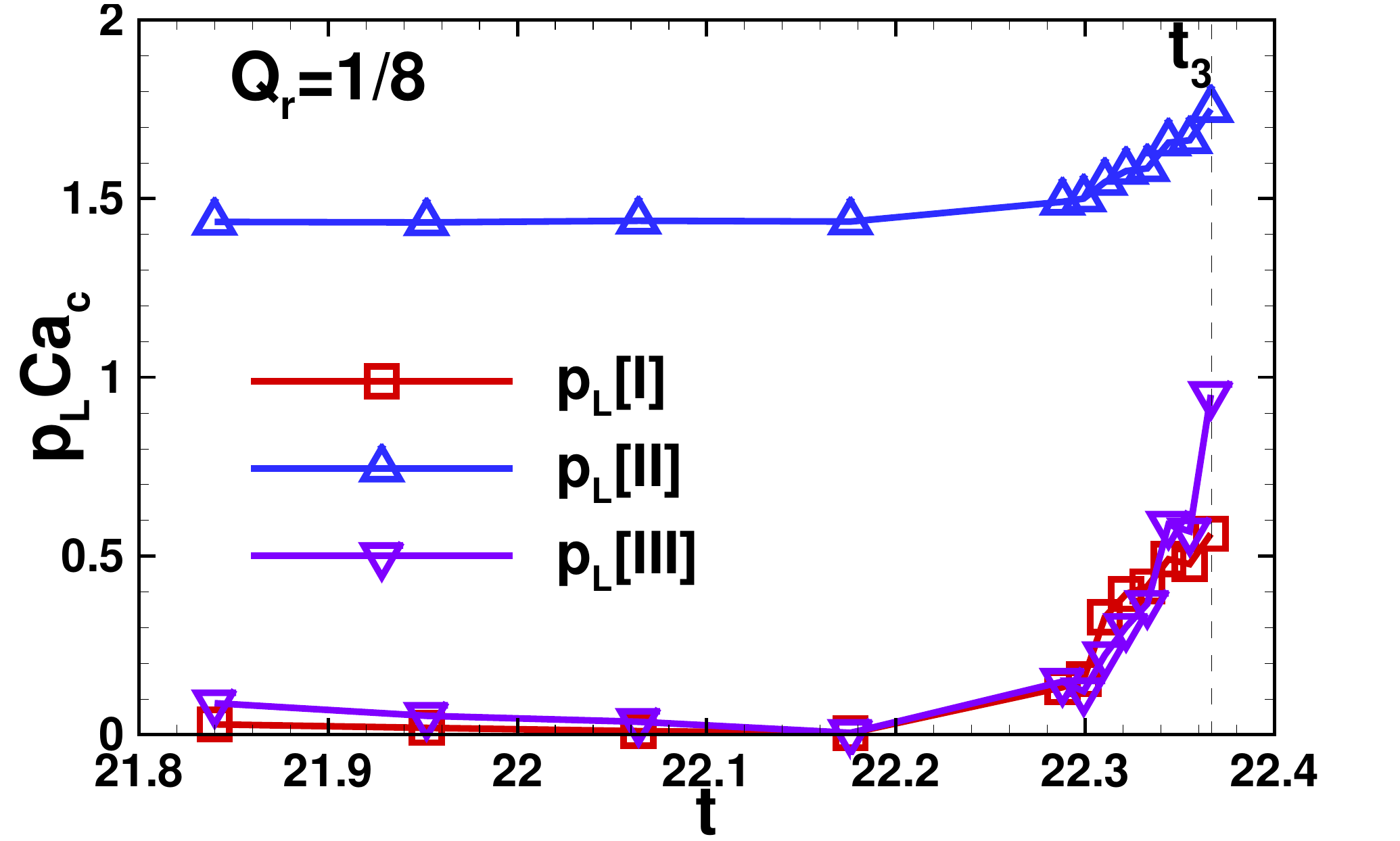}}
	\subfloat[$\qr=1/10$]{\includegraphics[width=0.45\linewidth]{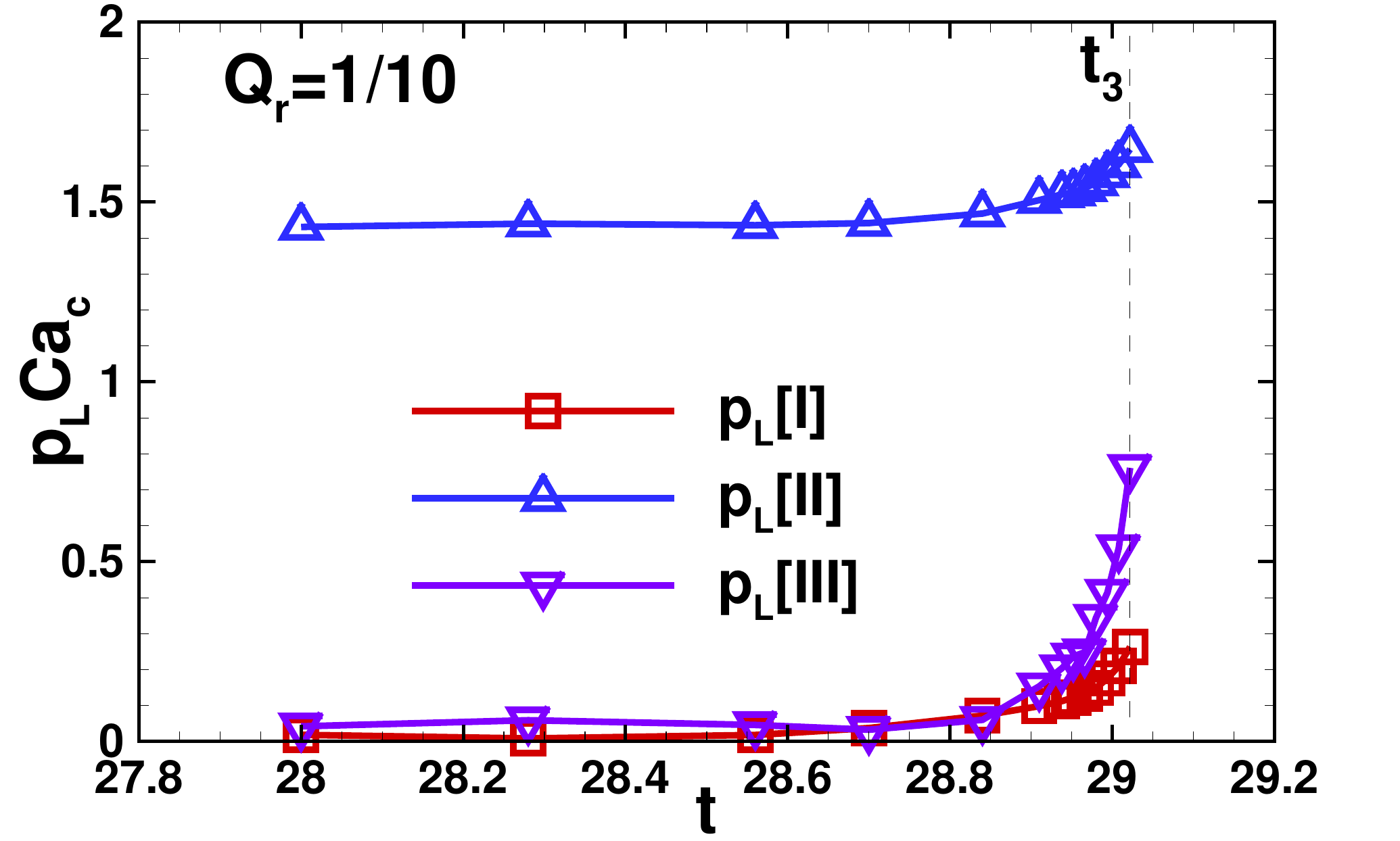}}\\
	\caption{Instantaneous Laplace pressure ($p_{\text{L}}$), acting on the neck during the pinch-off stage, evaluated using [I] \eqn(\ref{eqn:Pdc}), [II] \eqn(\ref{eqn:PRfRr}), and [III] \eqn(\ref{eqn:PRcmin}) at $\cac=10^{-4}$.}
	\label{fig:13}
\end{figure}
%
\subsection{The droplet pinch-off mechanism}\label{sec:dpm}
%
\noindent
Understanding the role of the evolution of curvature of the interface during the droplet pinch-off stage is very essential to the droplet dynamics.
%
%
The Laplace pressure ($p_{\text{L}}$) profiles discussed in preceding section display important condition ($p_{\text{L}} = p_{\text{cp}}$) of transiting stages of the droplet formation, and thus, the evolution of interface curvature.
In this section, the mechanism of the droplet pinch-off stage is presented and analyzed in terms of the instantaneous evolution of the Laplace pressure ($p_{\text{L}}$) acting on the neck, local minimum radius of curvature ($\rcmin$),  the neck width ($2r$) as a function of $\qr$ and $\cac$.

\noindent
The Laplace pressure ($p_{\text{L}}$) is defined as the product of the interfacial tension ($\sigma$) and the curvature of the interface ($\kappa$). As the pinch-off stage is spontaneous, the droplet shape is related to the curvature of the interface formed at the end of the squeezing stage.
The instantaneous Laplace pressure ($p_{\text{L}}$), acting on the neck during the pinch-off stage, is thus analyzed here to gain a further understanding of the droplet formation mechanism.
\fig\ref{fig:13} compares the instantaneous Laplace pressure ($p_{\text{L}}$) evaluated by using the following three approaches.
\begin{enumerate}
\item[I.]
The Laplace pressure ($p_{\text{L}}$) is obtained by measuring the pressure in continuous and dispersed phases by placing the discrete pressure sensors at `cp' and `dp' locations (\eqn\ref{eqn:Pdc}).

\item[II.]
The Young – Laplace (Y-L) equation, based on the radius of the curvature \citep{Volkert2009,Glawdel2012a,Wong2015}, is used to obtain the Laplace pressure ($p_{\text{L}}$) as follow.
\begin{gather}
p_{\text{L}} = \Delta p_{\rc}=\sigma \left[ \frac{1}{R_{\text{f}}}-\left(-\frac{1}{R_{\text{r}}} \right) \right]=\sigma \left( \frac{1}{R_{\text{f}}}+\frac{1}{R_{\text{r}}} \right)
	\label{eqn:PRfRr}
\end{gather}
where, $R_{\text{f}}$ and $R_{\text{r}}$ are the radius of front and rear sides of the interface curvature, as shown in \fig\ref{fig:1c}. Since the rear side of interface curvature points outward, its radius becomes negative, as shown in \eqn(\ref{eqn:PRfRr}). In this work, the numerically data for the interface curvature has been statistically approximated and the radius of interface curvature ($R_{\text{f}}$ and $R_{\text{r}}$) is obtained by using the Taubin's iterative method \citep{ GabrielTaubin1991,Ali2009,Jianfeng2019, NikolaiChernov2021} for robust and stable circle fitting.

\item[III.]
The Young – Laplace (Y-L) equation, based on the minimum local radius of the curvature ($\rcmin$), is used to obtain the Laplace pressure ($p_{\text{L}}$) as follow.
\begin{gather}
	p_{\text{L}} = \Delta p_{\rcmin}=\sigma \left( \frac{1}{\rcmin} \right)
	\label{eqn:PRcmin}
\end{gather}
\end{enumerate}
The numerically data for the interface curvature has been statistically approximated by the most accurate (i.e., 7th order) polynomial curve (refer \figs\ref{fig:3} and \ref{fig:2}) to determine the minimum local radius of the curvature ($\rcmin$).
\noindent
The instantaneous Laplace pressure ($p_{\text{L}}$) calculated using all three approaches (\eqns \ref{eqn:Pdc} to \ref{eqn:PRcmin}) display qualitatively similar trends (\fig\ref{fig:13}) for all values of $\qr$. For instance, $p_{\text{L}}$ increases with time ($t$) and attains highest value at the pinch-off point ($t_3$). Quantitatively, the instantanous values of $p_{\text{L}}$ obtained using second approach (\eqn \ref{eqn:PRfRr}) are comparatively higher than the values  obtained using the first and third approaches (\eqns \ref{eqn:Pdc} and \ref{eqn:PRcmin}).
\begin{figure}[htbp]
	\centering
	\subfloat[$\qr=10$]{\includegraphics[width=0.45\linewidth]{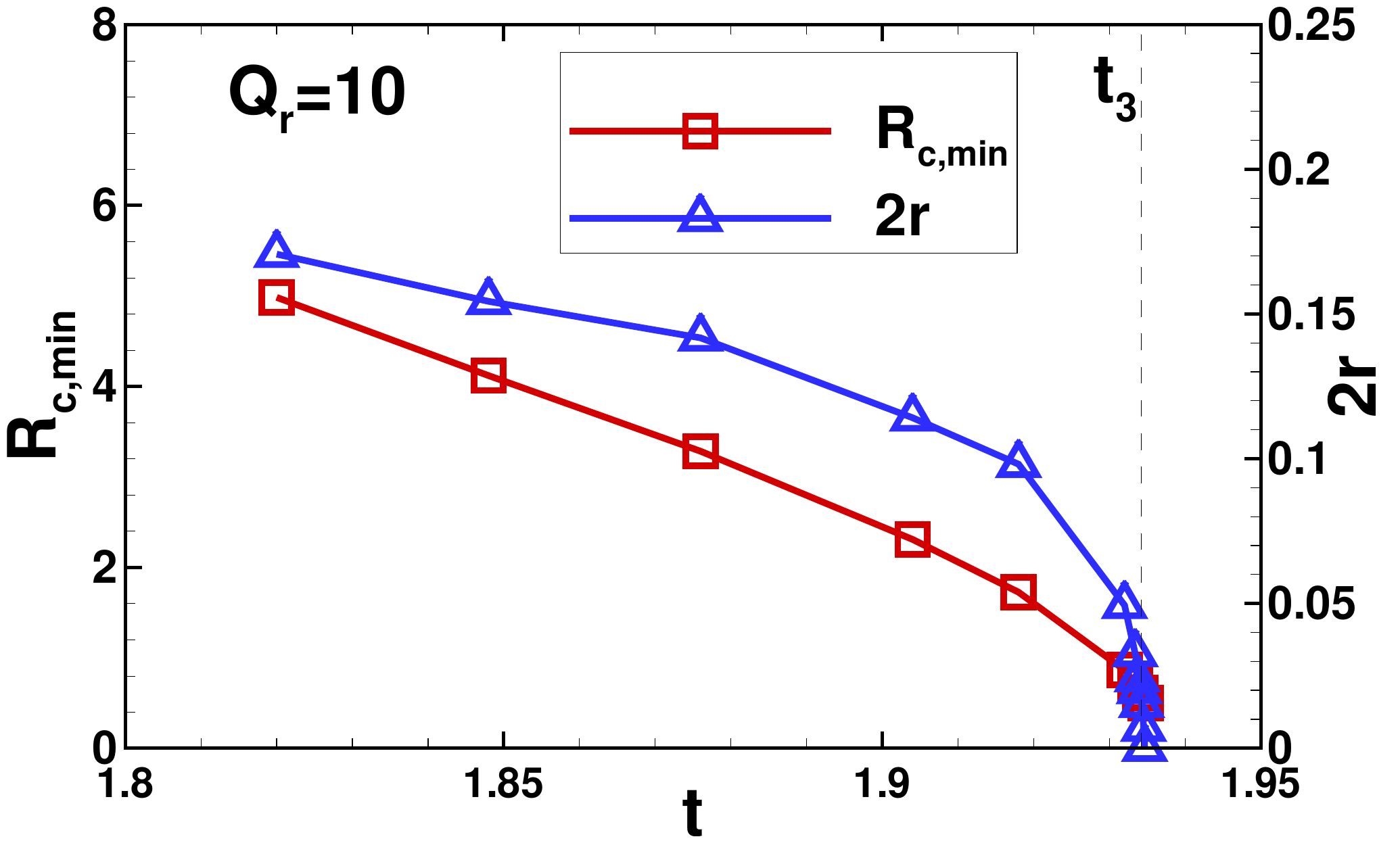}\label{fig:14a}}
	\subfloat[$\qr=5$]{\includegraphics[width=0.45\linewidth]{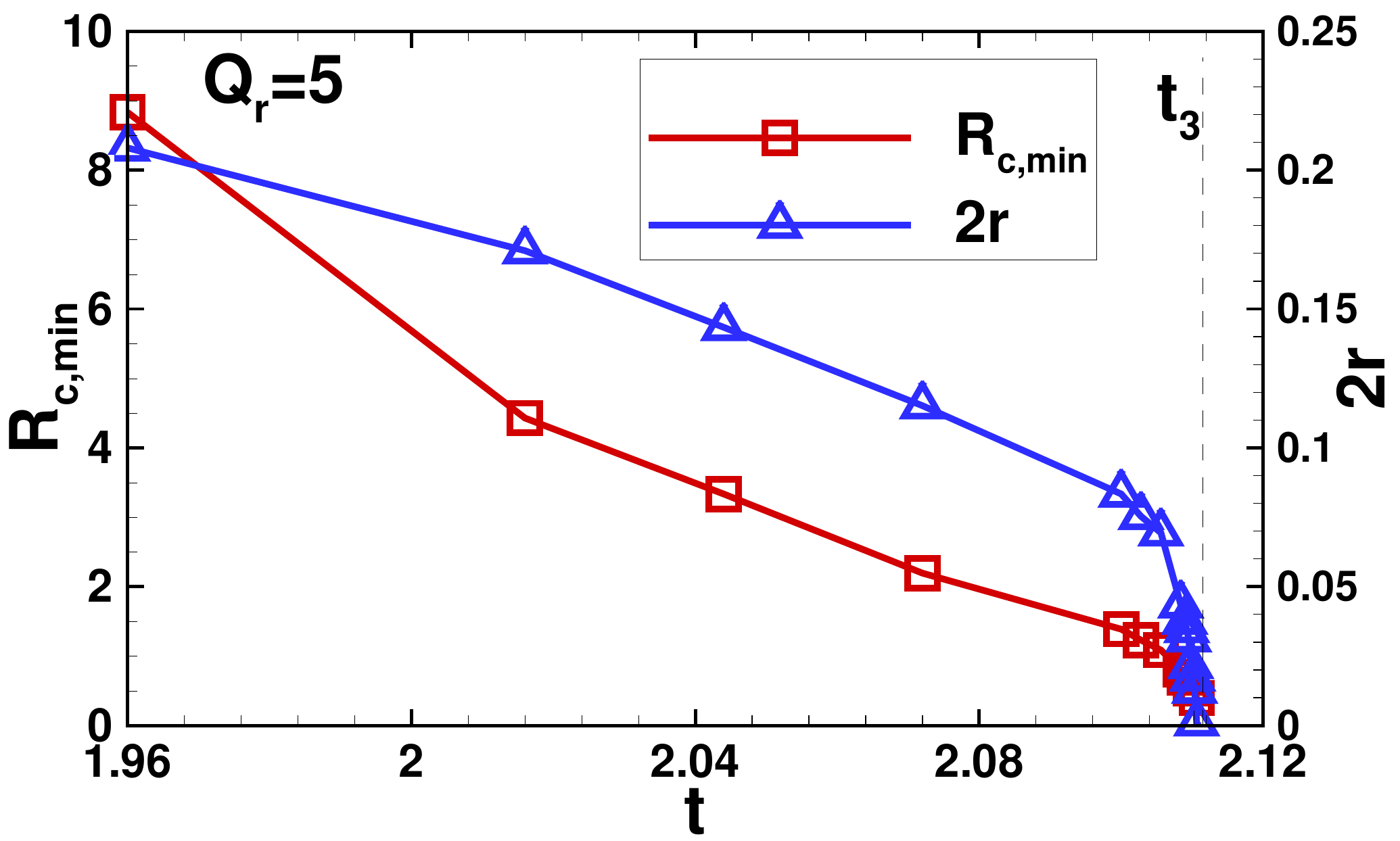}\label{fig:14b}}\\
	\subfloat[$\qr=2$]{\includegraphics[width=0.45\linewidth]{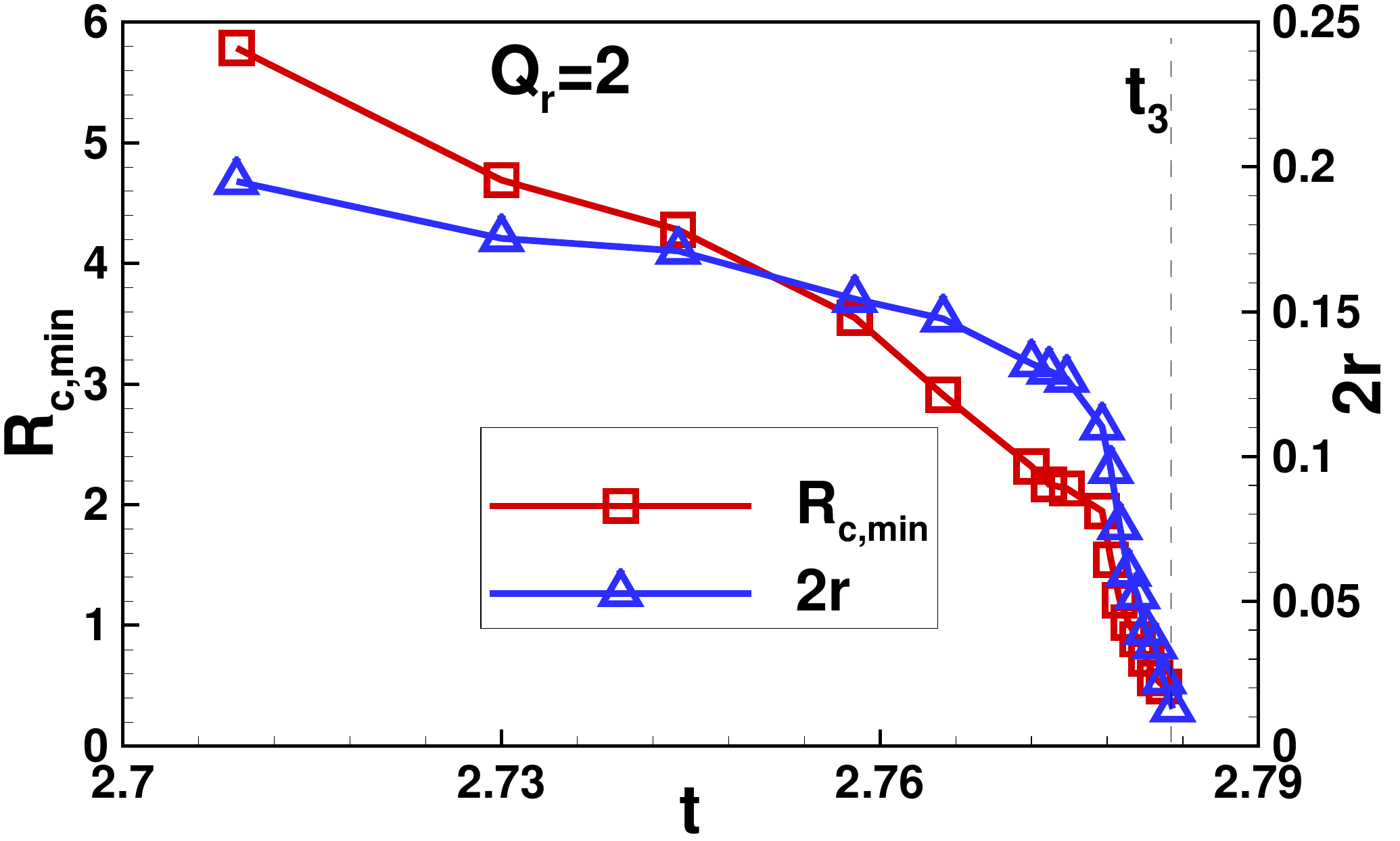}\label{fig:14c}}
	\subfloat[$\qr=1$]{\includegraphics[width=0.45\linewidth]{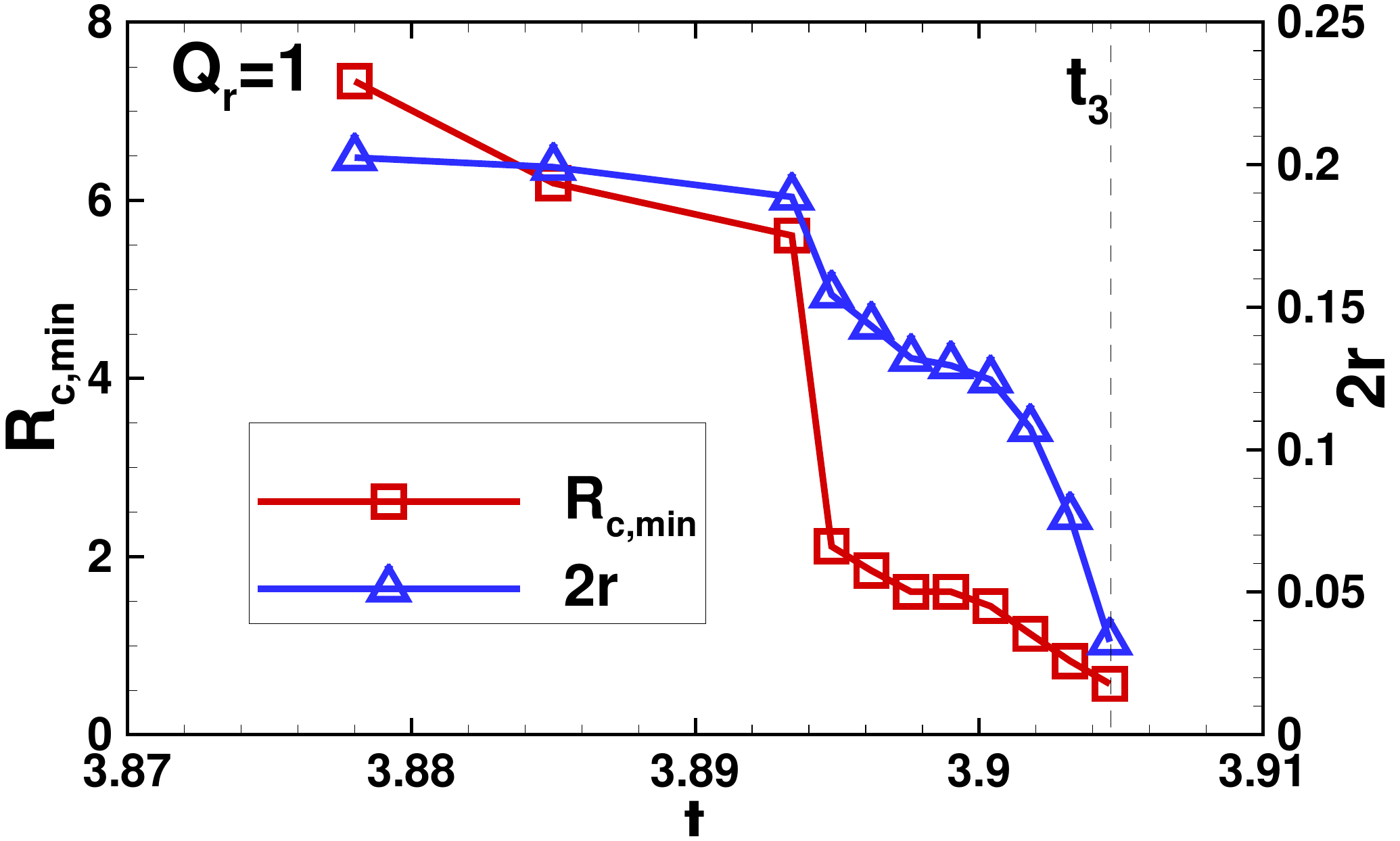}\label{fig:14d}}\\
	\subfloat[$\qr=1/2$]{\includegraphics[width=0.45\linewidth]{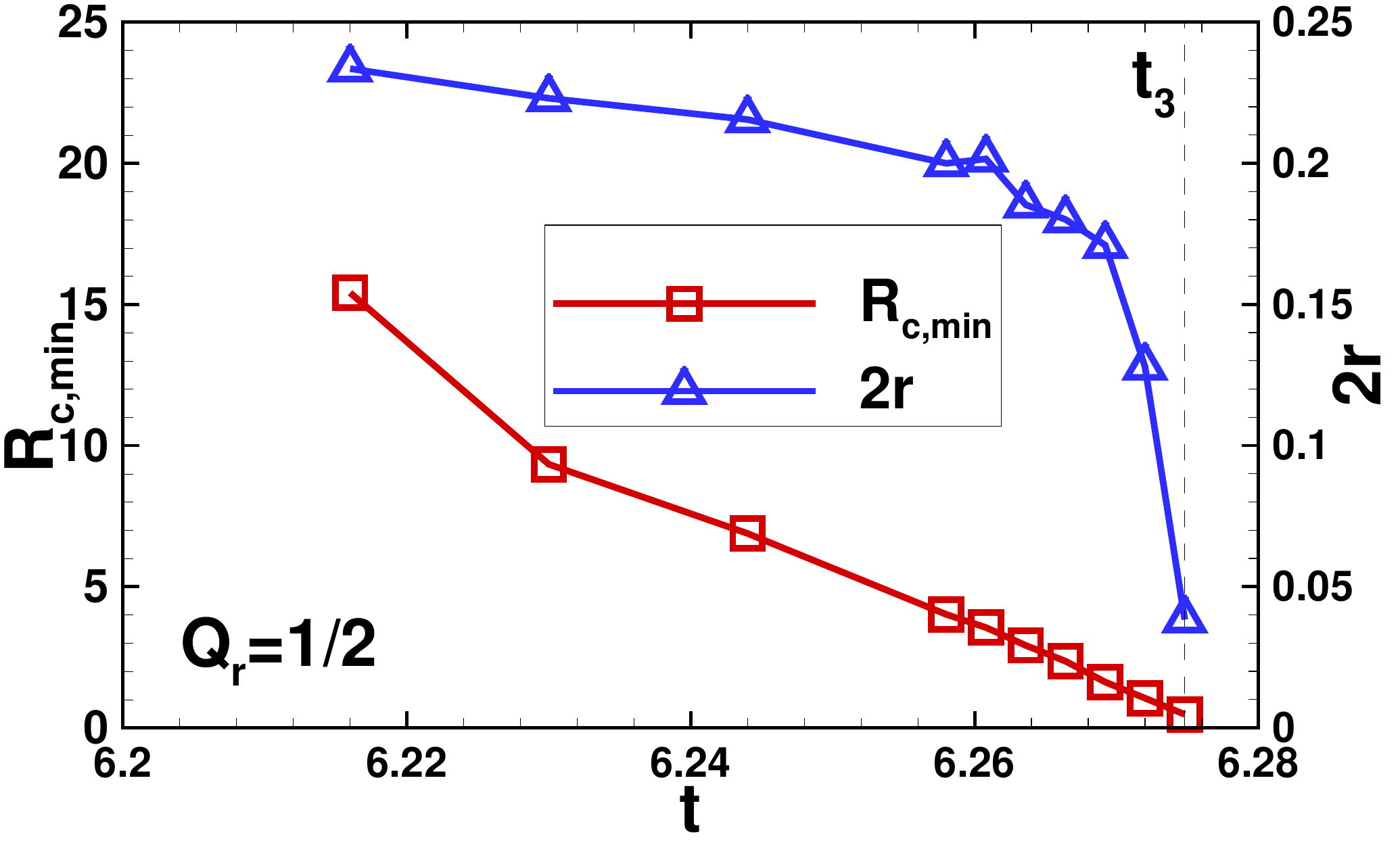}\label{fig:14e}}
	\subfloat[$\qr=1/4$]{\includegraphics[width=0.45\linewidth]{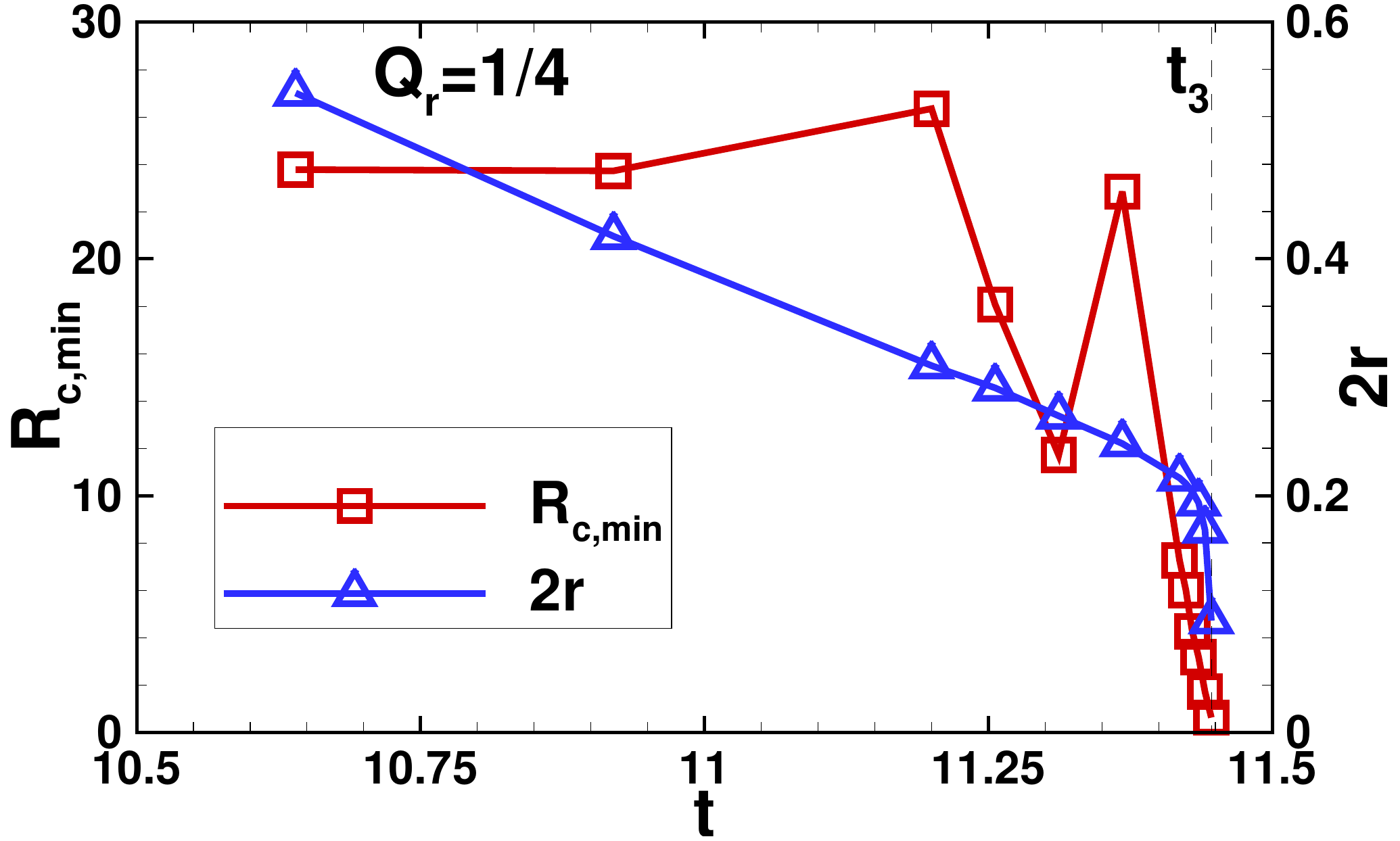}\label{fig:14f}}\\
	\subfloat[$\qr=1/8$]{\includegraphics[width=0.45\linewidth]{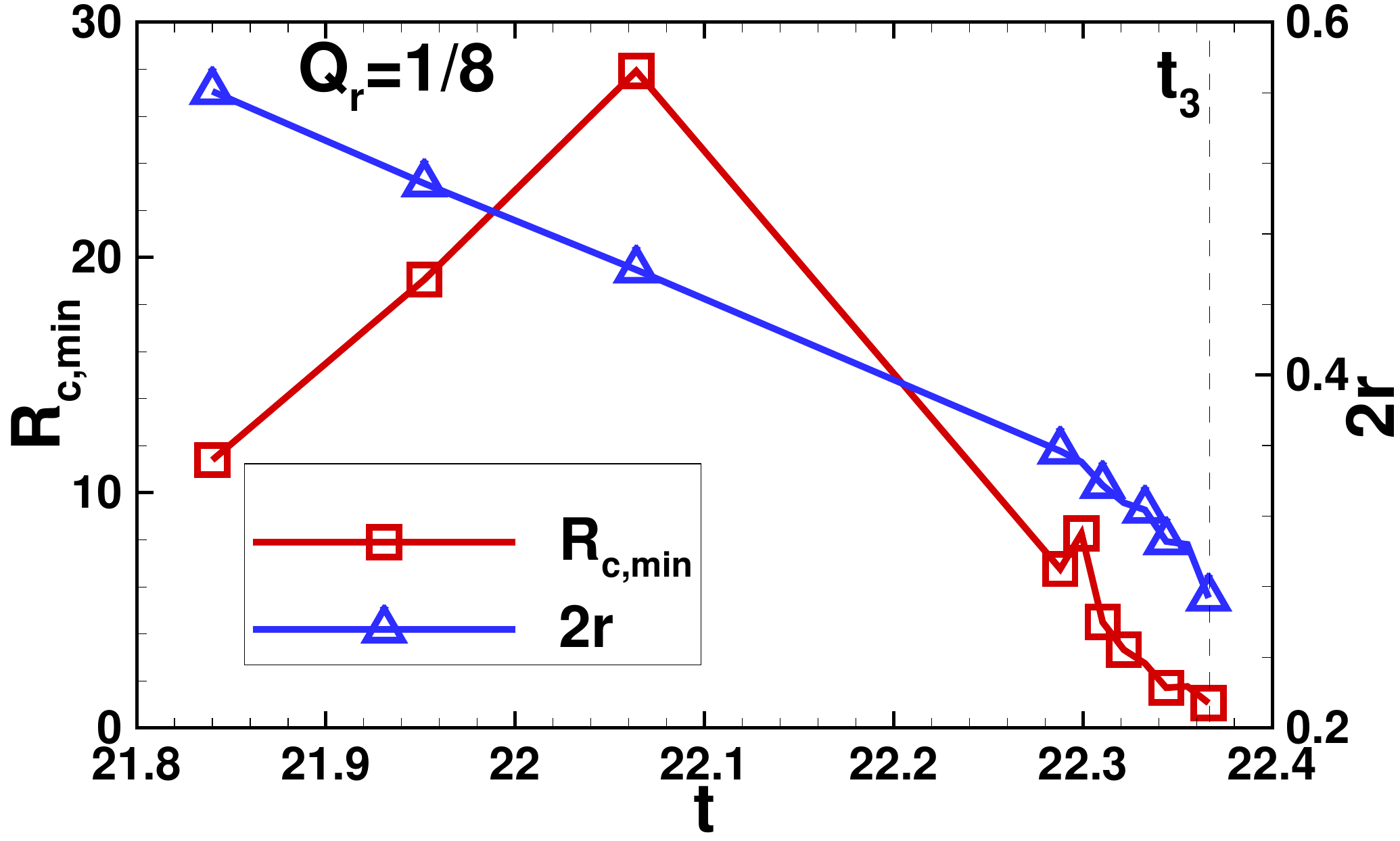}\label{fig:14g}}
	\subfloat[$\qr=1/10$]{\includegraphics[width=0.45\linewidth]{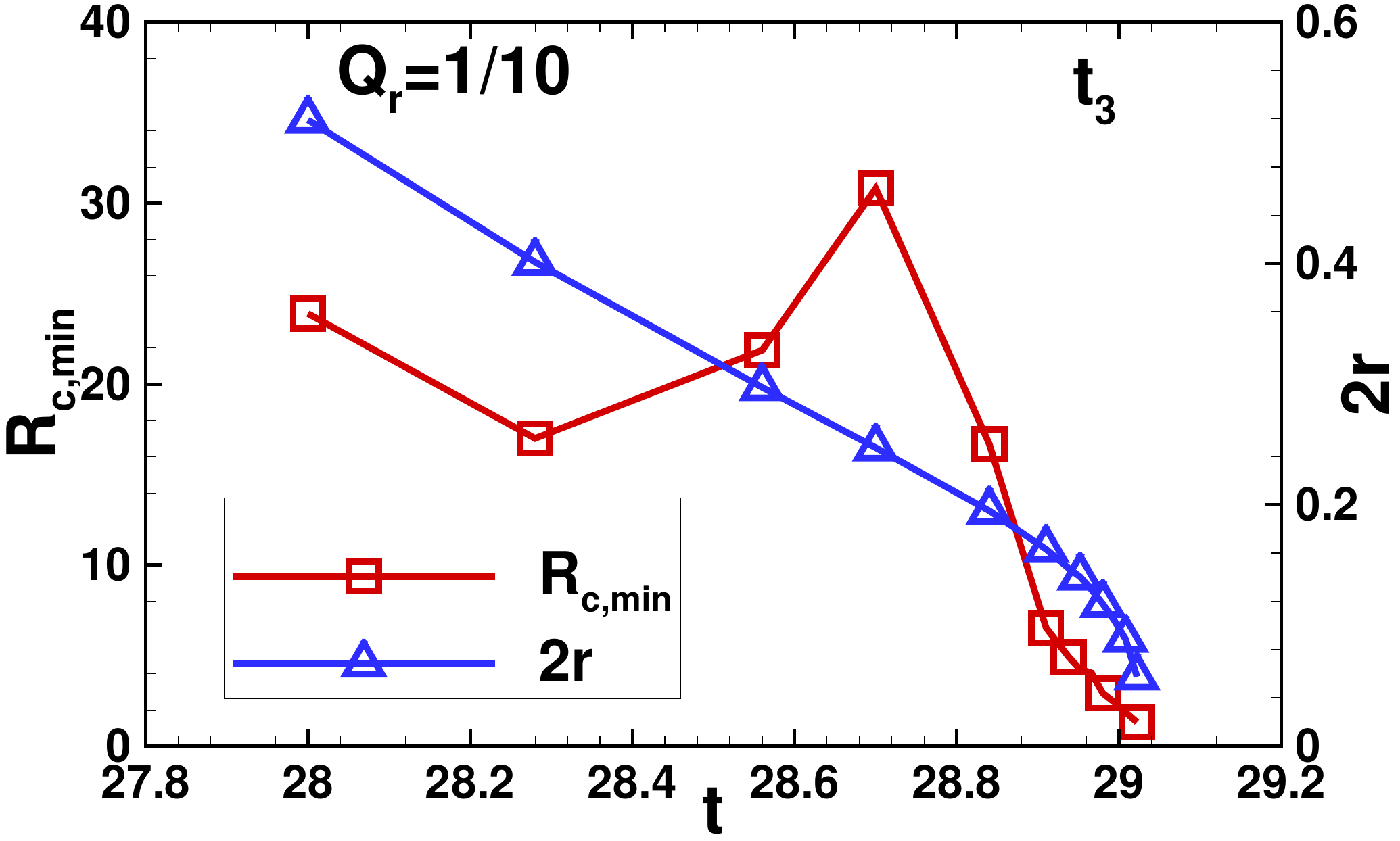}\label{fig:14h}}\\
	\caption{The dynamics of radius of curvature during the pinch-off stage at $Ca_{c}=10^{-4}$.}
	\label{fig:14}
\end{figure}
This deviation is possible because of the evolution of the sharp angled concave shape curvature in the rear side of the interface, and consideration of all points of interface curvature position is quite tricky and impossible, especially at the pinch-off point (refer \fig\ref{fig:8}), in circle fitting using the Taubin's method. It results in smaller radius ($R_{\text{r}}$) of rear interface curvature and thereby larger $p_{\text{L}}$ as front radius $R_{\text{f}}$ is minimally influenced. However, instantaneous Laplace pressure ($p_{\text{L}}$) values obtained using \eqns(\ref{eqn:Pdc}) and (\ref{eqn:PRcmin}) are found to be closer to each other.  It is mainly due to the accurate determination of the local radius of the curvature ($\rcmin$) using the most accurate polynomial fitting approach. Nevertheless, a consistent method can accurately determine the Laplace pressure ($p_{\text{L}}$) either based on the interface evolution profiles or using pressure sensors in both continuous and dispersed phases.

%
%
\noindent
Furthermore, understanding the role of the evolution of the radius of curvature during the droplet pinch-off stage is essential to droplet dynamics. The local minimum radius of curvature ($\rcmin$) and neck width ($2r$) are two critical dynamic characteristics of the interface curvature governing the droplet breakup. \fig\ref{fig:14} depicts the instantaneous profiles of $\rcmin$ (labelled on primary Y-axis) and $2r$  (labelled on secondary Y-axis) for all time ($t$) instants in the droplet break-up stage for $0.1\le \qr\le 10$. Both $\rcmin$ and $2r$ shows, qualitatively, similar trends (i.e., decrease) with time for all $\qr$. Quantitatively, they initially differ by order of magnitude and attain the minimum possible ($\approx 0$) values at the pinch-off time ($t_3$).
As the interface gradually bends to a concave shape with time,  $\rcmin$ decreases solely due to the shear exerted by the continuous phase on the interface.
Notably, $\rcmin$ decreases smoothly and approaches minimum value at the pinch-off time for $\qr \ge 1/2$ (refer \figs\ref{fig:14a}-\ref{fig:14e}). However, there is sudden increase then decrease and sharp fall in $\rcmin$ at the pinch-off point for  $\qr < 1/2$  (refer \figs\ref{fig:14f}-\ref{fig:14h}) because of abrupt change in the interface evolution. On the other hand, the neck width ($2r$) decreases smoothly for all the values of $\qr$.  Nevertheless, at the pinch-off time ($t_3$), both $\rcmin$ and $2r$ achieve the minimum values.
\begin{figure}[t]
	\centering
	\subfloat[$\qr=10$]{\includegraphics[width=0.5\linewidth]{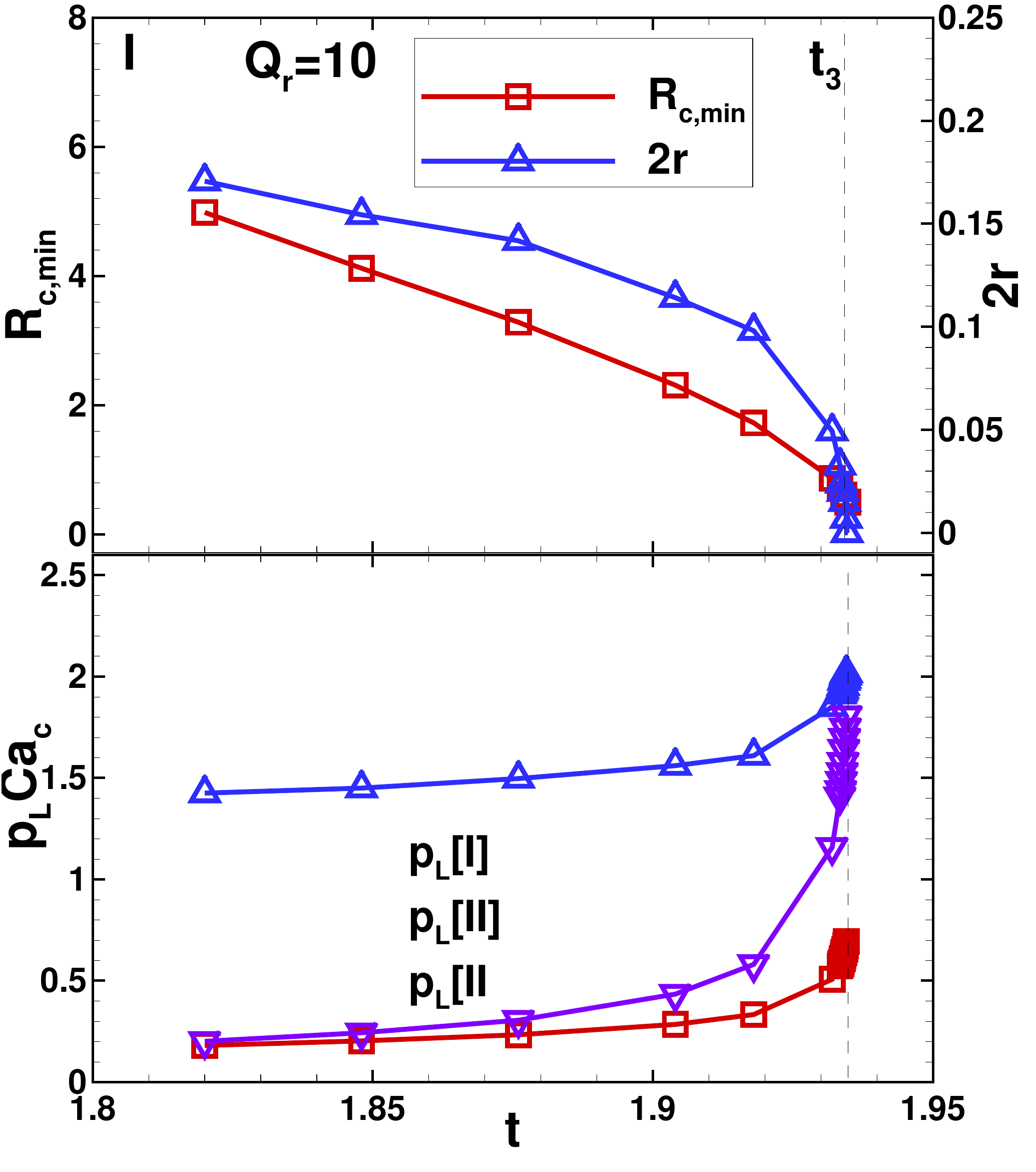}}
	\subfloat[$\qr=1/10$]{\includegraphics[width=0.5\linewidth]{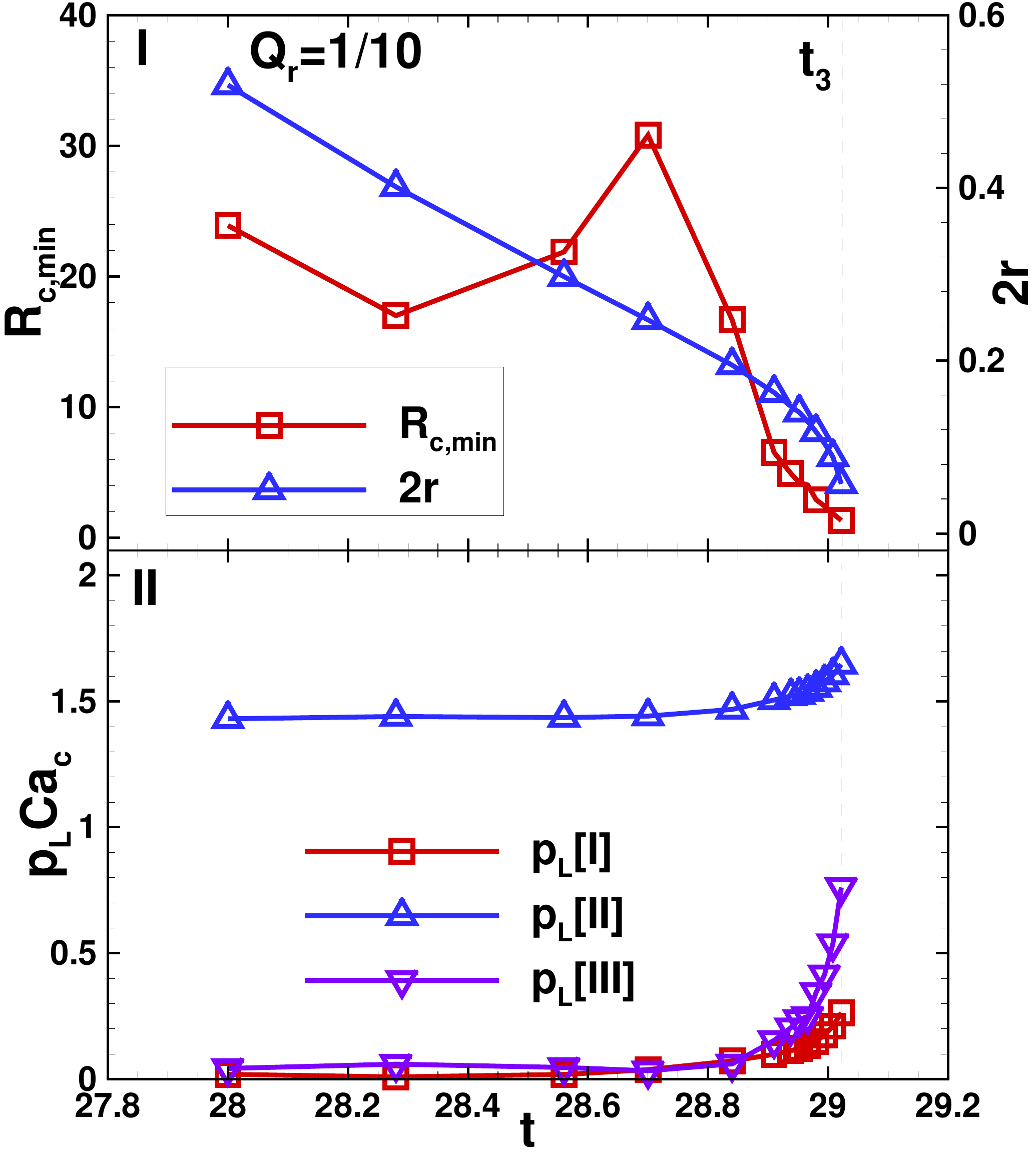}}
	\caption{(I)The dynamics of the radius of curvature during the pinch-off stage, (II) The Laplace pressure acting on the interface curvature during the pinch-off stage at $Ca_{c}=10^{-4}$.}
	\label{fig:15}\label{fig:16}
\end{figure}
It is also noted that the Laplace pressure ($p_{\text{L}}$) obtained using all approaches (\eqns \ref{eqn:Pdc} to \ref{eqn:PRcmin})  is highest at the pinch-off time. The pinch-off point can also be obtained by setting the Laplace pressure ($p_{\text{L}}$), i.e., difference between the pressure acting on the droplet's neck ($R_{\text{r}}$) and the tip ($R_{\text{f}}$), equals to zero.

\noindent
Subsequently, \fig\ref{fig:15} describes the relation of local minimum radius of curvature ($\rcmin$) and neck width ($2r$) with the Laplace pressure ($p_{\text{L}}$) acting on the interface at the pinch-off time ($t_3$). A dashed vertical line (\figs\ref{fig:15}I and \ref{fig:15}II) relates the two features of the droplet pinch-off point.
The value of $p_{\text{L}}$ is responsible for the droplet pinch-off, and the breakup location and time depends upon the values of $\rcmin$ and $2r$. At the pinch-off time ($t_3$), the interface curvature is becoming infinitely large (i.e.,  $R_{\text {r}} \rightarrow 0$) and $\rcmin$ reduces spontaneously and approaches to zero, thereby, results in the pinch-off or breakup of the droplet.
Similarly, the neck width ($2r$) is also approaching zero at the pinch-off point (refer \fig\ref{fig:15}I) and the corresponding Laplace pressure ($p_{\text{L}}$) during the pinch-off stage is shown in \fig\ref{fig:15}II.
The values of  $p_{\text{L}}$ are increasing and sudden shoot up at the pinch-off point as the curvature becomes infinitely large (shown with a dotted line). Therefore, the pinch-off time and location can be known based on the Laplace pressure ($p_{\text{L}}$) acting on the interface curvature by drawing a line as shown in \fig\ref{fig:15}.
\begin{figure}[htbp]
	\centering
	\subfloat[neck width ($2r_{\text{min}}$)]{\includegraphics[width=0.5\linewidth]{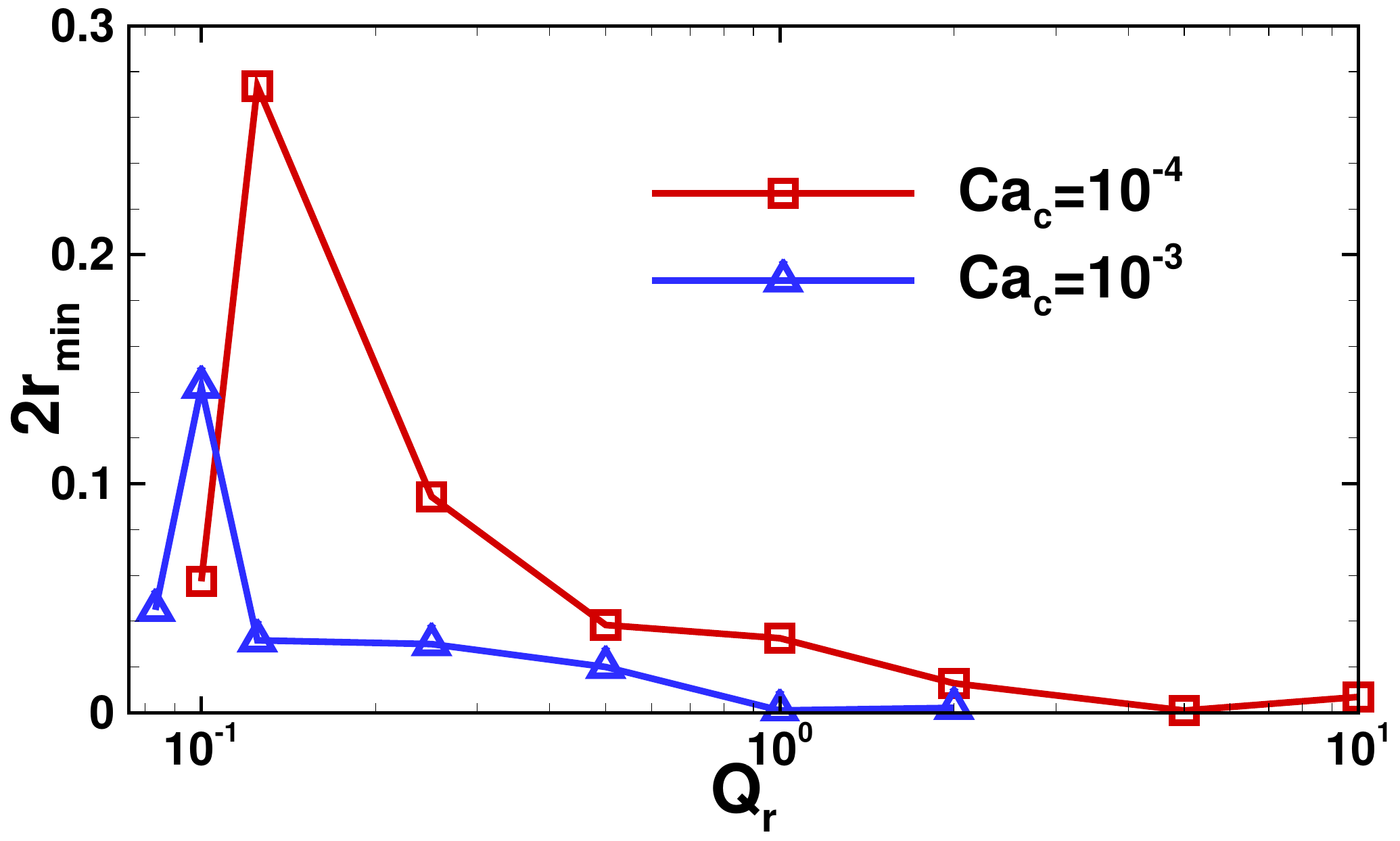}\label{fig:17a}}
	\subfloat[local minimum radius of curvature ($\rcmin$)]{\includegraphics[width=0.5\linewidth]{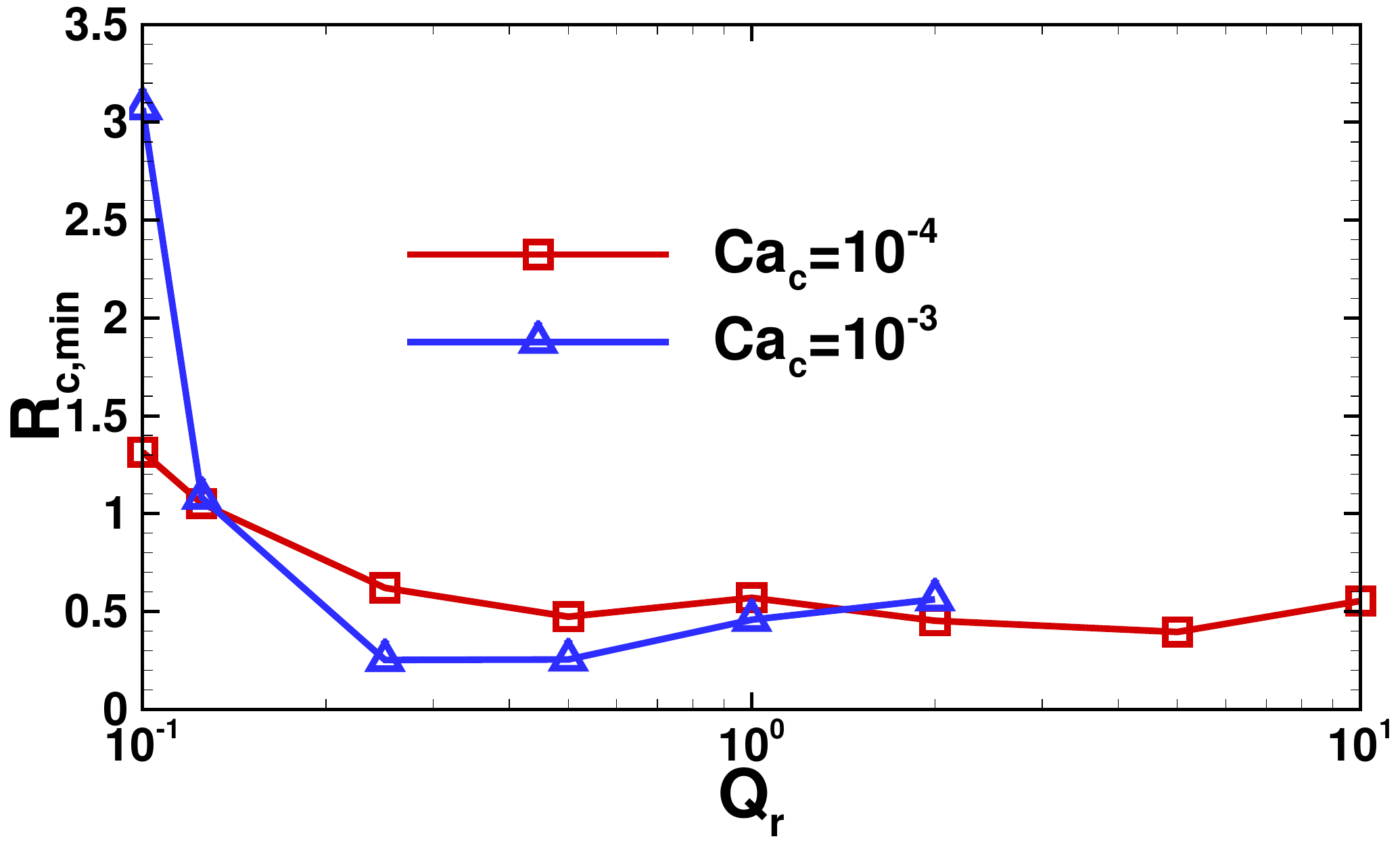}\label{fig:17b}}\\
	\caption{The dynamics of the radius of curvature at the pinch-off point.}
	\label{fig:17}
\end{figure}

\noindent
Finally, a graph is plotted between minimum neck width ($2r_{\text{min}}$) as a function of flow rate ratio ($\qr$) and capillary number ($\cac$) at the pinch-off point ($t_3$) in \fig\ref{fig:17a}. The values of $2r_{\text{min}}$ are found to be inversely proportional to $\qr$ whereas linearly proportional to $\cac$ at pinch-off. The statistical analysis has correlated the numerical data as follows.
\begin{gather}
	\text{At } t_3:\qquad 2r_{{\text{min}}}=A\qr^{B}
	\label{eqn:2rmin}
\end{gather}
where the correlation coefficients are obtained as $A=(0.0165-16.55\cac)$ and $B=-(5235.8\cac+0.872)$ with $R^{2}=0.94$ for $1/8 \leq \qr \leq 10$.
Similarly, a graph is plotted between $\rcmin$, at the pinch-off time, as a function of $\qr$ and $\cac$  is shown in \fig\ref{fig:17b}. A complex relation of $\rcmin$ is observed with $\qr$, whereas it varies linearly with $\cac$. The statistical analysis has resulted in the following correlation to predict $\rcmin$ at the pinch-off point.
\begin{gather}
\text{At } t_3:\qquad \rcmin=A+(B/\qr)+(C/\qr^{1.5})
\label{eqn:Rcmin}
\end{gather}
where, the fitted coefficients are $A=(0.0.4276+631.33 \cac)$, $B=(0.0524-1030.1 \cac)$ and $C=(0.0063+358.82 \cac)$ with $R^{2}=0.96$ for $1/10 \leq \qr \leq 10$.

\noindent
In summary, the dynamics of interface evolution and pinch-off of the droplet in two-phase flow through T-junction microfluidic device are strongly dependent on the flow rate ratio ($\qr$) under otherwise identical conditions ($\cac$, $\theta$, $\rhor$, and $\mur$). Various stages of droplet formation result from the interplay of the responsible local instantaneous forces due to viscous, inertial, and interfacial tension. The channel wall confinement also plays a significant role in the droplet formation process.
\rev{The fundamental insights gained in this work can reliably be used to design the model and prototype the microfluidic devices for droplet generation. The predictive correlations have been proposed for the time required for each droplet stage (\eqns\ref{eqn:tf}-\ref{eqn:tsd}), neck width (\eqn\ref{eqn:2rmin}), and minimum radius of interface curvature (\eqn\ref{eqn:Rcmin}) for their practical use in design and engineering. While the accurate measurements of velocity and pressure fields experimentally are challenging, they are essential in the optimal design of the microfluidic device. The present correlation to predict maximum pressure in the continuous phase (\eqn\ref{eqn:Pmax}) can suitably be used to select the microfluidic pump. The three approaches (\eqns\ref{eqn:Pdc}-\ref{eqn:PRcmin}) explain the droplet breakup mechanism for accurately predicting the pinch-off moment.
Further, the present results have applicability limited to the equal density of the phases, constant physical properties throughout the process, isothermal fluids, similar cross-sectional areas of the inlets and outlet channels, and the fixed contact angle. It nevertheless makes the scope for future investigation.}
	
\section{Concluding remarks}
\noindent
In the present framework, the dynamics of interface evolution and droplet pinch-off in two-phase incompressible flow through T-junction cross-flow microfluidic device have been modelled by the Navier-Stokes equations in conjunction with the conservative level set method. The mathematical model has been solved by using the finite element method for the wide range of the flow rate ratio ($0.1\le\qr\le10$) under the squeezing flow regime ($\cac\rev{<} 10^{-2}$) for \rev{a} fixed contact angle ($\theta=135^{\text{o}}$). The instantaneous cyclic stages of droplet formation are characterized as initial, filling, squeezing, pinch-off and stable droplet. The time required ($t_0$ to $t_4$) for completion of each stage is a complex function of $\qr$ and $\cac$. The evolutions of interface profiles are presented to understand the pinch-off mechanism of the droplet. The interface profiles are fitted with the highest (seventh) order accurate polynomial function to determine the local minimum radius of the rear side interface curvature ($\rcmin$). Taubin's method is used to obtain an instant radius of interface curvature ($R_{\text{f}}$ and $R_{\text{r}}$). Dynamics of the pinch-off stage is gained through Laplace pressure ($p_{\text{L}}$) evaluated using three approaches (a) by measuring pressure in dispersed and continuous phases, (b) by using $R_{\text{f}}$ and $R_{\text{r}}$, and (c) by using $\rcmin$. The relation of $2r$, and $\rcmin$ with $p_{\text{L}}$ has also been established as a function of $\qr$ and $\cac$ at the pinch-off point. 
\rev{The present predictive correlations are developed by using the numerical data obtained with the highly refined mesh (i.e., sufficiently large number of discrete points in the flow domain, $\Delta x\ =\Delta y\ =\ 10\ \mu$m), and the very small ($\Delta t=10 \mu$s) time step.  
Experimentally, it is undoubtedly quite complex and difficult to obtain such an accurate data as it would require a high-speed (capturing speed of $10^5$ fps) and high-resolution (10\ $\mu$m pixel size) digital camera mounted with microscope. Undoubtedly, the measurement of the velocity and pressure profiles experimentally with accuracy is complex. On the other hand, they are essential parameters that help in optimizing the design of microfluidic devices and describing the physics of droplet formation.}

%
\section*{Declaration of Competing Interest}
\noindent
The authors declare that they have no known competing financial interests or personal relationships that could have appeared to influence the work reported in this paper.
\section*{Acknowledgments}
\noindent
R.P. Bharti would like to acknowledge Science and Engineering Research Board (SERB), Department of Science and Technology (DST), Government of India (GoI) for providence of MATRICS grant (File No. MTR/2019/001598).
%
%
{
\noindent 
\nomenclature[z0]{\textit{Abbreviations}}{}
\nomenclature[g0]{\textit{Greek letters}}{}
\nomenclature[d0]{\textit{Dimensionless groups}}{}
%
%
 \nomenclature[zbdf]{BDF}{backward differentiation formula}
 \nomenclature[zcfd]{CFD}{computational fluid dynamics}
 \nomenclature[zcp]{CP}{continuous phase}
 \nomenclature[zdp]{DP}{disperse phase}
 \nomenclature[zdae]{DAE}{differential algebraic equations}
 \nomenclature[zlsm]{LSM}{level set method}
 \nomenclature[zfem]{FEM}{finite element method}
%
%
\nomenclature[aDt]{$\mathbf{D}$}{rate of strain tensor (\eqn\ref{eq:tauD}), s$^{-1}$}
\nomenclature[aFsigma]{$\mathbf{F}_{\sigma}$}{interfacial tension force (\eqn\ref{eq:ift}), N}
\nomenclature[aFi]{${F}_{\text{i}}$}{magnitude of inertial force, N} 
\nomenclature[aFv]{${F}_{\text{v}}$}{magnitude of viscous force, N} 
\nomenclature[aFs]{${F}_{\sigma}$}{magnitude of interfacial tension force, N} 
\nomenclature[aLu]{$L_{\text{u}}$}{upstream length of the main channel, m} 
\nomenclature[aLd]{$L_{\text{d}}$}{downstream length of the main channel, m}
\nomenclature[aLm]{$L_{\text{m}}$}{length of the main channel, m}
\nomenclature[aLs]{$L_{\text{s}}$}{length of the side channel, m}
\nomenclature[aP]{$p$}{pressure, Pa} 
\nomenclature[aPcp]{$p_{\text{cp}}$}{pressure in CP at point `cp', -} 
\nomenclature[aPcp]{$p_{\text{dp}}$}{pressure in DP at point `dp', -} 
\nomenclature[aPl]{$p_{\text{L}}$}{Laplace pressure, -} 
\nomenclature[aQc]{$Q_{\text{c}}$}{flow rate of CP, m$^{3}$/s}
\nomenclature[aQd]{$Q_{\text{d}}$}{flow rate of DP, m$^{3}$/s}
\nomenclature[aQr]{$Q_{\text{r}}$}{flow rate ratio (\eqn\ref{eq:8}), -}
\nomenclature[atf]{$t_{\text{f}}$}{filling (S-1) stage time, -}
\nomenclature[ats]{$t_{\text{s}}$}{squeezing (S-2) stage time, -}
\nomenclature[atb]{$t_{\text{b}}$}{breakup (S-3) stage time, -}
\nomenclature[atsd]{$t_{\text{sd}}$}{stable droplet (S-4) stage time, -}
\nomenclature[aRec]{$Re_{\text{c}}$}{Reynolds number for CP (\eqn\ref{eq:8}), -}
\nomenclature[aCac]{$Ca_{\text{c}}$}{capillary number for CP (\eqn\ref{eq:8}), -}
\nomenclature[aU]{$\mathbf{u}$}{velocity vector, m/s}
\nomenclature[awc]{$w_{\text{c}}$}{width of the main channel, m}
\nomenclature[awd]{$w_{\text{d}}$}{width of the side channel, m}
\nomenclature[awr]{$w_{\text{r}}$}{channel width ratio (\eqn\ref{eq:8}), -}
\nomenclature[a2r]{$2r$}{neck width, -}
\nomenclature[aRcmin]{$\rcmin$}{local minimum radius of curvature, -}
\nomenclature[ax]{$x$}{stream-wise coordinate}
\nomenclature[ay]{$y$}{transverse coordinate}
%
%
\nomenclature[ggamma]{$\gamma$}{re-initialization or stabilization parameter  (\eqn\ref{eqn:lsm}), m/s}
\nomenclature[gepsilon]{$\epsilon_{\text{ls}}$}{interface thickness controlling parameter  (\eqn\ref{eqn:lsm}), m}
\nomenclature[gmuc]{$\mu_{\text{c}}$}{viscosity of CP, Pa.s}
\nomenclature[gmud]{$\mu_{\text{d}}$}{viscosity of DP, Pa.s}
\nomenclature[gmur]{$\mu_{\text{r}}$}{viscosity ratio (\eqn\ref{eq:8}), -}
\nomenclature[grhoc]{$\rho_{\text{c}}$}{density of CP, kg/m$^3$}
\nomenclature[grhod]{$\rho_{\text{d}}$}{density of DP, kg/m$^3$}
\nomenclature[grhor]{$\rho_{\text{r}}$}{density ratio (\eqn\ref{eq:8}), -}
\nomenclature[gsigma]{$\sigma$}{interfacial tension, N/m}
\nomenclature[gtau]{$\tau$}{extra stress tensor (\eqn\ref{eq:tauD}),  N/m$^2$}
\nomenclature[gphi]{$\phi$}{level set function, dimensionless}
\nomenclature[gkappa]{$\kappa$}{curvature of the interface, m}
\nomenclature[gtheta]{$\theta$}{contact angle, degrees}
\nomenclature[gdelma]{$\delta_{\text{max}}$}{maximum percent relative error, -}
\nomenclature[gdelmi]{$\delta_{\text{min}}$}{minimum percent relative error, -}
\nomenclature[gdelav]{$\delta_{\text{avg}}$}{average percent relative error, -}
%
%
\nomenclature[dRe]{$Re$}{Reynolds number (\eqn\ref{eq:8}), -}
\nomenclature[dCa]{$Ca$}{Capillary number (\eqn\ref{eq:8}), -}
%
%
\renewcommand{\nompreamble}{\vspace{1em}\fontsize{10}{8pt}\selectfont}
{\printnomenclature[5em]}}
%
%
\bibliographystyle{elsarticle-harv}
\biboptions{authoryear}
%
\bibliography{references}
%
%
%
%
%
\addcontentsline{toc}{section}{References}
\clearpage
%
%
%
%
%
%
%
%
%
\end{document}